\documentclass[12pt]{article}
\pdfoutput=1

\usepackage{putex}

\usepackage{cancel}
\usepackage{simplewick}
\usepackage{multirow}
\usepackage{comment}

\usepackage{graphicx}
\usepackage{caption}
\usepackage{amsmath}
\usepackage{array}
\usepackage{subcaption}
\usepackage{epstopdf}
\usepackage{enumerate}
\usepackage{cite}
\usepackage{youngtab}
\usepackage{tensor}
\usepackage{slashed}
\usepackage[aligntableaux=center]{ytableau}
\usepackage[utf8]{inputenc}
\usepackage{rotating}
\usepackage{bigfoot}

\usepackage[
      colorlinks=true,
      linkcolor=blue,
      urlcolor=blue,
      filecolor=black,
      citecolor=red,
      ]{hyperref}

\usepackage[compat=1.1.0]{tikz-feynman}
\usetikzlibrary{decorations.markings}

\numberwithin{equation}{section}

\def \( {\left(}
\def \) {\right)}
\def \< {\left<}
\def \> {\right>}
\def \eps {\varepsilon}

\newcommand{\be}{\begin{equation}} \newcommand{\ee}{\end{equation}}

\newcommand\tpind[5]{\langle {\cal O}_{#1}^{#4}\left(#2\right) {\cal O}_{#1}^{#5}\left(#3\right) \rangle}
\newcommand\tp[3]{\tpind{#1}{#2}{#3}{}{}}
\newcommand\thpind[7]{\langle {\cal O}_{#1}\left(#2\right) {\cal O}_{#3}\left(#4\right) {\cal O}_{#5}^{#7}\left(#6\right) \rangle}
\newcommand\thp[6]{\thpind{#1}{#2}{#3}{#4}{#5}{#6}{}}

\newcommand\thpnormind[5]{\langle {\cal O}_{\Delta_0}\left(#1\right) \hat{\cal O}_{\Delta_0}\left(#2\right) {\cal O}_{#3}^{#5}\left(#4\right) \rangle}
\newcommand\thpnorm[4]{\thpnormind{#1}{#2}{#3}{#4}{}}
\newcommand\thptildind[5]{\langle {\cal O}_{\tilde\Delta_0}\left(#1\right) \hat{\cal O}_{\tilde\Delta_0}\left(#2\right) {\cal O}_{#3}^{#5}\left(#4\right) \rangle}
\newcommand\thptild[4]{\thptildind{#1}{#2}{#3}{#4}{}}

\tikzfeynmanset{
	O/.style = {very thick}
}
\tikzfeynmanset{
	Ot/.style = {scalar}
}
\tikzfeynmanset{
	Th/.style = {double distance=2pt}
}
\tikzfeynmanset{
	dO/.style = {fermion}
}

\newcommand{\es}[2] {\begin{equation} \label{#1} \begin{split} #2 \end{split} \end{equation}}

\begin{document}

\institution{WZ}{Department of Particle Physics and Astrophysics, Weizmann Institute of Science, Rehovot, Israel}

\title{A Derivation of AdS/CFT for Vector Models}

\authors{Ofer Aharony\worksat{\WZ}\footnote{e-mail: {\tt ofer.aharony@weizmann.ac.il}}, Shai M.~Chester\worksat{\WZ}\footnote{e-mail: {\tt iahs81@gmail.com}} and Erez Y.~Urbach\worksat{\WZ}\footnote{e-mail: {\tt erez.urbach@weizmann.ac.il}}}

\abstract{
We explicitly rewrite the path integral for the free or critical $O(N)$ (or $U(N)$) bosonic vector models in $d$ space-time dimensions as a path integral over fields (including massless high-spin fields) living on ($d+1$)-dimensional anti-de Sitter space. Inspired by de Mello Koch, Jevicki, Suzuki and Yoon and earlier work, we first rewrite the vector models in terms of bi-local fields, then expand these fields in eigenmodes of the conformal group, and finally map these eigenmodes to those of fields on anti-de Sitter space. Our results provide an explicit (non-local) action for a high-spin theory on anti-de Sitter space, which is presumably equivalent in the large $N$ limit to Vasiliev's classical high-spin gravity theory (with some specific gauge-fixing to a fixed background), but which can be used also for loop computations. Our mapping is explicit within the $1/N$ expansion, but in principle can be extended also to finite $N$ theories, where extra constraints on products of bulk fields need to be taken into account.
}
\date{November 2020}

\maketitle

\tableofcontents

\section{Introduction and Summary}
\label{intro}

The AdS/CFT correspondence \cite{Maldacena:1997re,Witten:1998qj,Gubser:1998bc} is a general relation between conformal field theories (CFTs) in $d$ space-time dimensions and quantum gravity theories on $(d+1)$-dimensional asymptotically anti-de Sitter (AdS) space-times. For $d>1$ the quantum gravity theories do not have any independent non-perturbative definition, so the correspondence should be viewed as defining them in terms of the corresponding CFT. When the CFT has an appropriate large $N$ limit, its $1/N$ expansion may be identified with a perturbative expansion of the bulk theory, such that this theory is classical in the large $N$ limit. One can then check the correspondence by comparing this perturbative expansion of the bulk theory (which is well-defined, for instance, if the bulk is a superstring theory) with the $1/N$ expansion of the CFT. This perturbative expansion starts with some classical solution of the bulk theory (this can be a solution of string theory, or, in an appropriate limit, a solution of its supergravity approximation) and quantizes the theory around it. Many checks of the correspondence have been done within this expansion (see, for instance, \cite{Chester:2019jas,Binder:2019mpb,Gromov:2009zb}), and in special cases it has even been proven to all orders in this expansion \cite{Eberhardt:2019ywk,Dei:2020zui,Eberhardt:2020akk}.

Beyond perturbation theory in $1/N$, the bulk theory is implicitly defined by the CFT, but it would be nice to understand what this definition means for the bulk fields. Is there still for finite $N$ any description of the theory as a sum over bulk gravity configurations, or is the gravity description an effective description that is only valid within the context of the $1/N$ expansion? Do off-shell gravitational configurations have any meaning? Even in the semi-classical limit, it is not clear if we need to sum over all possible gravitational solutions with appropriate boundary conditions, or if there are some limitations. This question arises, for instance, when considering a CFT on a disconnected space-time, where in some cases gravitational solutions connecting two components of the space-time through the bulk exist, but including them in the partition function would contradict locality in the CFT; in the $d=1$ case such configurations are included on the (well-defined) $2d$ gravity side, related to the fact that the dual CFT is an ensemble of theories rather than a specific CFT \cite{Saad:2019lba,Almheiri:2019qdq}, but the situation in higher dimensions is not clear.

In this paper we would like to suggest answers to these questions, in the context of one of the simplest examples of the AdS/CFT correspondence, the duality between bosonic vector models (with $d>2$) and high-spin gravity theories \cite{Klebanov:2002ja}. The CFT in this case is just the theory of $N$ complex scalar fields $\phi_I$ ($I=1,\cdots,N$), with a projection to $U(N)$-invariant states and operators (alternatively one can consider $N$ real scalar fields $\varphi_I$ and project to $O(N)$ invariants). In general it is subtle to impose this projection while keeping the CFT local (though for $d=3$ this can be done by coupling to a $U(N)$ Chern-Simons theory at infinite level \cite{Giombi:2009wh,Giombi:2011kc,Aharony:2011jz}), but in this paper we focus just on correlation functions in flat space, where this subtlety does not arise. We will map the $U(N)$-singlet sector of these CFTs to bulk fields (with all integer spins) living on a fixed AdS space, obtaining a description of the theory as a path integral over these bulk fields. Within the $1/N$ expansion we will write down an explicit (albeit non-local) action for the bulk theory, and we will argue that it allows for perturbative computations, with all loop divergences canceling when appropriate counter-terms (which we write down explicitly) are introduced. For finite $N$ we will argue that the theory can still be written as a path integral over the same bulk fields, but that these fields obey complicated non-local constraints involving products of $(N+1)$ bulk fields, such that most configurations of the bulk fields are not included in the path integral (indeed, even anti-de Sitter space itself turns out not to be one of the allowed bulk configurations for finite $N$). Interestingly, we do not see any sign of a sum over different topologies in the bulk, but only of continuous field configurations (including fluctuations of the metric) on a fixed anti-de Sitter space.

We focus on studying this case of free CFTs (and their deformations) since it is the simplest case on the field theory side, where we can easily perform explicit computations. In this case the dual gravitational theories are only understood classically, where they are believed to be described by the equations of motion of Vasiliev's high-spin gravity \cite{Vasiliev:1990en,Vasiliev:1992av,Vasiliev:1995dn}. No action is known for these theories,\footnote{See \cite{Bekaert:2014cea,Bekaert:2016ezc,Skvortsov:2018uru,Sleight:2016dba} for attempts to compute some terms in the action. It is not clear if Vasiliev's equations of motion are really well-defined, since infinities show up when performing various (classical) computations \cite{Boulanger:2015ova,Sleight:2017pcz}; we will not discuss these issues here. We thank Evgeny Skvortsov for discussions on these issues.} which would allow one to compare the partition functions of different solutions, or to compute loop corrections by summing over off-shell modes. Thus, an additional motivation to study this case is that our formalism provides a bulk action for these gravitational theories that allows these computations to be performed, and thus constructs a quantum completion for high-spin gravity theories. Vasiliev's equations have a huge gauge symmetry, while our bulk fields live on a fixed anti-de Sitter space and have no gauge redundancy. We show that the physical degrees of freedom following from our action match those of Vasiliev's high-spin gravity. We believe, though we have not shown this, that there is a specific gauge-fixing of Vasiliev's equations which will reduce them to the classical equations of motion that follow from our bulk action. Note that the huge gauge symmetry of Vasiliev's equations allows them to be written in a formalism which has a fixed background metric \cite{Neiman:2015wma}, as we have.

The formalism we use to construct the bulk theory is based on the idea of bi-local holography \cite{Das:2003vw,Koch:2010cy,Koch:2014aqa,Koch:2014mxa,deMelloKoch:2012vc}, and in particular inspired by the recent paper \cite{deMelloKoch:2018ivk}. 
We begin in section \ref{sec:local_to_bilocal} by changing variables in our CFT path integral from the fundamental fields $\phi_I(x)$ to the bi-local $U(N)$-invariant fields 
\be \label{bi_local}
G(x_1,x_2) \equiv \frac{1}{N} \sum_{I=1}^N \phi_I^*(x_1) \phi_I(x_2).
\ee
Correlation functions of these fields capture all the $U(N)$-invariant information of the theory, so that we do not lose any information (or any $U(N)$-invariant deformations) by changing to these variables. Performing this change of variables requires introducing a cutoff in the field theory, which can be thought of as having a finite number $V$ of space-time points. In the large $N$ limit $N\gg V$ we can perform the change of variables explicitly, and most of our paper will focus on this limit and on the resulting $1/N$ expansion. For finite $N$ in the continuum limit, the change of variables can still be performed but the bi-local variables obey complicated constraints, that will map in our formalism to complicated constraints on the bulk fields.

In order to map the bi-local fields to the bulk, we first expand them in a basis of eigen-modes of the conformal group $SO(d+1,1)$ (we work in Euclidean space throughout this paper for simplicity), given by three-point functions $\thpnorm{x_1}{x_2}{\Delta,J}{y}$ (where the first two operators are scalar operators with the free field dimension $\Delta_0=(d-2)/2$, and the last one has dimension $\Delta$ and spin $J$) \cite{Dobrev:1977qv,Dobrev:1976vr}. This is the topic of section \ref{sec:bi_local_to_conformal}, where we rewrite our CFT action in variables $C_{\Delta,J}(y)$ which are the coefficients of these eigen-modes. In section \ref{sec:mapping_to_the_bulk} we show (see also \cite{deMelloKoch:2018ivk}) that modes with precisely the same quantum numbers arise when considering symmetric traceless transverse fields $\Phi_J(X)$ of spin $J$ in $AdS_{d+1}$, with all non-negative-integer spins $J$. In this case the eigen-modes are bulk-to-boundary propagators $G_{\Delta,J}(X,y)$ \cite{Costa:2014kfa}, connecting a point $X$ in the bulk of AdS space with a point $y$ on its boundary. It is then natural to identify the coefficients $C_{\Delta,J}(y)$ of the eigen-modes in the CFT with the coefficients $C_{\Delta,J}^{bulk}(y)$ of the bulk fields, up to a possible constant $f_{\Delta,J}$ (which can depend on the mode). This gives us an explicit linear mapping between the fluctuation $\eta(x_1,x_2)$ around the bi-local field \eqref{bi_local} and the bulk fields $\Phi_J$:\footnote{A similar linear mapping was suggested in \cite{deMelloKoch:2018ivk}, based on earlier works on the bi-local formalism. However, the mapping was only presented explicitly there for the case of $d=2$, for which our analysis does not hold because the free scalar $\phi_I$ is not a primary field in this case. Many of our results are closely related to those of \cite{deMelloKoch:2018ivk}, so one can view our formalism as an explicit realization of their framework for $d>2$.}
\es{introMap}{
\Phi_J(X)=\int_{\gamma_J}\frac{d\Delta}{2\pi i}\frac{f_{\Delta,J}}{N_{\Delta,J}}\int d^dy G_{\Delta,J}(X,y) \int d^dx_1d^dx_2 \eta(x_1,x_2) \thptildind{x_1}{x_2}{\tilde\Delta,J}{y}\,,
}
where $\tilde\Delta\equiv d-\Delta$, the normalization constant $N_{\Delta,J}$ will be defined in \eqref{eq:CFT_norm}, we suppressed the spin indices for simplicity, and in most cases the contour $\gamma_J$ goes over the principal series $\Delta=\frac{d}{2}+is$ (the precise contour $\gamma_J$ will be described in the main text)\footnote{
Note that this mapping is off-shell and exact, and that it involves bi-local CFT operators; this is very different from the HKLL-type mapping \cite{Hamilton:2006az} which writes (on-shell) bulk fields in terms of local CFT operators.}.
We can similarly write an inverse map from $\Phi_J(X)$ to $\eta(x_1,x_2)$.

We can then use this map to write down the bulk action for the higher spin fields, which turns out to be explicitly non-local. In section \ref{sec:quadratic_action} we show that for a specific choice of the constant in the mapping we can obtain a local quadratic term in the bulk action, albeit one that is quartic in derivatives, of the schematic form
\begin{equation}\label{eq:bulk_S2_local_schematic}
\begin{split}
	S^{(2)}_\text{local}[\Phi_J] \simeq \sum_{J=0}^\infty &
	\int dX \Phi_{J}(X) \left(\nabla_{X}^{2}-M_{d+J-2,J}^{2}\right)\left(\nabla_{X}^{2}-M_{d+J,J}^{2}\right) \Phi_{J}(X)\,,
\end{split}
\end{equation}
where $M_{\Delta,J}^2 \equiv \Delta (\Delta-d) - J$.
Quantizing this action leads to two particles of each spin $J$ in the bulk. One of these has a positive propagator and matches with the expected physical particles in the bulk theory (including massless high-spin particles, that are dual to the conserved currents with $J=1,2,3,\cdots$ of the free CFT). The other has a negative propagator, and we interpret it as an unphysical `ghost' mode (in some cases we can show that the quantum numbers of these `ghosts' match with the ones arising from a specific gauge-fixing of Vasiliev's equations of motion). 

In section \ref{sec:bulk_theory} we discuss the Feynman rules for performing computations with our bulk action, and argue that all loop divergences cancel so that we can perform computations to all orders in $1/N$. In principle this is automatic in our formalism since we directly map the field theory to the bulk, so we are guaranteed to get finite results to all orders in $1/N$ which agree with the CFT, but one has to be careful about several regularizations that are needed in order to get sensible results.

In the final two sections we apply our formalism to two interesting deformations of the CFT. In section \ref{sec:massDef} we discuss a mass deformation where we give a mass to the fields $\phi_I$, such that the field theory is still free but is no longer conformal. We show that in our formalism there is a corresponding classical solution of the bulk equations of motion, with the appropriate boundary condition. This solution still lives on anti-de Sitter space, but there is a scalar field turned on which breaks the isometries of AdS. In section \ref{sec:bulk_deform} we discuss the deformation of the free theory to the critical $U(N)$ theory, which is a non-trivial interacting CFT. As expected \cite{Klebanov:1999tb,Giombi:2011ya}, the critical theory is described by the same bulk action but with a different boundary condition for the bulk scalar field, which leads to non-trivial (but finite) loop corrections.

Several appendices contain some technical results.

\subsection{Future directions}

There are many remaining open questions and future directions, and we list some of them here.

A very intriguing question is whether our main result, that we can rewrite our CFT as a bulk path integral over fields living on a fixed anti-de Sitter background, generalizes to more complicated theories. We derived this result for vector models on $\mathbb{R}^d$, and it seems likely that even for these theories on more complicated backgrounds, such as backgrounds including circles, this will need to be modified (for instance, the vector models on a circle exhibit phase transitions in the large $N$ limit \cite{Shenker:2011zf}, which are expected to map to transitions between different topologies in the bulk). Putting the vector models on more complicated backgrounds requires adding $U(N)$ (or $O(N)$) gauge fields (in particular holonomies around circles), and it would be interesting to understand if this can be done in our framework. Generalizing our framework to matrix models (such as one would get, in particular, in the presence of dynamical $U(N)$ gauge fields) seems very complicated, due to the large number of independent $U(N)$ invariants that exist in this case (beyond the bi-local fields \eqref{bi_local}). However, perhaps some general lessons may still apply also in that case.

Some more specific questions are :

\begin{itemize}

\item It would be nice to prove that in the classical limit our bulk theories are identical to Vasiliev's high-spin gravity theories in some gauge; we expect this to be the case since these theories are claimed to be unique.

\item It would be interesting to understand the relation between our approach and earlier approaches to derive the AdS/CFT correspondence in this case of free vector models, such as approaches based on identifying the radial direction with a renormalization group scale \cite{Douglas:2010rc,Leigh:2014tza,Leigh:2014qca,Mintun:2014gua}.

\item The action we obtain on anti-de Sitter space is manifestly non-local. On general grounds we do not expect the gravitational dual to free field theories to be local at distances shorter than the anti-de Sitter radius, but we do expect it to be local at much longer distance scales. It would be nice to confirm that the actions that we obtain have this property.

\item In our approach we do not directly use the high-spin symmetry of the free field theories, partly because we are interested in applications like the finite $N$ critical models where it is broken. The presence of massless high-spin fields in the bulk implies that our bulk theory is a gauge-fixed version of a theory in which this high-spin symmetry is gauged. It would be interesting to understand the implementation of this symmetry in our formalism and whether it can be made more explicit.

\item It would be interesting to understand better the constraints on our bulk fields at finite $N$, and whether there is a nice way to write these constraints in the bulk language. More speculatively, non-local relations between quantum gravity fields (which we find) may be related to the black hole information paradox.

\item In this paper we analyzed only two solutions of the classical bulk theory (dual to the large $N$ CFT), corresponding to the undeformed CFT and to its mass deformation. It would be interesting to study other solutions. 


\item It is natural to generalize our analysis to have $N$ fermions instead of $N$ scalars, and work on this is in progress \cite{tomer}. One advantage of using fermions is that (unlike free scalars) they are primary fields also for $d=2$, so one can include this case in the analysis. Another generalization, to the $U(N)$ singlet sector of $F$ flavors of scalars (each with $N$ components), is straightforward, and just requires replacing our bulk fields by $F\times F$ matrices (with the bulk action involving a single trace over products of these fields).
It is also interesting to generalize our results to theories having both fermions and scalars, and, in particular, to supersymmetric vector models.

\item In this paper we focus on $d>2$, but our methods can be used (with some modifications) also for the $d=1$ case of conformal quantum mechanics. In this case it would be interesting to relate our results to many results on the SYK model (see \cite{Rosenhaus:2018dtp} for a review), which can also be written in bi-local variables that are mapped to the bulk \cite{Jevicki:2016bwu}, and to various recent results on gravitational theories on $AdS_2$.

\item It would be interesting to analyze vector models on other space-times. In particular, the theory on $S^d$ is related by a conformal transformation to the theory on $\mathbb{R}^d$, so it should be easy to generalize our analysis to this case, and to perform a bulk computation of the $S^d$ partition function of these theories. Attempts to do this in high-spin gravity at one-loop order appeared in \cite{Giombi:2013fka,Giombi:2014iua,Skvortsov:2017ldz}; our framework will allow us to compute also the leading order partition function (without which the one-loop result is not really meaningful), and it will lead to a different one-loop result than the one in \cite{Giombi:2013fka,Giombi:2014iua,Skvortsov:2017ldz} (because we have different quadratic terms in the bulk). 

\item It would be interesting to generalize our analysis to $SU(N)$ (or $SO(N)$) gauge theories, instead of $U(N)$ ($O(N)$). The big difference in this case is the appearance of extra baryon-like operators, where $N$ fields are contracted with an epsilon symbol, and it would be interesting to understand how to take these operators into account in our framework.

\item In the $d=3$ case, a natural generalization of our models is by coupling them to a $U(N)_k$ Chern-Simons theory, as this does not add any new dynamical fields. The resulting theories still have a high-spin symmetry in the large $N$ limit (with fixed `t~Hooft coupling $N/k$), though the symmetry is broken at finite $N$, and in the large $N$ limit the dual gravitational theory is believed to be a continuous deformation of the Vasiliev high-spin gravity theories by turning on a single extra coupling there, related to the `t Hooft coupling \cite{Giombi:2011kc,Aharony:2011jz}. In our framework the modification seems to be much more drastic, since the bi-locals \eqref{bi_local} are no longer gauge-invariant (they can be made gauge-invariant by inserting a Wilson line connecting $x_1$ and $x_2$), but the results on the classical gravity dual at infinite $N$ suggest that at least within the $1/N$ expansion there should be a simple way to implement this generalization. 

\item 
It would be interesting to generalize our analysis to Lorentzian space, and to understand how the Wick rotation affects our mappings and the bulk physics.\footnote{Note that in this paper we use covariant bi-local fields (1.1); there is also a Hamiltonian approach using as the basic variable the equal-time bi-local field $\sum_{I=1}^N \phi^*_I(\vec{x_1},t) \phi_I(\vec{x_2},t)$ \cite{Koch:2010cy,Koch:2014aqa}. The two approaches should be equivalent, and it would be interesting to understand the relation between our results and previous results in the equal-time bi-local approach.} In particular it would be nice to understand the Hilbert space in the bulk, and the role of the ``ghost''-like fields. It would also be interesting to look for Lorentzian classical solutions corresponding to finite energy density states of the CFT, which we expect to map to black holes on the gravity side, though the free theory does not thermalize so the dynamics of these black holes is quite different from standard ones. 

\item Some other Euclidean vector models are believed to be dual to gravitational theories on de Sitter space \cite{Anninos:2011ui}, and it would also be interesting to generalize our analysis to these theories, and see if they shed any light on these mysterious gravitational theories.

\end{itemize}

\section{From the scalar theory to the bi-local theory} \label{sec:local_to_bilocal}

In this section we map the theory of $N$ free massless real scalar fields in $d$ flat infinite Euclidean space-time dimensions to bi-local $O(N)$-invariant variables \cite{Jevicki:1979mb,Das:2003vw}, and the theory of $N$ complex fields to $U(N)$-invariant variables, in preparation for later mapping these variables to a theory living on anti-de Sitter space. This section does not contain any new results, so it can be skipped by readers familiar with this mapping. 

For $N$ complex scalar fields $\phi_I(x)$ ($I=1,\cdots,N$), all $U(N)$-invariants may be written in terms of the bi-local field\footnote{Note that this is not true if we require only $SU(N)$ invariance, since then we would have also ``baryonic'' operators made using an epsilon symbol, such as $\epsilon^{I_1 I_2 \cdots I_N} \phi_{I_1}(x_{I_1}) \cdots \phi_{I_N}(x_{I_N})$, which we would need to consider as well.}
\be \label{eq:bilocal}
G(x_1,x_2) \equiv \frac{1}{N} \sum_{I=1}^N \phi_I^{*}(x_1) \phi_I(x_2).
\ee
In particular, the local $U(N)$-invariant operators are all descendants of spin $J$ operators ($J=0,1,2,\cdots$) of the schematic form $O_J(x) \sim \sum_I \phi_I^{*}(x) \partial^J \phi_I(x)$, which can arise by Taylor expansion of the bi-local fields \eqref{eq:bilocal} near $x_2=x_1$. Similarly, we can take $N$ real fields $\varphi_I(x)$ and write all $O(N)$ invariants (but not all $SO(N)$ invariants) in terms of bi-locals $G(x_1,x_2) \equiv \frac{1}{N} \sum_I \varphi_I(x_1) \varphi_I(x_2)$, whose Taylor expansion includes local operators of all even spins ($J=0,2,4,\cdots$).

The $U(N)$-invariant correlation functions close on themselves, but projecting just to $U(N)$ invariants does not give a local theory (e.g. the theory on a torus would not be modular invariant). We can obtain a local theory of $U(N)$ invariants by gauging $U(N)$, but in general this would force us to introduce gauge fields and modify the theory. There is one case where this projection can be done locally without adding more fields, which is the $d=3$ case. In this case we can couple the theory to a $U(N)$ Chern-Simons theory at level $k$ and take $k\to \infty$, and this leads to a complete decoupling of the gauge fields \cite{Giombi:2009wh}. In this paper we will not worry about obtaining a local theory, we will concern ourselves only with correlation functions on $\mathbb{R}^d$ and work in any dimension. A similar procedure should be possible on $S^d$ (which is related by a conformal transformation to our analysis), but not on manifolds containing circles, where making the theory local by gauging $U(N)$ would necessarily add non-trivial holonomies.

We can write the generating function for all $U(N)$-invariant correlation functions as
\es{eq:gen_func}{
	Z_\text{free}[J] &= \int \prod_{I=1}^N D\phi_I(x) \exp \left( -\sum_{I=1}^N \int d^d x | \vec \nabla \phi_I(x) | ^2 -N \int d^d x_1 d^d x_2 J(x_1,x_2) G(x_1,x_2) \right)\,,
}
or equivalently
\es{eq:zfree}{
	Z_\text{free}[J] &= \int \prod_{I=1}^N D\phi_I(x) \exp \left( - S[G,J] \right)\\
}
where 
\es{eq:sgj}{
	S[G,J] &= N \( \int d^d x_1 \vec \nabla_1\cdot \vec \nabla_2 G(x_1,x_2) \mid_{x_2=x_1} + \int d^d x_1 d^d x_2 J(x_1,x_2) G(x_1,x_2) \) .\\
}
The expressions for the $O(N)$ case are similar, with a $\frac{1}{2}$ in front of the first term in the action.

Our goal, following \cite{Jevicki:1979mb,Das:2003vw}, is to rewrite \eqref{eq:zfree} as a path integral over the bi-local field $G(x_1,x_2)$. The problem is that not all components of this field are independent of each other (though they become independent as $N\to \infty$). In particular, if we consider the $G(x_i,x_j)$ for some set of $K$ points $x_i$, and view them as a $K\times K$ matrix with indices $i$ and $j$, then the rank of this matrix is bounded from above by $N$, leading to various relations between its elements (the simplest relations arise for $N=1$, when we have for instance $G(x_1,x_2) G(x_3,x_4) = G(x_1,x_4) G(x_2,x_3)$; in general we have constraints on products of $(N+1)$ $G$'s or more). In the continuum limit at infinite volume, these constraints are very complicated. In order to simplify the analysis, we put in both an IR and a UV cutoff by assuming that our field theory lives on a lattice with $V$ points, and perform the analysis with finite $V$, and take it to infinity only at the end. With this cutoff, $G(x_1,x_2)$ can be considered as a Hermitean $V\times V$ matrix (or as a real symmetric matrix for the $O(N)$ case), and its definition implies that the rank of this matrix, and of any sub-block of it, is not larger than $N$. In addition, the definition of $G$ implies that it is a non-negative matrix. This matrix notation will be useful also for traces over matrices (defined by ${\rm Tr}(G) \equiv \int d^d x G(x,x)$) and for matrix multiplication ($(GH)(x_1,x_2) \equiv \int d^d x_3 G(x_1,x_3) H(x_3,x_2)$). Note that the translation from the matrix to continuum integrals involves some power of the UV cutoff (the lattice spacing), which we keep implicit in our notation.

When $N < V$, which is relevant for finite $N$ theories in the continuum limit, the constraints on the possible $G$'s are quite complicated, and it is not known how to solve them explicitly. There is some path integral over the sub-space of $G$'s solving the rank and non-negativity constraints which is equivalent to \eqref{eq:zfree}, but its form is not known explicitly. However, when $N \geq V$, which is relevant in particular for the $1/N$ expansion, there are no rank constraints, and the only constraint is that $G$ must be non-negative. In this case we will work out the precise mapping to the bi-local variables below.

\subsection{The bi-local action}


In order to compute the Jacobian for the change of variables from $\phi_I(x)$ to $G(x_1,x_2)$ for $N \geq V$,
for a general $S[G]$ of the form \eqref{eq:zfree}, 
we begin by adding a path integral over $G$ by:
\begin{equation}\begin{split}
	Z &= \int D\phi_{I}(x) \exp\left(-S[G\left(\phi_I\right)]\right) \cr &= \int DG(x_1,x_2) D\phi_{I}(x) \exp\left(-S[G] \right)\prod_{x_1,x_2} \delta\left(G(x_1,x_2)-\frac{1}{N}\sum_I \phi^{*}_{I}(x_1)\phi_{I}(x_2)\right) .
\end{split}\end{equation}
Next, we write the delta function as an integral over an auxiliary field $\Sigma(x_1,x_2)$ (using the matrix notation described above, and considering $\phi_I(x)$ as a vector in this notation):
\begin{equation}\begin{split}
	Z & = \int DG(x_1,x_2) \exp\left(-S[G]\right)\,\int D\Sigma(x_1,x_2) D\phi_{I}(x) \exp\left(\text{Tr}\left(i\Sigma\,G-\frac{i}{N} \sum_I \phi_{I}\Sigma\phi^{*}_{I}\right)\right).
\end{split}\end{equation}
We can now integrate over the $\phi_I(x)$ (Gaussian integrals), obtaining (up to an $N$-dependent constant):
\begin{equation}\begin{split}
	Z &=\int DG(x_1,x_2) \exp\left(-S[G]\right)\,\int D\Sigma(x_1,x_2) \left|{\rm det}(\Sigma)\right|^{-N}\exp\left(i \text{Tr}\left(\Sigma\cdot G\right)\right).
\end{split}\end{equation}
The integral over $\Sigma$ can be performed by the change of variables $\tilde{\Sigma}=\Sigma\cdot G$, with a Jacobian $D\Sigma=\left|{\rm det}(G)\right|^{-V}D\tilde{\Sigma}$; the path integral over $\tilde{\Sigma}$ is then a constant and we find (up to an overall constant depending on $N$ and $V$)
\begin{equation}\begin{split} \label{eq:gaction}
	Z &= \int DG(x_1,x_2) \exp\left(-S[G]\right)\left|{\rm det}(G)\right|^{(N-V)} \cr
	  &= \int DG(x_1,x_2) \exp(-S[G]+(N-V)\text{Tr}(\log(G))).
\end{split}\end{equation}
This correctly incorporates the Jacobian for the change of variables from $\Phi$ to $G$. In addition, we need to require that the path integral is only over non-negative matrices $G$ (we implicitly assumed this in the last line of \eqref{eq:gaction}, when we dropped the absolute value from the determinant).

If we were doing the $O(N)$ case, the matrix $G$ would be real and symmetric instead of Hermitean, and the coefficient in the exponent would be $(N-V-1)/2$ instead of $(N-V)$. (Of course in the large $V$ limit we can ignore the shift of $V$ by one.)

We see that for $N\geq V$ the only modification of the action when we change to the bi-local variables is a single extra term ${\rm Tr}(\log(G))$, with a coefficient that depends on $N$ and on our cutoff $V$ (which can be thought of as $V = {\rm Tr}(1)$). However, because of the logarithm of the matrix $G$, this term is highly non-local. 

In general, the second term in \eqref{eq:gaction} means that the bi-local theory is interacting and strongly coupled. However, if $S[G]$ is proportional to $N$ (as in our free scalar theory \eqref{eq:zfree}), we have in \eqref{eq:gaction} an action proportional to $N$ with a cutoff-dependent counter-term, so we can perform a perturbative expansion in $1/N$. We will see below how these perturbative computations give us the expected results, and in the continuum large-volume limit of infinite $V$, are independent of $V$ and depend only on $N$. We expect similar results to hold also in other regularization methods for the theory, with the value of $V$ depending on the details of the UV and IR regulators. Our computations will be done taking $N$ to infinity first and only then taking $V$ to infinity.


Formally the derivation of the Jacobian above applies also for $N<V$, but in this case it gives nonsensical results (the integral over $G$ in \eqref{eq:gaction} generally diverges), so in this case we have to find a different way to impose the rank constraints on $G$. We expect that the transition between the regime of $N\geq V$ (where we can use \eqref{eq:gaction} and perform our large $N$ expansion) and the regime of $N<V$ should be smooth. \footnote{Note that the projection to $U(N)$ or $O(N)$ invariants is expected to give a unitary theory for integer $N$, but for non-integer $N$ it does not \cite{Maldacena:2011jn,Binder:2019zqc}, and one can construct (using a number of fields of order $N$) $O(N)$-invariant operators that have negative two-point functions; but this may not visible in our action \eqref{eq:gaction} which is valid for large $N$.}


\subsection{Perturbative expansion in {$1/N$} -- the Feynman rules}
\label{sec:pert_bilocal}

In this section we perform the pertburbative expansion of \eqref{eq:gaction} in the $U(N)$ case (the analyis for $O(N)$ is very similar).
The field $G(x_1,x_2)$ has a non-zero expectation value in the free scalar theory, and it is natural to expand around it. In $d$ dimensions the scaling dimension of a free scalar field is $\Delta_0 \equiv \frac{d-2}{2}$, and it is clear from the free scalar action that (in the continuum limit)
\be \label{eq:gvev}
\< G(x_1,x_2)\>  = G_0(x_1,x_2) \equiv \frac{\Gamma(d/2-1)}{4\pi^{d/2}}|x_1-x_2|^{-2\Delta_0}
\ee
(this is true except for $d=2$, where it should be replaced by a logarithm; we will not consider the $d=2$ case here).
As discussed above, in general it is not clear how to compute with \eqref{eq:gaction}, but in the large $N$ limit (viewing the term proportional to $V$ as a one-loop counter-term independent of $N$) the same expectation value \eqref{eq:gvev} arises from the classical equation of motion for $G$ in \eqref{eq:gaction}. Note that the Laplacian action on \eqref{eq:gvev} gives a delta function, which is the identity matrix in our notation, so the inverse of the matrix $G_0$ is the Laplacian operator. In particular, the matrix $G_0$ has maximal rank (for finite $N$ and $V$) even though as discussed above for $N<V$ the allowed matrices $G$ all have rank $\leq N$; but there is no problem in averaging matrices of rank $\leq N$ to get a matrix of maximal rank.

We can now expand around $G_0$:  
\begin{equation} \label{eq:eta_def}
	G(x_1,x_2)= G_{0}(x_1,x_2)+\frac{1}{\sqrt{N}}\eta(x_1,x_2),
\end{equation}
where we chose a convenient normalization for the large $N$ expansion that we will perform below. Note that 
this expansion is not really sensible at finite $N$, since as discussed above $G_0$ is not an allowed configuration there, so $\eta=0$ is not allowed; but it is fine in perturbation theory in $1/N$. Note also that the constraint on $\eta$ coming from non-negativity of $G$ is complicated, but in the large $N$ limit this constraint disappears, since all eigenvalues of $G_0$ are non-negative (and independent of $N$).

The free bi-local action \eqref{eq:gaction} in terms of $\eta$ reads (up to integration by parts and additive constants)
\be \label{eq:bilocalaction}
S\left[\eta\right]={\sqrt{N}}\text{Tr}\left( G_{0}^{-1}\eta\right)-\left(N-V\right)\log\left(1+\frac{1}{\sqrt{N}} G_{0}^{-1}\eta\right).
\ee
Expanding in powers of $\eta$, we find that the linear term of order $\sqrt{N}$ vanishes as it should, and we can write the action as a sum of a bare action and counter-terms:
\begin{align}
-S_{bare}[\eta] & =-\frac{1}{2}\text{Tr}\left(\left( G_{0}^{-1}\eta\right)^{2}\right)+\sum_{n=3}^{\infty}\frac{\left(-1\right)^{n+1}}{n}N^{1-\frac{n}{2}}\text{Tr}\left(\left( G_{0}^{-1}\eta\right)^{n}\right), \label{eq:free_pert_bilocal_action}\\
-S_{ct}[\eta] & =-\frac{V}{\sqrt{N}}\text{Tr}\left( G_{0}^{-1}\eta\right)+\sum_{n=2}^{\infty}\frac{\left(-1\right)^{n}}{n}VN^{-\frac{n}{2}}\text{Tr}\left(\left( G_{0}^{-1}\eta\right)^{n}\right)\label{eq:free_pert_bilocal_action_ct}.
\end{align}
So the Feynman rules for a perturbative expansion in $1/N$ are :
\begin{enumerate}
	\item The contraction/propagator is
		\begin{equation} \label{eq:etaprop}
			\contraction{\ }{\eta(x_1,x_2)}{\eta(x_3,x_4)} 
			\qquad \qquad \eta(x_1,x_2) \eta(x_3,x_4) = G_0(x_1,x_4) G_0(x_2,x_3), 
		\end{equation}
	\item The $n$-vertex ($n\ge 3$) is
	\begin{equation}
		\frac{\left(-1\right)^{n+1}}{n}N^{1-\frac{n}{2}}\text{Tr}\left(\left( G_{0}^{-1}\eta\right)^{n}\right),
	\end{equation}
	\item The counter-term $n$-vertex ($n\ge 1$) is
	\begin{equation}
		\frac{\left(-1\right)^{n}}{n}VN^{-\frac{n}{2}}\text{Tr}\left(\left( G_{0}^{-1}\eta\right)^{n}\right),
	\end{equation}
	\item The symmetry factor should be taken with regard to the ordering in the
loop up to cyclic transformations.
\end{enumerate}

\subsection{Explicit 1-loop Calculations} \label{sec:bilocal_1_loop}

The free scalar theory has the special property that all connected correlation functions of $(\sqrt{N}\eta)$'s are proportional to $N$ (they are given by contractions of the free scalar fields around a single loop). From the point of view of our perturbative expansion in $1/N$, this means that the connected correlators of $\eta$'s should all be given by their classical answers, and all loop corrections should vanish. In this section we show explicitly how this happens in a few examples.

\subsubsection{The one-point function of {$\eta$}}

There is no contribution to the one-point function of $\eta$ at tree level. At 1-loop order, we have a tadpole diagram from the cubic interaction, and also the linear counter-term:
\begin{equation}\begin{split}
\left\langle \eta_{1,2}\right\rangle _{one-loop} & =\qquad 
\vcenter{\hbox{\includegraphics[scale=.05]{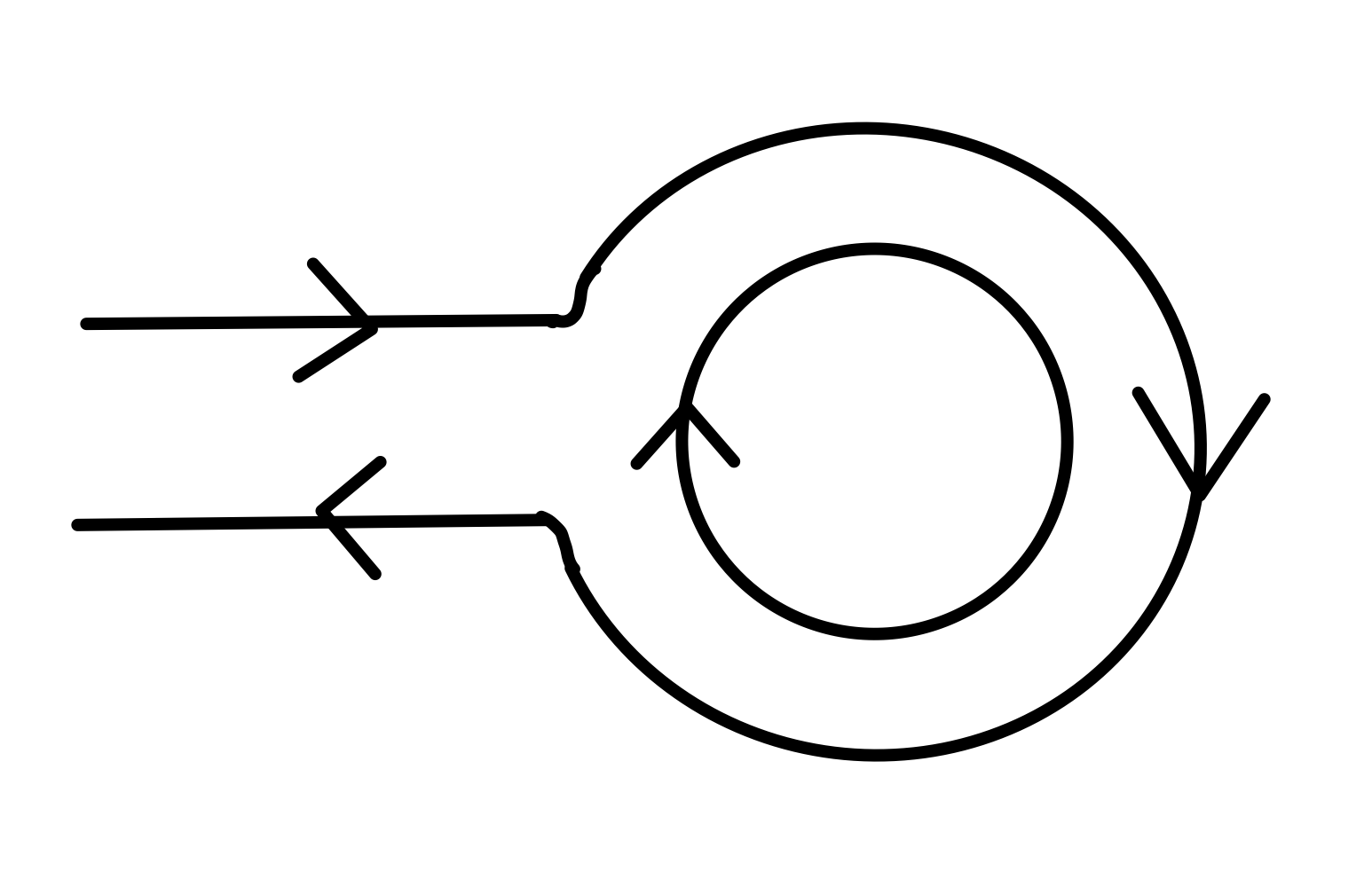}}} 
+
\vcenter{\hbox{\includegraphics[scale=.03]{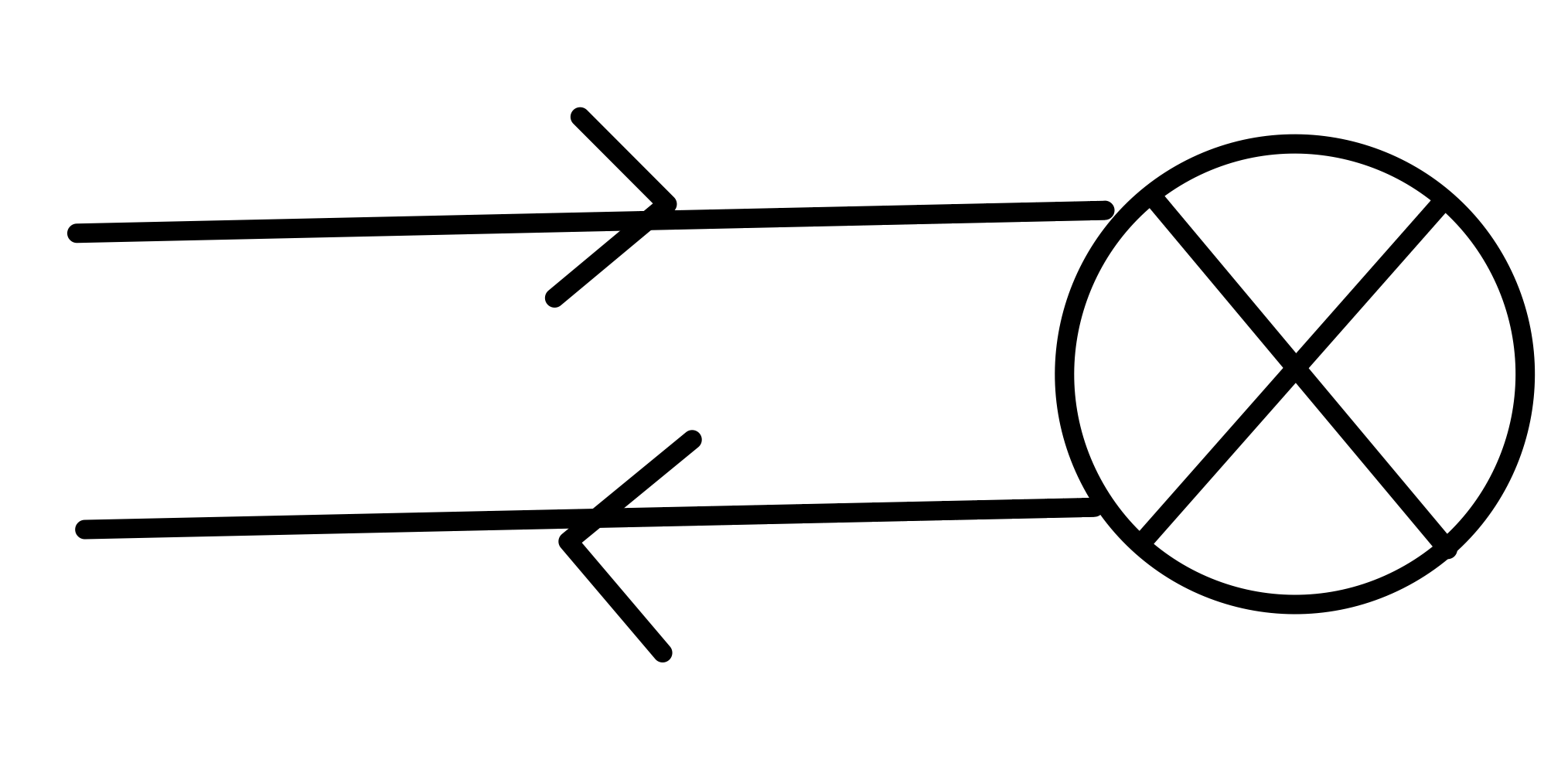}}}
\cr
 & =\left\langle \eta_{1,2}\frac{1}{3\sqrt{N}}\text{Tr}\left(\left( G_{0}^{-1}\eta\right)^{3}\right)\right\rangle +\left\langle \eta_{1,2}\frac{-V}{\sqrt{N}}\text{Tr}\left( G_{0}^{-1}\eta\right)\right\rangle \cr
 & =\frac{3V}{3\sqrt{N}} (G_0)_{1,2}-\frac{V}{\sqrt{N}} (G_0)_{1,2}=0,
\end{split}\end{equation}
where we got one factor of $V$ from ${\rm Tr}(1)=V$ in the loop.

\subsubsection{The two-point function of {$\eta$}}

At tree-level, the two-point function of $\eta$ \eqref{eq:etaprop} is the same as the connected four-point function of the scalar fields, which gives the full correct answer in the free scalar theory.

Next we
calculate the connected 1-loop contributions to $\left\langle \eta_{12}\eta_{34}\right\rangle $.
The first contribution is from the \includegraphics[scale=.02]{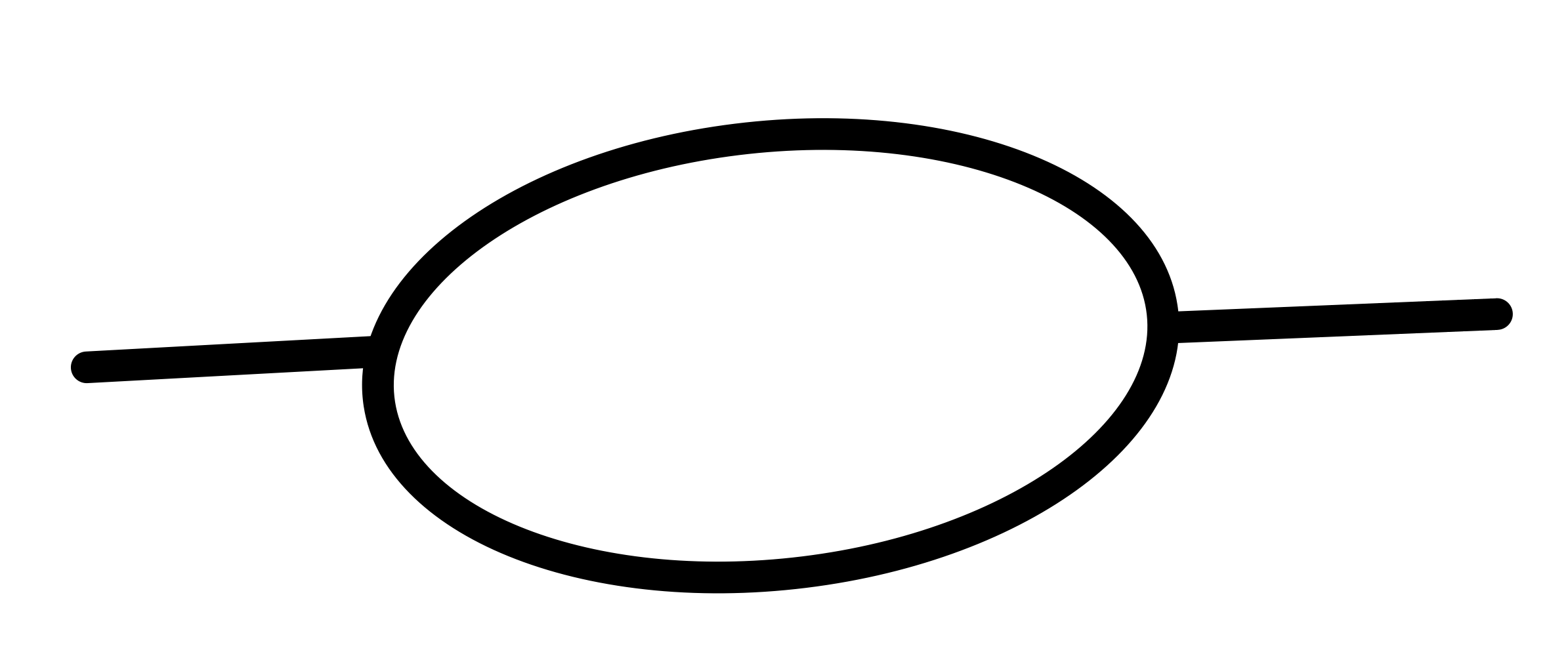} diagram,
which in terms of index contractions (using a hopefully obvious double-line notation) has two diagrams\footnote{For the $O(N)$ theory we would have the same diagrams without arrows and with a symmetrization over each pair of lines.}
that give
\begin{equation}
\begin{split}
\left\langle \eta_{12}\eta_{34}\frac{1}{3^{2}N}\cdot\frac{1}{2}\text{Tr}\left(\left( G_{0}^{-1}\eta\right)^{3}\right)^{2}\right\rangle _{conn}
& =
\vcenter{\hbox{\includegraphics[scale=.06]{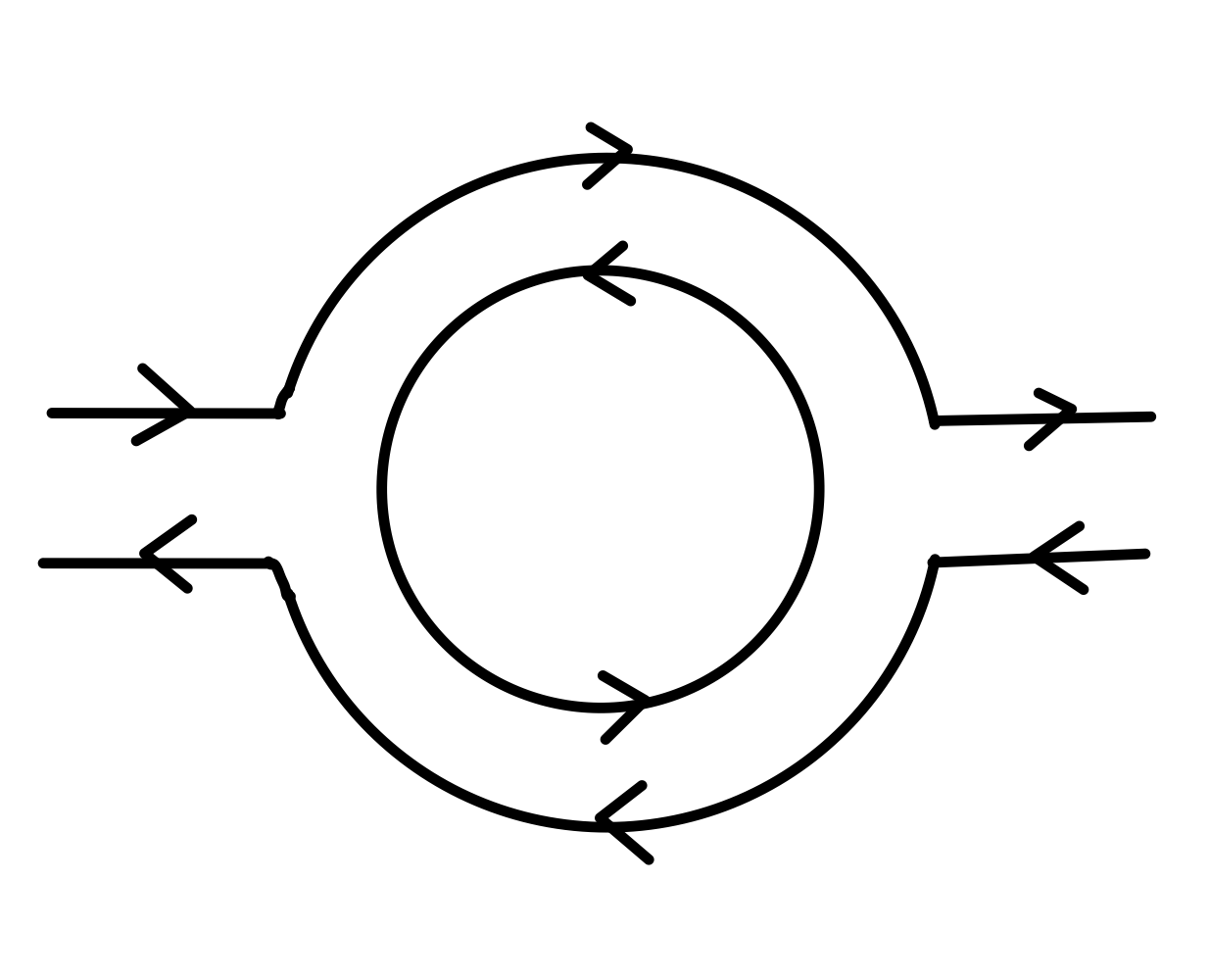}}} 
+
\vcenter{\hbox{\includegraphics[scale=.05]{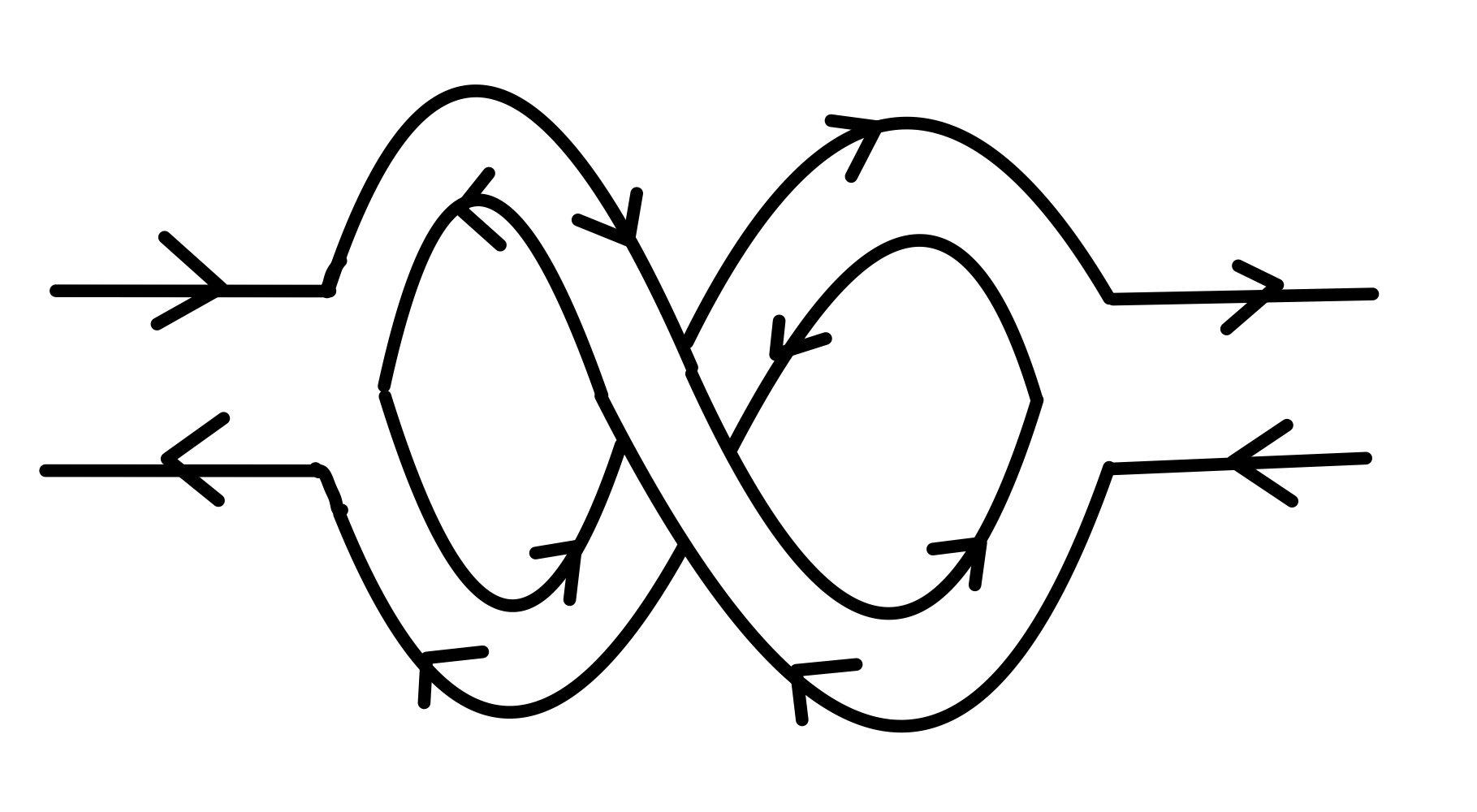}}} \\
& = \frac{1}{N}\left[V\cdot  (G_0)_{14} (G_0)_{23} + (G_0)_{12} (G_0)_{34}\right].
\end{split}
\end{equation}
Second is the 4-vertex diagram \includegraphics[scale=.02]{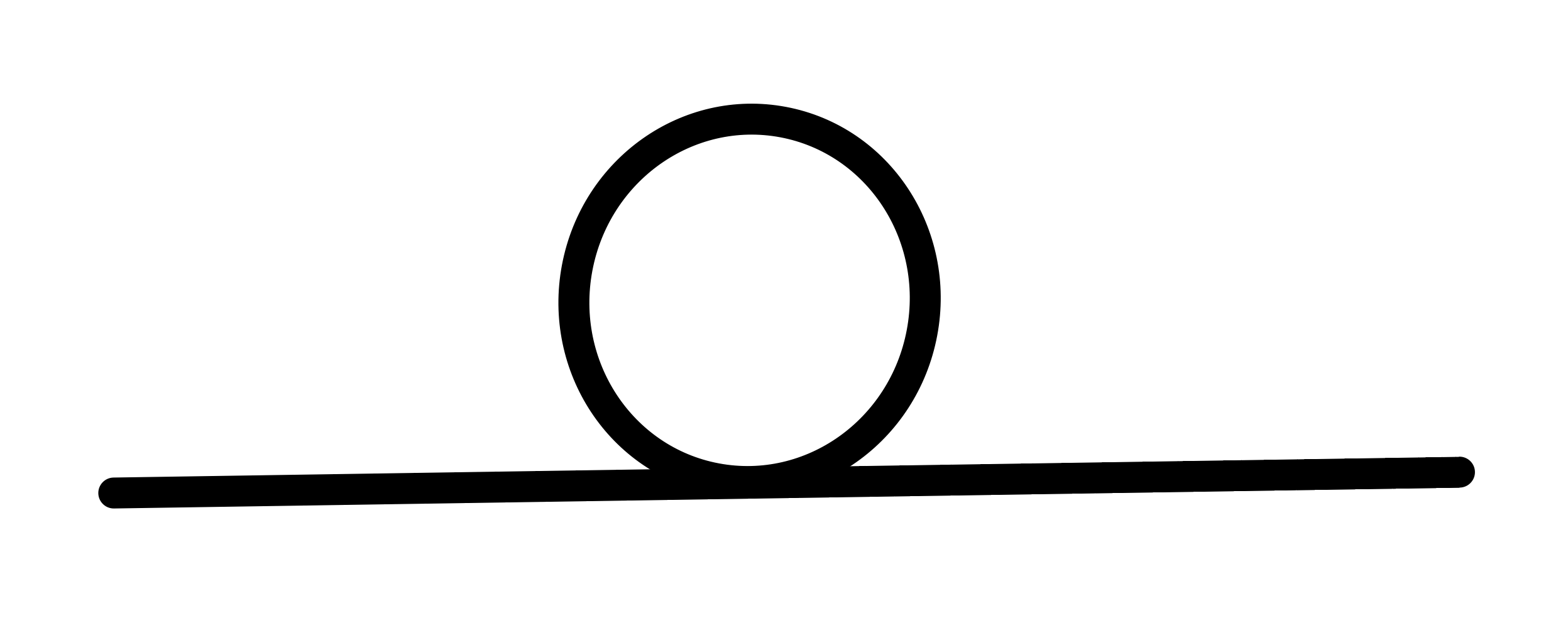},
which in terms of index contractions also includes two diagrams:
\begin{equation}
\begin{split}
\left\langle \eta_{12}\eta_{34}\frac{-1}{4N}\text{Tr}\left(\left( G_{0}^{-1}\eta\right)^{4}\right)\right\rangle _{conn}&=
\vcenter{\hbox{\includegraphics[scale=.06]{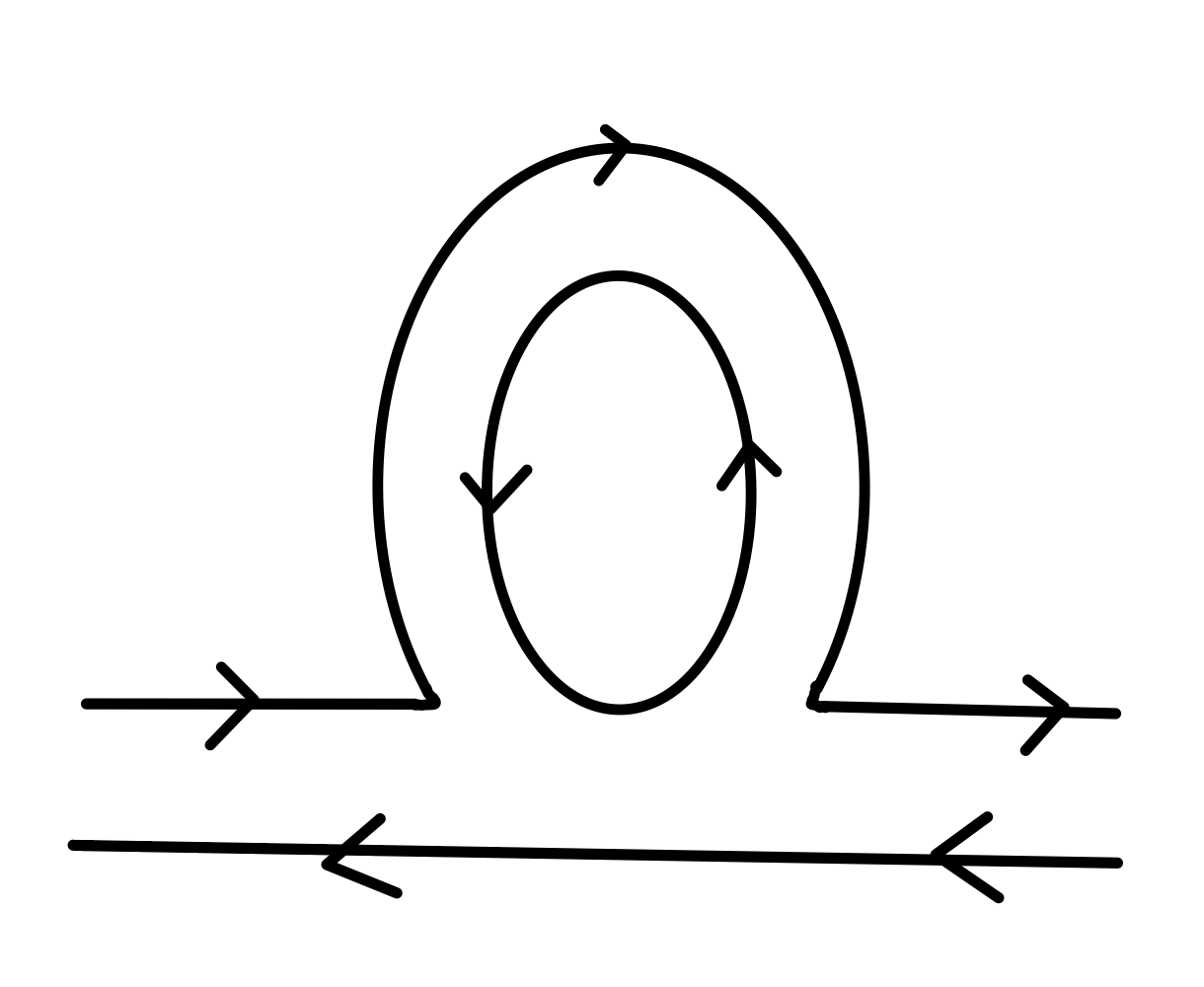}}} 
+
\vcenter{\hbox{\includegraphics[scale=.06]{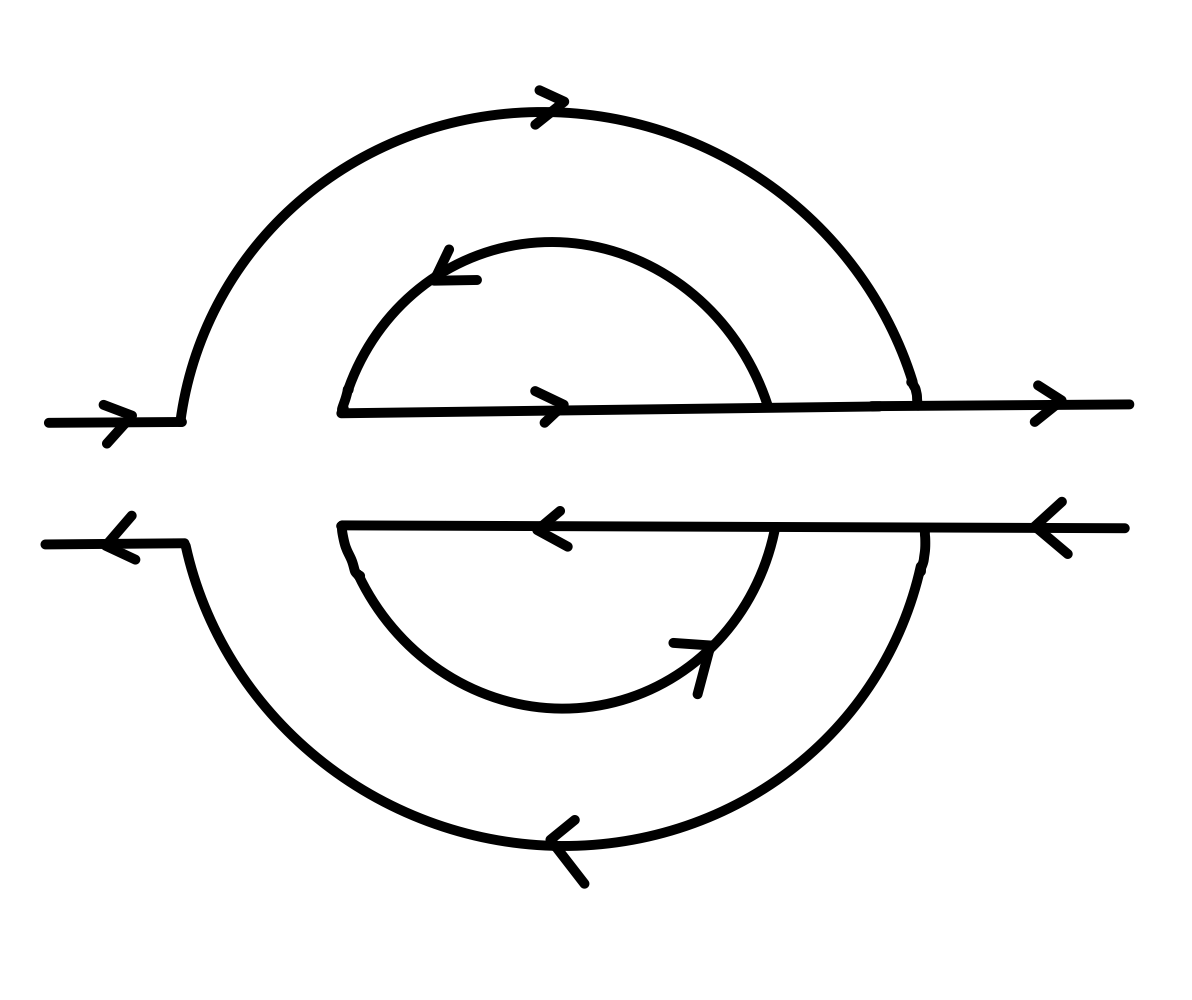}}} \\
& = -\frac{1}{N}\left[2V (G_0)_{14} (G_0)_{23}+ (G_0)_{12} (G_0)_{34}\right].
\end{split}
\end{equation}
Note that the first diagram has a symmetry factor of 2. Finally the counter-term diagram \includegraphics[scale=.05]{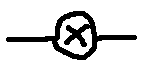}
is
\be
\left\langle \eta_{12}\eta_{34}\frac{V}{2N}\text{Tr}\left(\left( G_{0}^{-1}\eta\right)^{2}\right)\right\rangle _{conn}=\frac{V}{N} (G_0)_{14} (G_0)_{23}.
\ee
Adding everything together we get
\be
\left\langle \eta(x_1,x_2) \eta(x_3,x_4) \right\rangle_\text{1-loop} = 0
\ee
as expected.

\subsubsection{Generalities}

The possible terms that can appear in connected diagrams at any loop order are just free contractions of
the external points (e.g. $ (G_0)_{ij} (G_0)_{jk} \cdots (G_0)_{li}$), with some power of $V$ (coming from the counter-terms and from the extra loop lines). The vanishing
of loop corrections amounts to the combinatorical calculation of the
coefficients of the possible external contractions (including the
loop power $V$).

We were not able to prove explicitly the cancellation to all orders in perturbation theory, though it implicitly follows from our derivation of the action. For instance, it seems that there is no simple way to regard a contraction of two
vertices inside some generic diagram as one bigger vertex (which would have been useful for this). As an example, one of the contractions of
\includegraphics[scale=.05]{2_pt_1_loop_d_1.jpeg} can be connected
into a 4-vertex of \includegraphics[scale=.05]{2_pt_1_loop_d_3.jpeg} . But
there are two ways to do this, that together correspond to the symmetry
factor of the second diagram. So it is difficult to relate the computations in a simple way. Still it is true that both have the
same $ G_0$-dependence and opposite signs.

%

\section{From the bi-local theory to the conformal basis} \label{sec:bi_local_to_conformal}

In the previous section we rewrote the free theory partition function as a path integral over bi-local fields $\eta(x_1,x_2)$ with interacting action \eqref{eq:bilocalaction}. In order to map this to AdS, it will be convenient to expand this bi-local field in a basis that diagonalizes the Cartan subalgebra of the conformal algebra, involving functions transforming as conformal primaries of dimension $\Delta$ and spin $J$. This ``conformal basis'', whose elements are 3-point functions, was introduced in \cite{Dobrev:1976vr,Dobrev:1977qv} (see \cite{Karateev:2018oml} for a recent discussion). In this and subsequent sections we will work in the embedding space formalism, which we review in Appendix \ref{app:emb_form}. We denote the coefficients obtained by expanding $\eta(x_1,x_2)$ in the conformal basis as $C_{\Delta,J}(P,Z)$, where $P$ is an embedding space position coordinate (equivalent to a $d$-dimensional position $x$ in the CFT), and $Z$ is a null vector that keeps track of the spin indices. We can then rewrite the functional integral $D\eta(P_1,P_2)$ as a functional integral over these coefficients, $DC_{\Delta,J}(P,Z)$. To show this, we first introduce the basis and its precise completeness relation. Next, we write the quadratic action in terms of the $C_{\Delta,J}(P,Z)$ (we will discuss the higher order terms later). In an appendix we repeat the tree-level calculation of the 2-point function of $\eta$ by calculating the Gaussian integral for the $C_{\Delta,J}(P,Z)$'s explicitly, which confirms the validity of the change of variables. While we are interested in the free scalar theory, it will be useful at this stage to perform most of the analysis for a generalized free field theory (GFFT), where the generalized free field $\phi_I$ has conformal dimension $\Delta_0$. The free theory results can then be recovered by setting $\Delta_0=\frac{d-2}{2}$.

\subsection{The conformal 3-point basis} \label{sec:conformal_basis}

In order to make the conformal symmetry explicit at the level of the path integral, we would like to use a basis that diagonalizes the Cartan subalgebra of the conformal algebra. In a GFFT, $\eta(x_1,x_2)$ is in the tensor product of two representations with a scalar primary of dimension $\Delta_0$. Because we want to decompose it into irreducible representations of the conformal group, the basis is by definition the Clebsch-Gordan coefficients of the conformal group. By known harmonic analysis \cite{Dobrev:1977qv}, the coefficients are exactly three-point functions with OPE coefficient set to one. For a representation with a primary of spin $J$ (a symmetric traceless tensor) and dimension $\Delta$, these take the form
\es{eq:3_point}{
	\thpnorm{P_1}{P_2}{\Delta,J}{P_3,Z}
	=\frac{\left( 
	\left( Z \cdot P_1\right)\left( -2P_2 \cdot P_3\right)
	-\left( Z \cdot P_2\right)\left( -2P_1 \cdot P_3\right)\right)^J}
	{(-2P_1\cdot P_2)^{\frac{2\Delta_0-\Delta+J}{2}}
	(-2P_2\cdot P_3)^{\frac{\Delta+J}{2}}
	(-2P_3\cdot P_1)^{\frac{\Delta+J}{2}}}\,,
}
where $\mathcal{O}_{\Delta_0}$ and $\hat{\mathcal{O}}_{\Delta_0}$ denote scalar operators with scaling dimension $\Delta_0$. The functions \eqref{eq:3_point}, with an appropriate range of values of $\Delta$, form a basis in which functions of $P_1$ and $P_2$ may be expanded. This basis satisfies the completeness relation
\es{eq:3p_comp_rel}{
&	\delta(P_1,P_3)\delta(P_2,P_4)=\frac12\sum_{J=0}^\infty \int_{\gamma_J}\frac{d\Delta}{2\pi i} \int \frac{dP}{J!(\frac d2-1)_J}  \, 
	\frac{1}{N_{\Delta,J}}
	\\
	&\qquad\qquad\qquad\times\thpnorm{P_1}{P_2}{\Delta,J}{P,D_Z}\thptild{P_3}{P_4}{\tilde\Delta,J}{P,Z}\,,
}
where we defined the general notation $\tilde\Delta \equiv d -\Delta$, the contour $\gamma_J$ goes over the principal series $\Delta = \frac{d}{2}+is$ for real $s$ (with a small modification for $J=0$ that we will derive shortly), and we define the normalization 
\begin{equation}
	N_{\Delta,J}\equiv \frac{\pi^{\frac{3d}{2}}\Gamma(J+1)}
	{2^{J-1}\Gamma(\frac{d}{2}+J)}
	\frac{\Gamma\left(\Delta-\frac{d}{2}\right)}{\Gamma\left(\Delta-1\right)\left(\Delta+J-1\right)}\frac{\Gamma\left(\tilde{\Delta}-\frac{d}{2}\right)}{\Gamma\left(\tilde{\Delta}-1\right)\left(\tilde{\Delta}+J-1\right)}\,. \label{eq:CFT_norm}
\end{equation}
For the $O(N)$ theory, where we multiply a real primary of dimension $\Delta_0$ by itself, we only have even values of $J$; in this case we sum only over even $J$ on the right-hand side of the completeness relation \eqref{eq:3p_comp_rel}, and then the left-hand side becomes $\frac12[\delta(P_1,P_3)\delta(P_2,P_4)+\delta(P_1,P_4)\delta(P_2,P_3)]$, related to the different form of the 2-point functions of $\eta$ in this theory. 

The conformal basis \eqref{eq:3_point} also satisfies (for $\Delta,\Delta'$ in the principal series) the orthogonality relation
\begin{equation}\begin{split}
	&\int \frac{dP_1dP_2}{2\pi i N_{\Delta,J}}
	\thpnorm{P_1}{P_2}{\Delta,J}{P,Z}\thptild{P_1}{P_2}{\tilde\Delta^\prime,J^\prime}{P^\prime,Z^\prime}\\
	&  \qquad = \delta_{J,J^\prime}\left( \delta\left(\Delta-\Delta^\prime\right)\delta(P,P')(Z\cdot Z^\prime)^J + \frac{\delta\left(\Delta-\tilde \Delta^\prime\right)}{S^{(\tilde\Delta,J)}_{\Delta_0,\Delta_0} } \left\langle O_{\Delta,J}(P,Z) O_{\Delta,J}(P^\prime,Z^\prime) \right\rangle\right)\,, \label{eq:orthogonality}
\end{split}\end{equation}
where the shadow coefficient is defined as
\begin{equation}
	S^{(\tilde\Delta,J)}_{\Delta_0,\Delta_0} \equiv \frac{\pi^{\frac{d}{2}} \Gamma\left( \tilde \Delta -\frac{d}{2} \right)\Gamma\left(\tilde\Delta+J-1\right)\Gamma^2\left(\frac{\Delta+J}{2}\right)}
	{\Gamma\left(\tilde\Delta-1\right)\Gamma\left(\Delta+J\right)\Gamma^2\left(\frac{\tilde\Delta+J}{2}\right)}\,.
	\label{eq:shadow_coeff_2}
\end{equation}
The reason for the second term in \eqref{eq:orthogonality} is because the basis elements for $\Delta$ and for $\tilde \Delta$ are not independent of each other, but are related by the shadow transform\footnote{Algebraically this means there exists an intertwining operator between the $\Delta$ representation and the $\tilde \Delta$ representation \cite{Dobrev:1977qv}.}:
\es{eq:shadow_trans}{
&	\thpnorm{P_1}{P_2}{\Delta,J}{P,Z}\\
&	= \frac{1}{S^{(\tilde\Delta,J)}_{\Delta_0,\Delta_0} } 
	\int  \frac{dP^\prime}{J!(\frac d2-1)_J}
	\tp{\Delta,J}{P,Z}{P^\prime,D_{Z^\prime}} \thpnorm{P_1}{P_2}{\tilde\Delta,J}{P^\prime,Z^\prime}\,.
	}

We can now decompose $\eta(P_1,P_2)$ into this basis using the completeness relation \eqref{eq:3p_comp_rel}, as
\begin{equation}
	\eta(P_1,P_2) = \sum_{J=0}^\infty \int_{\gamma_J}\frac{d\Delta}{2\pi i} \int \frac{dP}{J! \left(\frac{d}{2}-1\right)_J}C_{\Delta,J}\left(P,D_Z\right) 
	\thpnorm{P_1}{P_2}{\Delta,J}{P,Z}\,,
	\label{eq:eta_decomp}
\end{equation}
which holds for the $O(N)$ theory by simply restricting the sum to even $J$, and where the coefficients $C_{\Delta,J}\left(P,Z\right)$ in both theories are given by
\begin{equation}
	C_{\Delta,J}(P,Z) = \frac12\frac{1}{N_{\Delta,J}}\int dP_1 dP_2 \, \eta\left(P_1,P_2\right)
	\thptild{P_1}{P_2}{\tilde\Delta,J}{P,Z}\,. \label{eq:coeff}
\end{equation}
Since $\eta(P_1,P_2)$ is Hermitian for the $U(N)$ theory, on the principal series we have
\begin{equation}
	C_{\Delta,J}^*(P,Z) = (-1)^J C_{\tilde\Delta,J}(P,Z)\,, \label{eq:cdelta_const}
\end{equation}
which also holds for the $O(N)$ case with only even $J$, since in that case $\eta(P_1,P_2)$ is real. Note that due to the shadow relation \eqref{eq:shadow_trans}, only half of the coefficients $C_{\Delta,J}(P,Z)$ along the contour $\gamma_J$ are independent. We choose those with ${\rm Im}( \Delta)\geq 0$ to be the actual dynamical variables. To extend the integration in \eqref{eq:eta_decomp} to the full principal series, we would then like \eqref{eq:coeff} to hold also for ${\rm Im}(\Delta)<0$. Using \eqref{eq:shadow_trans} we should define the coefficients in this range by:
\begin{equation}
	C_{\tilde\Delta,J}\left(P,Z\right) = \frac{1}{ S^{(\tilde\Delta,J)}_{\Delta_0,\Delta_0} } 
	\int \frac{dP^\prime }{J! \left(\frac{d}{2}-1\right)_J}
	\tp{\Delta,J}{P,Z}{P^\prime,D_{Z^\prime}}
	C_{\Delta,J}\left(P^\prime,Z^\prime\right)\,, \label{eq:basis_extension}
\end{equation}
and they will need to satisfy the reality condition \eqref{eq:cdelta_const}. This extension of the basis will be useful (but not necessary) later.  

Note that our mapping \eqref{eq:eta_decomp} to the conformal basis implicitly assumed that the bi-local field obeys the following constraints:
\begin{enumerate}
	\item $\lim_{P_2\to P_1}\eta(P_1,P_2)$ should be finite.
	\item At large $|P_1+P_2|$ (and fixed difference) $\eta(P_1,P_2)$ should decay.
	\item At large $|P_1|$ (or $|P_2|$) and fixed $P_2$ ($P_1$)
	\begin{equation}
		\eta(P_1,P_2) \sim \left|P_{1}\right|^{-2\Delta_{0}}\cdot \text{Power series in }\ensuremath{\frac{1}{|P_1|}}\label{eq:bc_eta}
	\end{equation}
	\item $\eta(P_1,P_2)$ must be smooth.
\end{enumerate}
The first and third constraints follow in a GFFT from the definition of $\eta$, the second is a sufficient condition for the completeness relation to hold \cite{Dobrev:1977qv} \footnote{In fact, the completeness relation applies to a slightly more general class of bi-local functions $f(x_1,x_2)$, which need only satisfy the conformally invariant condition \cite{Dobrev:1977qv}:
\es{dobrevCon}{
\infty>\int d^d x_1d^d x_2d^d x'_1d^d x'_2 f(x_1,x_2)f(x'_1,x'_2)\langle\mathcal{O}_{\tilde\Delta_0}(x_1)\mathcal{O}_{\tilde\Delta_0}(x'_1)\rangle\langle\mathcal{O}_{\tilde\Delta_0}(x_2)\mathcal{O}_{\tilde\Delta_0}(x'_2)\rangle\,.
} 
This condition is automatically satisfied if $f(x_1,x_2)$ decays at infinity.
} (but we will discuss later the extension of our formalism also to functions that do not obey it), and the fourth is a necessary condition for the completeness relation.

An important subtlety of this construction is that the completeness relation \eqref{eq:3p_comp_rel} was shown in the literature to hold on the principal series $\Delta_0=\frac{d}{2}+i s$, with the contour of integration $\gamma_J$ including only the principal series, while we would like to use it for real values of $\Delta_0$, by analytically continuing it to these values. In order to do this we need to carefully check for which values of $\Delta,\Delta_0$ the integral in \eqref{eq:coeff} still converges. The conditions for convergence are:
\begin{enumerate}
	 \item Around $P_1\sim P_2$ we have (assuming a finite limit $\eta(P_1,P_1)$) a power of $|P_1-P_2|$ equal to $d+{\rm Re}(\tilde\Delta)-2{\rm Re}(\tilde\Delta_0)=2{\rm Re}(\Delta_0)-{\rm Re}(\Delta)$. For $\Delta=\frac{d}{2}+i s$, the condition for convergence is thus $\Delta_0 > \frac{d}{4}$.
	 \item In the limit of large $P_1$ (or large $P_2$) we have (using \eqref{eq:bc_eta}) a power of $|P_1|$ equal to $d-2{\rm Re}(\Delta_0)-2{\rm Re}(\tilde\Delta_0)=-d$, which is convergent for any $\Delta_0$.
\end{enumerate}
Various subtleties occur for odd $d$ and for $\Delta_0\leq \frac d4$, which we now discuss.

\paragraph{Subtlety for odd $d$:}

Plugging \eqref{eq:coeff} into \eqref{eq:eta_decomp}, we can close the contour of integration over $\Delta$ on the real-positive half-plane, and write the integral as a sum over poles. For the completeness relation to make sense, we expect the $\frac{1}{N_{\Delta,J}}$ factor not to contribute any residue. But as we can see in \eqref{eq:CFT_norm}, $N_{\Delta,J}=0$ for $\Delta = d-1,d,d+1,\cdots$ at odd dimension $d$ (because of the $\Gamma\left(\tilde{\Delta}-1\right)$ term, except at $\Delta=d+J-1$; for even values of $d$ these zeros are canceled by the $\Gamma\left(\tilde{\Delta}-\frac{d}{2}\right)$ term in the numerator). So naively we expect to need to deform the contour so that these poles won't contribute.~\footnote{These poles (also called the discrete basis) are related to a discrete basis contributing to the Plancherel measure of the group at odd dimensions.~\cite{Dobrev:1977qv}} But in fact, the analytical continuation of the $dP$ integral has its own zeros. Below we discuss this integral and its poles (see \eqref{eq:CPW_integral}).
For $d>1$ it can be shown to be proportional to $\Gamma\left(\tilde{\Delta}-1\right)$, thus removing the poles from $\frac{1}{N_{\Delta,J}}$.\footnote{This corresponds to the fact that while the discrete representations of the conformal group exist for any odd $d$, they don't contribute to the Clebsch-Gordan coefficients of two scalars for $d>1$.~\cite{Dobrev:1977qv}} To conclude, we need to deform the contour $\gamma_J$ to exclude the discrete poles at integer values of $\Delta$ \emph{only for $d=1$.} In this paper we only discuss $d>2$ so we can ignore this subtlety.

\begin{figure}[!ht]
\centering
	\includegraphics{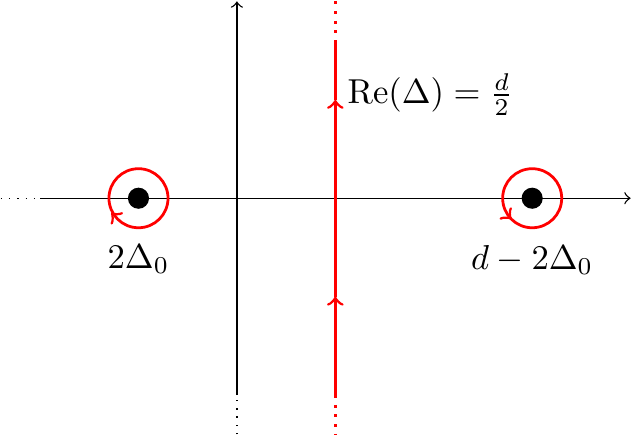}
	\caption{The deformation of the contour $\gamma_0$ from the principal series $\text{Re}(\Delta)=\frac{d}{2}$ to include $\Delta=2\Delta_0$ and exclude $d-2\Delta_0$, when $2\Delta_0<d/2$. In general we should deform the contour $\gamma_J$ for any $2\Delta_0+2n+J<\frac{d}{2}$ and $n=0,1,\cdots$.}
	\label{fg:Delta_contour}
\end{figure}

\paragraph{Subtlety for $\Delta_0<\frac{d}{4}$:}
As we just found out, the expression for $C_{\Delta,J}$ \eqref{eq:coeff} is convergent around the principal series only for $\Delta_0 > \frac{d}{4}$. This is an actual problem for us, as for the free scalar theory $\Delta_0=\frac{d-2}{2} < \frac{d}{4}$ for $d < 4$. To solve this problem we define a new function $\tilde \eta(x_1,x_2)$ that is the double shadow-transform~\cite{Karateev:2018oml} of $\eta$: it is implicitly defined by
\begin{equation}
	\eta(P_1,P_2) = \int dPdP^\prime
	\tp{\Delta_0}{P_1}{P}\tp{\Delta_0}{P_2}{P^\prime} \tilde{\eta}\left(P,P^\prime\right). \label{eq:shadow_eta}
\end{equation}
The idea is that $\tilde{\eta}\left(P_{1},P_{2}\right)$ transforms under the tensor product of two $\tilde{\Delta}_{0}$ representations, and also has the same boundary conditions as $\eta(P_1,P_2)$ described in \eqref{eq:bc_eta}, except replacing $\Delta_0\leftrightarrow\tilde\Delta_0$. Since for any $\Delta_0<\frac{d}{4}$, its shadow has $\tilde \Delta_0 > \frac{d}{4}$, we have no problem decomposing $\tilde\eta(P_1,P_2)$ as
\begin{equation}
	\tilde\eta(P_1,P_2) = \sum_{J=0}^\infty \int_{P.S.}\frac{d\Delta}{2\pi i} \int \frac{dP}{J! \left(\frac{d}{2}-1\right)_J}  \, \tilde C_{\Delta,J}\left(P,D_Z\right)
	\thptild{P_1}{P_2}{\Delta,J}{P,Z}, \label{eq:shadow_decomp}
\end{equation}
where the coefficients $\tilde C_{\Delta,J}$ are now defined by
\begin{equation}
	\tilde C_{\Delta,J}\left(P,Z\right)=\frac12\frac{1}{N_{\Delta,J}}\int dP_{1}dP_{2}\, \tilde\eta\left(P_{1},P_{2}\right)
	\thpnorm{P_1}{P_2}{\tilde\Delta,J}{P,Z}. \label{eq:shadow_eta_coeff}
\end{equation}
This integral is now convergent both for principal series $\Delta_0$ and for our $\Delta_0<\frac{d}{4}$. We can then substitute \eqref{eq:shadow_decomp} inside our definition for $\tilde\eta$ \eqref{eq:shadow_eta}, and obtain
\begin{equation}\label{eq:eta_decomp_shadow_int}
\begin{split}
	\eta(P_1,P_2)  &= \int dP'dP''\tp{\Delta_0}{P_1}{P'}\tp{\Delta_0}{P_2}{P''} \tilde{\eta}\left(P',P''\right)\\
	& = \sum_{J=0}^\infty \int_{P.S.}\frac{d\Delta}{2\pi i} \int \frac{dP}{J! \left(\frac{d}{2}-1\right)_J}  \, \tilde C_{\Delta,J}\left(P,D_Z\right) \int dP'dP''
	\tp{\Delta_0}{P_1}{P'}\\
	&\qquad\qquad\times\tp{\Delta_0}{P_2}{P''}\thptild{P'}{P''}{\Delta,J}{P,Z}\\
	& =\sum_{J=0}^\infty \int_{P.S.}\frac{d\Delta}{2\pi i} \int \frac{dP}{J! \left(\frac{d}{2}-1\right)_J}  \, 
	\frac{16\pi^d \Gamma^2(\frac{d}{2}-\Delta_0)}{\Gamma^2(\Delta_0)\lambda_{\Delta,J}}	\tilde C_{\Delta,J}\left(P,D_Z\right) 
	\thpnorm{P_1}{P_2}{\Delta,J}{P,Z}\,,
\end{split}
\end{equation}
where in the last equality we twice used the shadow transform
\begin{equation}
\begin{split}
	\int dP &
	\tp{\tilde\Delta_1}{P}{P_1} \thp{\Delta_1}{P}{\Delta_2}{P_2}{\Delta_3,J}{P_3,Z}\\
	&\qquad\qquad\qquad\qquad= S^{\Delta_1}_{\Delta_2,(\Delta_3,J)} 
	\thp{\tilde\Delta_1}{P_1}{\Delta_2}{P_2}{\Delta_3,J}{P_3,Z}
\end{split}
\end{equation}
with shadow coefficient
\begin{equation}
	S^{\Delta_1}_{\Delta_2,(\Delta_3,J)} \equiv \frac{\pi^{\frac{d}{2}}\Gamma\left(\Delta_{1}-\frac{d}{2}\right)\Gamma\left(\frac{\tilde{\Delta}_{1}+\Delta_{2}-\Delta_{3}+J}{2}\right)\Gamma\left(\frac{\tilde{\Delta}_{1}+\Delta_{3}-\Delta_{2}+J}{2}\right)}{\Gamma\left(\tilde{\Delta}_{1}\right)\Gamma\left(\frac{\Delta_{1}+\Delta_{2}-\Delta_{3}+J}{2}\right)
	\Gamma\left(\frac{\Delta_{1}+\Delta_{3}-\Delta_{2}+J}{2}\right)}\,, \label{eq:shadow_coeff}
\end{equation}
and we defined for future convenience
\begin{equation}
	\lambda_{\Delta,J} 
	\equiv \frac{\frac{16\pi^d \Gamma^2\left(\frac{d}{2}-\Delta_0\right)}{\Gamma^2\left(\Delta_0\right)}}{S^{\tilde{\Delta}_{0}}_{\tilde{\Delta}_{0},\left(\Delta,J\right)} 
	S^{\tilde{\Delta}_{0}}_{\Delta_{0},\left(\Delta,J\right)}}
	= 16 \frac
	{\Gamma\left(\tilde{\Delta}_{0}-\frac{\Delta-J}{2}\right)\Gamma\left(\tilde{\Delta}_{0}-\frac{\tilde{\Delta}-J}{2}\right)}
	{\Gamma\left(\Delta_{0}-\frac{\Delta-J}{2}\right)\Gamma\left(\Delta_{0}-\frac{\tilde{\Delta}-J}{2}\right)}\,. \label{eq:lambda_def}
\end{equation}
Note that the integral in the last line of \eqref{eq:eta_decomp_shadow_int} is divergent for $\Delta_{0}<\frac{d}{4}$, if we plug into it the definition of $\tilde{C}_{\Delta,J}\left(P,Z\right)$. On the other hand, as we change $\Delta_{0}$ this definition \eqref{eq:shadow_eta_coeff} is still convergent. Therefore the only problem in analytically continuing the final expression \eqref{eq:eta_decomp_shadow_int} for $\eta\left(P_{1},P_{2}\right)$ as a function of $\Delta_{0}$ are the poles coming from $1/\lambda_{\Delta,J}$.
For a given $J$, the function $1/\lambda_{\Delta,J}$ has poles at $\Delta^{\left(n,J\right)}=2\Delta_{0}+2n+J$ and at $\tilde{\Delta}^{\left(n,J\right)}$. Note that the $\Delta^{\left(n,J\right)}$ are precisely the dimensions of the physical primary operators of the GFFT. When $\Delta_0$ is in the principal series, only (and all the) poles of first type are inside the integration contour, ${\rm Re}(\Delta^{\left(n,J\right)})>\frac{d}{2}$. Therefore, in order to analytically continue this integral from the principal series $\Delta_{0}$ to $\Delta_{0}<\frac{d}{4}$, we need to deform the contour such that new poles $\tilde{\Delta}^{\left(n,J\right)}$ won't come in, and to add the $\Delta^{\left(n,J\right)}$ that left the contour by hand (see figure \ref{fg:Delta_contour}). We denote this deformed contour as $\gamma_J$, in terms of which we decompose $\eta$ as
\begin{equation}
	\eta(P_1,P_2) = \sum_{J=0}^\infty \int_{\gamma_J}\frac{d\Delta}{2\pi i} \int \frac{dP}{J! \left(\frac{d}{2}-1\right)_J}  \, 
	C_{\Delta,J}\left(P,D_Z\right)
	\thpnorm{P_1}{P_2}{\Delta,J}{P,Z},
	\label{eq:eta_comp}
\end{equation}
where we now define 
\begin{equation}
	C_{\Delta,J}\left(P,Z\right) \equiv \frac{16\pi^d \Gamma^2(\frac{d}{2}-\Delta_0)}{\Gamma^2(\Delta_0)\lambda_{\Delta,J}} \,
	\tilde C_{\Delta,J}\left(P,Z\right). \label{eq:coeff_shadow_relation}
\end{equation}
Note that since the integral \eqref{eq:shadow_eta_coeff} is not convergent for ${\tilde C}_J$ except around the principal series, when we deform the contour in \eqref{eq:eta_comp} we are also taking the analytical continuation of its integrand as a function of $\Delta$.

The bottom line is that every time a ``physical primary" satisfies $\Delta^{(n,J)}<\frac{d}{2}$, we need to deform the contour so that $\eta$ will capture its residue, and not capture that of its shadow $\tilde\Delta^{(n,J)}$. This is parallel to the discussion of the different quantizations of fields in AdS space ~\cite{Klebanov:1999tb}. This is also similar to the discussion of contributions from operators with $\Delta < \frac{d}{2} $ to the 4-point function in \cite{Simmons-Duffin:2017nub}. Both parallels will have a direct connection in the following sections. 

\paragraph{Subtlety for $d=4$:}
In the special case of $\Delta_0=d/4$ (which arises in particular for free scalars in $d=4$) our translation of variables breaks down, both using $\eta$ and using $\tilde\eta$. We will not discuss this case in detail, but it seems that our final results for this case can be simply continued from $\Delta_0 > d/4$ (or $d>4$ for free fields). As discussed for instance in \cite{Caron-Huot:2017vep}, this effectively means taking half of the residue from the pole at $\Delta=2\Delta_0=d/2$.

\paragraph{The free scalar case:}
In the special case of the free scalar, namely $\Delta_0 = \frac{d-2}{2}$ (with $d < 4$), we can actually make our results more explicit. In \eqref{eq:shadow_eta} we assumed the existence of a unique $\tilde \eta$ such that its shadow is our $\eta$. If we try to construct $\tilde \eta$ by an inverse shadow transform, we find that it is divergent for the case we are trying to solve, for which $\Delta_0 < \frac{d}{4}$. However, for the free theory we can derive a non-integral relation:\footnote{This relation in fact implies that we can reconstruct $\eta$ out of its derivatives. Explicitly, substituting \eqref{eq:free_shadow_eta} and \eqref{eq:shadow_coeff} in \eqref{eq:eta_comp}:
\begin{equation}\label{eq:conf_comp_rel_free}
\begin{split}
	\eta(P_1,P_2) = & \sum_{J=0}^\infty \int_{\gamma_J}\frac{d\Delta}{2\pi i} \int \frac{dP}{J! \left(\frac{d}{2}-1\right)_J}  \, 	
	\frac{1}{2N_{\Delta,J}\lambda_{\Delta,J}}\, 
	\thpnorm{P_1}{P_2}{\Delta,J}{P,D_Z}\\
	& \cdot \int dP_3 dP_4 \nabla_{1}^{2}\nabla_{2}^{2}\eta\left(P_3,P_4\right)
	\thpnorm{P_3}{P_4}{\tilde\Delta,J}{P,Z}.	
\end{split}
\end{equation}
It is likely that because of the boundary condition of $\eta$ \eqref{eq:bc_eta}, it has the same number of modes as $\nabla_{1}^{2}\nabla_{2}^{2}\eta$. We can think of this equation as our version for the completeness relation.}
\begin{equation}
	\begin{split}
		\tilde\eta(P_1,P_2) & = \sum_{J=0}^\infty \int_{P.S.}\frac{d\Delta}{2\pi i} \int \frac{dP}{J! \left(\frac{d}{2}-1\right)_J}  \, \tilde C_{\Delta,J}\left(p,D_Z\right)
		\thptild{P_1}{P_2}{\Delta,J}{P,Z}\\
		&= \frac{\Gamma^2(\frac{d-2}{2})}{16\pi^d} \sum_{J=0}^\infty \int_{P.S.}\frac{d\Delta}{2\pi i} \int \frac{dP}{J! \left(\frac{d}{2}-1\right)_J}  \,
	C_{\Delta,J}(P,D_Z)\, 
	\lambda_{\Delta,J}	\thptild{P_1}{P_2}{\Delta,J}{P,Z}\\
	&= \left( \frac{\Gamma(\frac{d-2}{2})}{4\pi^{\frac{d}{2}}}\right)^2 \nabla^2_1\nabla^2_2 \eta(P_1,P_2)\,,
	\end{split}\label{eq:free_shadow_eta}
\end{equation}
where in the first equality we used \eqref{eq:shadow_decomp}, in the second we used \eqref{eq:coeff_shadow_relation}, and in the third we used \eqref{eq:eta_decomp} along with the action of the bilocal Laplacian
\begin{equation}
	\nabla^2_1\nabla^2_2
	\thpnorm{P_1}{P_2}{\Delta,J}{P,Z}
	=\lambda_{\Delta,J}
	\thptild{P_1}{P_2}{\Delta,J}{P,Z}\,, \label{eq:bi_laplacian_eigenvalue}
\end{equation}
where for the free theory we have 
\es{eq:free_lambda}{
	\lambda_{\Delta,J}=\left(M_{\Delta,J}^{2}-M_{d+J,J}^{2}\right)\left(M_{\Delta,J}^{2}-M_{d+J-2,J}^{2}\right)\,,\qquad M_{\Delta,J}^{2}\equiv \Delta\left(\Delta-d\right)-J\,.
	}
	So at least in this case we are sure that an appropriate $\tilde \eta$ exists. Note that
if we define $\tilde \eta$ from \eqref{eq:free_shadow_eta}, it is not obvious that it satisfies the constraint that $\lim_{P_2\to P_1}\tilde\eta(P_1,P_2)$ is finite. This in fact follows from the OPE for the free theory
\begin{equation}
	\eta(P,P+\eps) = \sum_p C_{\phi \phi O_p} \eps^{-2\Delta_0+\Delta_p} O_p(P)
\end{equation}
with $\Delta_p-2\Delta_0=2n$, from which we see that $\nabla^2_1\nabla^2_2 \eta(P_1,P_2)$ has a smooth limit (we included here only scalar operators but it is easy to generalize the argument to include all spins).

To summarize the discussion for the free scalar theory, we only need to deform the contour when $\Delta^{(n,J)} = d-2+2n+J < \frac{d}{2}$. This means that in any case we only need to take care of the poles of $\Delta^{(0,J)}$, and only for $J<2-\frac{d}{2}$. Therefore for $2< d <4$ the only deformation happens for $J=0$, and there we need to add to the contour a circle around the pole of $\Delta^{(0,0)} = d-2$, and to remove the pole at $\tilde \Delta^{(0,0)} = 2$ by adding to the contour a circle of opposite orientation around it (see figure \ref{fg:Delta_contour}). For $d>4$ or for $J>0$ we don't need to do any contour deformations. This defines the contour $\gamma_J$ that we will use from here on.

\subsection{The free scalar quadratic action} \label{quadFree}

We can now use the results of the previous section to write the action \eqref{eq:free_pert_bilocal_action} in terms of $C_{\Delta,J}$, where as usual we work with the $U(N)$ theory primarily, and comment on the differences for the $O(N)$ theory when they occur\footnote{Note that the change of variables from $\eta(x_1,x_2)$ to $C_{\Delta,J}$ is linear, so the measure just changes by an unimportant constant.}. Let us do this explicitly for the quadratic term in the action:
\begin{equation}\begin{split}
	S^{(2)}[C_{\Delta,J}] & =\frac{1}{2}\int dP_1 dP_2 \eta(P_1,P_2)  \nabla^2_2\nabla^2_1\eta(P_2,P_1)   \\
	& =\frac{1}{2}\sum_{J,J^\prime=0}^\infty(-1)^J \int_{\gamma_J}\frac{d\Delta}{2\pi i}\int_{\gamma_{J^\prime}}\frac{d\Delta^\prime}{2\pi i} 
	\int \frac{dP}{J! \left(\frac{d}{2}-1\right)_J} \int \frac{dP'}{J'! \left(\frac{d}{2}-1\right)_J'}
	  \, C_{\Delta,J}\left(P,D_Z\right)C_{\tilde\Delta^\prime,J^\prime}\left(P',D_{Z'}\right) \\
	  &  \times  \lambda_{\Delta^\prime,J^\prime} \int dP_1 dP_2
	  \thpnorm{P_1}{P_2}{\Delta,J}{P,Z}\, \thptild{P_1}{P_2}{\tilde\Delta^\prime,J^\prime}{P',Z'}\\
	& =\sum_{J=0}^\infty (-1)^J\int_{\gamma_J}\frac{d\Delta}{2\pi i}\int \frac{dP}{J! \left(\frac{d}{2}-1\right)_J} \, \lambda_{\Delta,J} N_{\Delta,J}\,\, C_{\Delta,J}(P,D_Z)C_{\tilde\Delta,J}(P,Z)\,,
	\label{eq:c_action}
\end{split}\end{equation}
where in the second equality we used the decomposition in \eqref{eq:eta_comp} as well as the action of the bi-local Laplacian on our basis with eigenvalue $\lambda_{\Delta,J}$ given in \eqref{eq:free_lambda}. In the third equality we used the orthogonality relation \eqref{eq:orthogonality}, where note that each term on the right-hand side of \eqref{eq:orthogonality} contributes equally under our contour. For the $O(N)$ theory, we have the exact same results except the sum now only runs over even $J$ and we have an extra factor of $\frac12$ due to that extra factor in the $O(N)$ version of \eqref{eq:free_pert_bilocal_action}.


In appendix \ref{sec:free_4_point} we use the quadratic action \eqref{eq:c_action} to compute the two-point function of $\eta$, and we confirm that we obtain the same result as in the previous section. This is a consistency check on our mapping to the $C_{\Delta,J}$ variables. By Taylor expanding the two-point function of $\eta$'s as some of the points approach each other, we can obtain the two-point functions of all the singlet local operators.

We can similarly write the higher order interaction terms in \eqref{eq:free_pert_bilocal_action} in terms of $C_{\Delta,J}$. Unlike the quadratic term, we do not expect these terms to be local in $P$ or $\Delta$, since we cannot apply the orthogonality relation anymore.\footnote{One might speculate about a generalized orthogonality relation that would collapse multiple three-point basis elements into pairs of $\Delta$ and $P$ delta functions. Unfortunately, it is unlikely such a relation exists, since a conformal integral of $n$ three-point functions should be related to both products of delta functions and $m$-point functions for $m\leq n$, but these latter are only orthogonal in $\Delta$ for $n=2$.} Nonetheless, one should still be able to compute $1/N$ corrections with them, and in particular show that (when including the counter-terms) all such corrections vanish for the free theory. Note that as in the previous section, we need to regularize the theory in order to be able to compute loop diagrams; any regularization (which cuts off both UV and IR divergences) should lead to some value of $V$ in \eqref{eq:free_pert_bilocal_action_ct}, such that after computing the diagrams all terms with positive powers of $V$ cancel, and we should be able to remove the regulator and obtain finite results.

\section{The AdS/CFT map} \label{sec:mapping_to_the_bulk}

We have completed our discussion of the CFT, and are now ready to discuss the bulk higher spin theory in AdS. We know that our CFT has one primary symmetric traceless local operator of every spin $J$ (only even spins for $O(N)$), and we anticipate that these will be related to bulk fields of spin $J$. Indeed, we will show that if we expand symmetric traceless spin $J$ bulk fields $\Phi_J$ in terms of a basis of bulk-to-boundary propagators, which diagonalizes the Cartan of the conformal algebra, then just like the CFT conformal basis, it is labeled by $\Delta$ and $J$. Thus we have variables with the same quantum numbers on both sides, and we can construct a mapping which identifies
the coefficients $C_{\Delta,J}(P,Z)$ of our CFT basis with the coefficients of the bulk basis, giving an explicit map from the CFT to the bulk (as anticipated in \cite{Das:2003vw,deMelloKoch:2018ivk}). We will use this map in particular to relate the spin $J$ singlet local operators in the CFT to the boundary limit of $\Phi_J$. As in previous sections, we explicitly consider the $U(N)$ theory, while similar results for the $O(N)$ theory can be found by simply restricting the sums over $J$ to even $J$ throughout.

For simplicity we set the AdS curvature radius to one everywhere. This means that in the bulk we measure everything in units of this radius. As usual, this implies that dimensionful bulk couplings (such as Newton's constant) will be given by the appropriate power of the AdS radius times an appropriate power of $N$, that comes from the fact that $n$-point vertices come with a factor of $N^{1-n/2}$.

\subsection{The isometric bulk-to-boundary propagator basis}

Bulk-to-boundary propagators $G_{\Delta,J}(X,P;W,Z)$ form a natural basis for spin $J$ bulk fields $\Phi_J(X)$, just as three-point functions formed a natural basis for bi-locals in the previous section. We express these propagators using the embedding space formalism for both bulk and boundary coordinates as reviewed in Appendix \ref{app:emb_form}, where $P$ is the embedding space boundary coordinate, $Z$ is a null vector that keeps track of boundary indices, $X$ is the embedding space bulk coordinate, and $W$ is a null vector that keeps track of bulk indices. The bulk-to-boundary propagator of a massive spin $J$ field on $AdS_{d+1}$ is this language is\footnote{In \cite{Costa:2014kfa} this bulk-to-boundary propagator is denoted by $\Pi_{\Delta,J}$ and includes an extra numerical factor.}
\begin{equation}
	G_{\Delta,J}(X,P;W,Z) = \frac{\left( (-2P\cdot X)(W\cdot Z) + 2(W\cdot P)(Z\cdot X)\right)^J}{(-2P\cdot X)^{\Delta+J}}\,, \label{eq:bulk_to_boundary_prop}
\end{equation}
which is the unique solution to the AdS$_{d+1}$ Laplace equation
\begin{equation}
	\nabla^2_X G_{\Delta,J}(X,P;W,Z) = M^2_{\Delta,J} G_{\Delta,J}(X,P;W,Z)\,,\qquad M_{\Delta,J}^{2}\equiv \Delta\left(\Delta-d\right)-J\,,\label{eq:bulk_boundary_DE}
\end{equation}
which satisfies the transversality condition
\begin{equation}
	\nabla_X\cdot K_W \, G_{\Delta,J}(X,P;W,Z) = 0\,, \label{eq:bulk_boundary_lorentz}
\end{equation}
and also the boundary condition (as $X$ approaches the boundary)\footnote{When $\Delta=\tilde\Delta$, as can happen for the scalar bulk field in $d=4$, the second term will have a $\log(z)$, which distinguishes its scaling in $z$ from the first term.}
 \begin{equation}\label{eq:2_point_limit}
 	G_{\Delta,J}\left(\frac{1}{z} P',P;\frac{1}{z} Z',Z\right)=z^{\Delta-J} 
 	\tp{\Delta,J}{P,Z}{P',Z'} + z^{d-\Delta-J} S_B ^{\Delta,J} \delta(P,P') (Z\cdot Z')^J + \cdots\,.
 \end{equation}
 Here, the bulk shadow coefficient is \cite{Costa:2014kfa}
 \be
	S_B^{ \Delta,J} 	\equiv \frac{\pi^{\frac{d}{2}} \Gamma( \Delta-\frac{d}{2})}
	{(J+ \Delta-1)\Gamma(\Delta-1)}\,, \label{eq:bulk_shadow_coeff}
\ee
 and it appears in the bulk analog of the shadow transform \eqref{eq:shadow_trans}: 
\es{eq:shadow_trans_bulk}{
	& G_{\Delta,J}(X,P;W,Z)  =  \frac{1}{S_B^{\tilde \Delta,J}}\int \frac{dP'}{J! (\frac{d}{2}-1)_J} G_{\tilde \Delta,J}(X,P',W,D_{Z'}) 
	\tp{\Delta,J}{P',Z'}{P,Z}\,.
}

 In \cite{Costa:2014kfa} it was shown that any appropriate\footnote{See the discussion about boundary conditions below.} bulk spin $J$ symmetric and traceless tensor can be spanned by the complete basis of gradients $\left(W\cdot\nabla\right)^{l}G_{\Delta,J-l}\left(X,P;W,Z\right)$ for $0\le l \le J$ and $\Delta = \frac{d}{2} + i s$ (the principal series). In order to write down the corresponding completeness relation, it is useful to define the AdS harmonic function
\begin{equation}
	\Omega_{\Delta,J}(X_1,X_2;W_1,W_2) =\int  \frac{dP}{J!\left(\frac{d}{2}-1\right)_{J}} \frac{G_{\Delta,J}\left(X_{1},P;W_{1},D_{Z}\right)G_{\tilde{\Delta},J}\left(X_{2},P;W_{2},Z\right)}{\alpha_J N_{\Delta,J}}
	\,,\label{eq:omega_def}
\end{equation}
where we already defined $N_{\Delta,J}$ in \eqref{eq:CFT_norm}, and
\es{eq:alpha_J_def1}{
 	\alpha_J  \equiv \frac{2^J \Gamma(\frac{d}{2}+J)}{\pi^{\frac{d}{2}}\Gamma(J+1)}\,.
 	} 
 The corresponding completeness relation then reads:\footnote{The bulk-to-boundary basis also obeys an orthogonality relation given in equation 227 of \cite{Costa:2014kfa}, which we will not use.}
\es{eq:bulk_comp_rel}{
 	\delta\left(X_{1},X_{2}\right)&(W_{12})^{J}=  \int_{P.S.}\frac{d\Delta}{2\pi i}\,\Omega_{\Delta,J}\left(X_1,X_{2};W_1,W_2\right)  \\
 	&+ \sum_{l=1}^{J}  {\left(W\cdot\nabla_1\right)^{l}\left(W\cdot\nabla_2\right)^{l} }  \int_{P.S.}\frac{d\Delta}{2\pi i} 
 	 {A_{\Delta,J,l}} \Omega_{\Delta,J-l}\left(X_1,X_{2};W_1,W_2\right)\,, \\
}
 where $W_{12} = W_1 \cdot W_2$, and
\es{eq:alpha_J_def}{
A_{\Delta,J,l} \equiv \frac{2^{l}\left(J-l+1\right)_{l}\left(\frac{d}{2}+J-l-\frac{1}{2}\right)_{l}}{l!\left(d+2J-2l-1\right)_{l}\left(d+J-l-\Delta\right)_{l}\left(\Delta+J-l\right)_{l}} \,,
 	} 
such that $A_{\Delta,J,0}=1$.
 As explained in \cite{Costa:2014kfa}, the role of the $l\geq1$ terms in \eqref{eq:bulk_comp_rel} is to give the non-transverse terms in the delta function, so that it is natural to define
 \es{eq:bulk_comp_rel_tt}{
 	\delta^{TT}\left(X_{1},X_{2}\right)\left(W_{12}\right)^{J}\equiv
\int_{P.S.}\frac{d\Delta}{2\pi i}\,\Omega_{\Delta,J}\left(X_1,X_{2};W_1,W_2\right)\,,\\
}
which is a delta function acting on traceless transverse functions.

We can now decompose $\Phi_J(X)$ into this basis using the completeness relation \eqref{eq:bulk_comp_rel} as
\begin{equation}
	\Phi_{J}\left(X,W\right)=\int_{P.S.}\frac{d\Delta}{2\pi i}\int 
	\frac{dP}{J!\left(\frac{d}{2}-1\right)_{J}}
	C^{bulk}_{\Delta,J}\left(P,D_Z\right)G_{\Delta,J}\left(X,P;W,Z\right)\,, \label{eq:bulk_comp}
\end{equation}
where the coefficients $C^{bulk}_{\Delta,J}\left(P,Z\right)$ are by the completeness relation 
	\begin{equation} \label{eq:bulk_coeff}
			C^{bulk}_{\Delta,J}\left(P,Z\right)=\frac{1}{\alpha_J \, N_{\Delta,J}}\frac{1}{\left(\frac{d-1}{2}\right)_{J}J!}\int dX\Phi_{J}\left(X,K_W\right)G_{\tilde{\Delta},J}\left(X,P;W,Z\right)\,.
		\end{equation}
Note that here we chose our bulk fields $\Phi_J(X,W)$ to be transverse
\begin{equation}\label{eq:bulk_transverse}
		\nabla_X \cdot K_W \, \Phi_{J}\left(X,W\right) = 0\,,
	\end{equation}
so that we did not need to include the $l\geq1$ basis functions from \eqref{eq:bulk_comp_rel}; we will see that we can map our CFT variables just to such transverse fields (if we want we can add in also non-transverse components by a bulk field redefinition, but we do not need them). 

\subsection{The CFT-to-AdS mapping} \label{sec:cft_to_ads}

We now have the same set of variables in the CFT and in the bulk AdS, so we can simply identify the $C_{\Delta,J}$ appearing in the expansion of $\eta(P_1,P_2)$ in the field theory in \eqref{eq:eta_decomp} with the $C^{bulk}_{\Delta,J}$ appearing in the expansion of $\Phi_J(X,W)$ in AdS in \eqref{eq:bulk_comp}. However, for each $\Delta$ and $J$ the conformal symmetry fixes this identification only up to a constant, so the general mapping between the two sides which is consistent with the conformal symmetry is given by
\begin{equation} \label{eq:mapping}
C^{bulk}_{\Delta,J}(P,Z) = f_{\Delta,J} C_{\Delta,J}(P,Z)
\end{equation}
for some normalization factor $f_{\Delta,J}$.
The only subtlety here is that for $\Delta_0 < d/4$ and $J=0$ our $C$'s in the CFT were defined to be on a different contour than the one we had above in AdS, so we need to modify the contour we use in \eqref{eq:bulk_comp} accordingly, and we will discuss the interpretation of this below.

The consistency of the identification \eqref{eq:mapping} imposes some constraints on the normalization factor $f_{\Delta,J}$. For instance, we can apply the CFT shadow relation for $C_{\Delta,J}$ in \eqref{eq:basis_extension} and the bulk shadow relation \eqref{eq:shadow_trans_bulk} to the bulk expansion \eqref{eq:bulk_comp}, and use the shadow invariance of the contour $\gamma_J$ to derive the consistency condition 
\begin{equation}
		\frac{f_{\Delta,J}}{f_{\tilde \Delta,J} } = \frac{S_B^{\tilde \Delta,J}}{S^{(\tilde\Delta,J)}_{\Delta_0,\Delta_0}} =
		\frac{ \Gamma\left(\Delta+J\right)\Gamma^2\left(\frac{\tilde\Delta+J}{2}\right)}
	 	{ \Gamma\left(\tilde\Delta+J\right)\Gamma^2\left(\frac{\Delta+J}{2}\right)}\,.
	 	\label{eq:bulk_norm_shadow_cond}
	\end{equation}
Also, the CFT hermiticity condition \eqref{eq:cdelta_const} implies that for real bulk fields $\Phi_J$ we need
\begin{equation}
		f_{\Delta,J}^* = (-1)^J f_{\Delta^*,J}\,, \label{bulk_norm_reality_cond}
	\end{equation}
	which can be easily fullfilled by taking the normalization $f_{\Delta,J}$ to be a real analytic function times (say) $(i)^J$.

We can in fact write an explicit mapping from $\eta(P_1,P_2)$ to $\Phi_J(X,W)$ by plugging the CFT definition of the $C_{\Delta,J}$ in \eqref{eq:coeff} into \eqref{eq:bulk_comp}. Recall from section  \ref{sec:bi_local_to_conformal} that the definition of $C_{\Delta,J}$ required subtle analytic continuations depending on the range of $\Delta_0$ (or of $d$ for the free theory), so to define this mapping we need to discuss each case separately. As in the CFT section, it is useful when performing these analytic continuations to consider the GFFT case with general real $\Delta_0$, and then restrict to the free theory we care about by setting $\Delta_0=\frac{d-2}{2}$. 

We start with the simplest case $\Delta_0 > \frac{d}{4}$ (i.e. $d>4$ for the free theory), where the integral in \eqref{eq:coeff} converges, so we can simply plug it into \eqref{eq:bulk_comp} to define the CFT-to-AdS mapping
\begin{equation}
\begin{split}\label{eq:cft_to_ads_1}
	\Phi_{J}\left(X,W\right)& =\frac12
	\int_{P.S.}\frac{d\Delta}{2\pi i}\frac{f_{\Delta,J}}{N_{\Delta,J}}\int \frac{dP}{J!\left(\frac{d}{2}-1\right)_{J}}\int dP_1 dP_2  \, G_{\Delta,J}\left(X,P;W,D_Z\right)
	\\
	&\qquad\qquad\qquad\quad\times \thptild{P_1}{P_2}{\tilde\Delta,J}{P,Z}\eta\left(P_1,P_2\right)\,.
\end{split}
\end{equation}

For $\Delta_0<\frac{d}{4}$ (i.e. $d<4$ for the free theory) recall that $C_{\Delta,J}$ is only well defined in terms of the auxiliary bilocal field $\tilde\eta(P_1,P_2)$ using \eqref{eq:shadow_eta_coeff} and \eqref{eq:coeff_shadow_relation}. We can plug these definitions into \eqref{eq:bulk_comp} and \eqref{eq:mapping} to get
\begin{equation}\label{eq:cft_to_ads_2}
\begin{split}
	\Phi_{J}\left(X,W\right)
		& = \frac12\frac{16\pi^d \Gamma^2(\frac{d}{2}-\Delta_0)}{\Gamma^2(\Delta_0)} \int_{\gamma_J} \frac{d\Delta}{2\pi i}
		\int  \frac{dP}{J!\left(\frac{d}{2}-1\right)_{J}} \int dP_1 dP_2 \frac{f_{\Delta,J}}{\lambda_{\Delta,J}N_{\Delta,J}}\\
		& \times G_{\Delta,J}\left(X,P;W,D_Z\right)
		\thpnorm{P_1}{P_2}{\tilde\Delta,J}{P,Z} \tilde{\eta}\left(P_1,P_2\right)\,.
\end{split}
\end{equation}
For the free theory we can use the explicit relation in \eqref{eq:free_shadow_eta} between $\tilde \eta$ and $\eta$ to then get
\begin{equation} \label{eq:free_cft_to_ads_2}
\begin{split}
	\Phi_{J}\left(X,W\right)& = \frac12 \int_{\gamma_J} \frac{d\Delta}{2\pi i}  \int \frac{dP}{J!\left(\frac{d}{2}-1\right)_{J}} \int dP_1 dP_2 \frac{f_{\Delta,J}}{\lambda_{\Delta,J}N_{\Delta,J}} \\
		& \times  
		G_{\Delta,J}\left(X,P;W,D_Z\right)
		\thpnorm{P_1}{P_2}{\tilde\Delta,J}{P,Z} \, \nabla^2_1\nabla^2_2\eta\left(P_1,P_2\right)\,.
\end{split}
\end{equation}
Note that naively we can integrate this by parts to obtain \eqref{eq:cft_to_ads_1} (just with a shifted contour), but if we try to do this we would get divergences near $P_1=P_2$ as discussed above.

We can think of the mappings \eqref{eq:cft_to_ads_1},\eqref{eq:free_cft_to_ads_2} as an explicit convolution
\begin{equation} \label{eq:explicit_cft_to_ads}
	\Phi_{J}\left(X,W\right) = \int dP_1 dP_2 {\cal M}_J\left(X,W\mid P_1,P_2\right) \eta(P_1,P_2)\,,
\end{equation}
and this will be useful in section \ref{sec:bulk_theory}.
These mappings involve various integrals, and we have to make sure that they are well-defined. Our definition of 
the CFT-to-AdS mappings in \eqref{eq:cft_to_ads_1} and \eqref{eq:free_cft_to_ads_2} is that the $\Delta$ integral with $\Delta=\frac{d}{2}+i s$ should have a large $s$ cutoff $|s|<s_\text{cutoff}$, and we take it to infinity only after performing all the other boundary and bulk coordinate integrals. As we will argue below, this procedure guarantees the convergence of the integrals. 
The right way to understand \eqref{eq:explicit_cft_to_ads} is to have different mappings ${\cal M}_J^{s_\text{cutoff}}$ as a function of the cutoff, and we take the limit only after performing the $P_1,P_2$ integrals. Thus ${\cal M}_J\left(X,W\mid P_1,P_2\right)$ must be understood as a distribution. Note that for any finite cutoff it is still invariant under the conformal symmetry.

We would like to understand why the $s_\text{cutoff}$ procedure gives finite results. We need to show that with a cutoff all the integrals are finite, and that the result does not diverge as we take $s_\text{cutoff}$ to infinity. In both  \eqref{eq:cft_to_ads_1} and \eqref{eq:free_cft_to_ads_2}, no singularities of the $P_1,P_2$ integrals (as functions of $P,\Delta$) are supposed to arise, as they were already taken care of by the construction of the $C_{\Delta,J}$ under the prescribed boundary conditions of $\eta$ (see \eqref{eq:bc_eta}). The $P$ integrand behaves as $|P|^{-2d}$ at large $|P|$ and thus converges. At $P=x$ (the tangential component of the bulk point $X$) the propagator has a finite limit as long as $z\ne0$ (where $z$ is the radial coordinate of $X$). For a finite cutoff the integral over $s$ ($\Delta=\frac{d}{2}+i s$) is also finite, as long as the normalization $f_{\Delta,J}$ is non-singular (or integrable) on the principal series. The reason is that both $\frac{1}{N_{\Delta,J}}$ (for $d>4$) and $\frac{1}{N_{\Delta,J}\lambda_{\Delta,J}}$ (for $d<4$), the propagators and the three-point functions in \eqref{eq:cft_to_ads_1},\eqref{eq:free_cft_to_ads_2} are continuous as a function of $s$. We conclude that indeed for a given cutoff the integrals are finite and commute with each other for any $d$.

The only possible divergence appears as we try to remove the cutoff. The reason is that at large $s$ we have (using \eqref{eq:CFT_norm}) $\frac{1}{N_{\Delta,J}} \sim s^d$ and therefore for $d>4$ \eqref{eq:cft_to_ads_1} naively diverges. For $d<4$ we may get away with it as (using \eqref{eq:CFT_norm} and \eqref{eq:lambda_def}) $\frac{1}{N_{\Delta,J}\lambda_{\Delta,J}} \sim s^{4\Delta_0-d}=s^{d-4}$. But this
analysis is lacking because of two reasons. First, we don't know the large $s$ scaling of $f_{\Delta,J}$. In addition, the rest of the integrand, i.e. the propagator and the three-point, oscillates rapidly at large $s$, and the $P,P_1,P_2$ integrals may generate a negative enough power of $s$ that will make the integral converge. We will first learn what is the correct large $s$ scaling of the $P,P_1,P_2$ integrals in both \eqref{eq:cft_to_ads_1} and \eqref{eq:free_cft_to_ads_2}, and then use it to constrain the behavior of $f_{\Delta,J}$ so that the limit converges.

We will use $y^i,y_1^i,y_2^i$ as the coordinates of $P,P_1,P_2$, and $(x^i,z)$ as the upper half space coordinates of $X$. For any $d$ and $J$, the $P,P_1,P_2$ integrand in  \eqref{eq:cft_to_ads_1},\eqref{eq:free_cft_to_ads_2} has the same $s$ dependence:
\begin{equation} \label{eq:large_s_dependence}
	\left( \frac{z}{(x-y)^2+z^2} \frac{|y-y_1||y-y_2|}{|y_1-y_2|}\right)^{i s}\,.
\end{equation}
The first term is from the bulk-to-boundary propagator, and the second is from the three-point function. We can find the large $s$ limit of the integrals by a stationary phase approximation of the integrals over $y$ and $y_+=\frac{1}{2}(y_1+y_2)$, keeping the integral over $y_- = y_1-y_2$. The saddle-point solution turns out to be $y_+=y=x$. The saddle-point value of \eqref{eq:large_s_dependence} is $\left| \frac{y_-}{z}\right|^{is}$, and the 1-loop determinant gives (as we integrated over $2d$ variables) $s^{-d}$ times an $s$-independent term. Therefore we have
\begin{equation}\label{eq:large_s_dependence_2}
	\int dP dP_1 dP_2 \dots \underset{\text{large }s}{\sim} \int d^d y_- (\cdots) \left| \frac{y_-}{z}\right|^{is} \cdot s^{-d}\,,
\end{equation}
Where the $\cdots$ on the right-hand side are some function that depends on $y_-,x,z$ and the spin indices of $\Phi$, and is linear in $\eta(x^i+\frac{y^i_-}{2},x^i-\frac{y^i_-}{2})$. 

Starting from the $d>4$ case \eqref{eq:cft_to_ads_1}, we have another overall factor of $\frac{f_{\Delta,J}}{N_{\Delta,J}} \sim s^d f_{\Delta,J}$ and so overall we have
\begin{equation}
	\Phi_J(X,W) \underset{\text{large }s}{\sim} \int^{\infty}_{-\infty} ds f_{\Delta,J} \int d^d y_- (\dots) \left| \frac{y_-}{z}\right|^{is}\,.
\end{equation}
Assuming at large $s$ that $f_{\Delta,J} \sim s^a e^{i  b s}$ for $a\in \mathbb{Z}$ and $b\in \mathbb{R}$, we get that the integral over $s$ converges to $a$ derivatives of a delta function that localizes $y_-$ to a sphere with a radius propotional to $z$, and thus a finite result (albeit with  ${\cal M}_J$ a distribution rather than a function if $a \geq 0$). For $d<4$ \eqref{eq:free_cft_to_ads_2} our outside factor is $\frac{f_{\Delta,J}}{N_{\Delta,J}\lambda_{\Delta,J}} \sim f_{\Delta,J} s^{d-4}$ and so overall
\begin{equation}
	\Phi_J(X,W) \underset{\text{large }s}{\sim} \int^{\infty}_{-\infty} ds f_{\Delta,J} s^{-4} \int d^d y_- (\dots) \left| \frac{y_-}{z}\right|^{is}.
\end{equation}
This has the exact same properties as the $d>4$ case up to a shift of $a$ by $4$,
and so is finite in the same sense. The bottom line is that as long as $f_{\Delta,J}$ has a power-law times an oscillating behavior at large $s$, instead of an exponentially large in $s$ behavior, all of the integrals involved in our mapping are finite, and we can safely take $s_\text{cutoff}$ to infinity at the end of our computations.

\subsection{The boundary conditions for $\Phi_J(X)$}\label{sec:bulk_bc}

So far we did not discuss the boundary conditions that $\Phi_J(X)$ needs to obey in order to have a decomposition \eqref{eq:bulk_comp}, and the boundary conditions we obtain (off-shell) by our mapping from the CFT, so let us discuss this now. For clarity, we take the bulk coordinates of $X$ to be upper half space coordinates $(x^i,z)$, and the boundary coordinate $P$ to be $(y^i)$, and our convention is that the $\Phi_J$ components always have lower indices. In these coordinates	
\begin{equation}
	G_{\Delta,J}(x,z|y)_{\mu_1,...,\mu_J|i_1,..,i_J} = \left(\frac{z}{(x-y)^2+z^2}\right)^\Delta \left(A_{\mu_1,i_1}\cdot...\cdot A_{\mu_J,i_J}-\text{traces}\right), \label{eq:explicit_prop}
\end{equation}
where $A_{i,j} = z^{-1}\Big({\delta_{i,j}-2\frac{(x-y)_i (x-y)_j}{(x-y)^2+z^2}}\Big)$ and $A_{z,i}=\frac{-2(x-y)_i}{(x-y)^2+z^2}$.

	\paragraph{Small $z$:}
	To find the small $z\sim 0$ limit of \eqref{eq:bulk_comp}, we should start with the small $z$ limit of the bulk-to-boundary propagator $G_{\Delta,J}$ (with all indices lowered). Using \eqref{eq:explicit_prop}, we get the expected scaling on the right-hand side of \eqref{eq:2_point_limit}. Similar to the discussion around \eqref{eq:bulk_shadow_coeff}, each term in \eqref{eq:2_point_limit} contributes the same amount inside the integral. The leading behavior depends on the contour of integration. In our decomposition \eqref{eq:bulk_comp} we integrate just over the principal series. Since we argued above that the integral over $s$ converges (perhaps giving some localized contributions), we see that $\Phi_J(X)$ decays as $z\to 0$ at least as $z^{\frac{d}{2}-J}$. In some cases (such as on-shell correlation functions) we may be able to close the contour and argue that there are only contributions from poles with ${\rm Re}(\Delta) > \frac{d}{2}$, such that the decay is even faster, but in general this is all we can say.

In our mapping from the CFT for $J=0$ and $d<4$, we use the deformed contour $\gamma_0$, which has an extra contribution near the allowed pole at $\Delta=2\Delta_0=d-2$. When the integrand has a pole at that value, it will give a contribution to $\Phi_0(X)$ going as $z^{2\Delta_0}$. Conversely, when we have such contributions, we need to modify the contour that we use in our expansion of the bulk field, in order to include them. We will see in the next subsection that we can expand the bulk fields with the modified boundary conditions that include such terms by modifying the contour, such that all of our analysis above still holds.

	\paragraph{Large $|x|$:}
	At large $|x|$ the bulk-to-boundary propagator at the principal series scales as $|x-y|^{-d}$ for $y^i\ne x^i$, and has a finite limit at $x^i=y^i$. Thus the integral over $y$ \eqref{eq:bulk_comp} in this limit gets contributions just from regions of large $y$, either near $x$ or far from $x$.
The large $|y|$ behavior of $C_{\Delta,J}(y^i)$ can be found using \eqref{eq:coeff} (or \eqref{eq:shadow_eta_coeff} for $d<4$), and similar arguments imply that it is governed by the behavior of $\eta(x_1,x_2)$ at large $x_+ = x_1+x_2$.
We conclude that the large $x_+$ behavior of $\eta$ dictates the large $|y|$ behavior of $C_{\Delta,J}(y)$, which in turn dictates the large $|x|$ behavior of $\Phi_J$. In particular, if $\eta$ decays at large $x_+$ as we assumed in \eqref{eq:bc_eta}, then $\Phi_J(x,z)$ will decay at large $|x|$. We will discuss non-decaying field configurations in section \ref{sec:massDef} below.


\subsection{The AdS-to-CFT mapping} \label{sec:ads_to_cft}

In the previous subsections, we defined the CFT-to-AdS mapping, by plugging the CFT definition of $C_{\Delta,J}$ in \eqref{eq:coeff} into the expansion of $\Phi_J(X,W)$ in \eqref{eq:bulk_comp} in terms of $C^{bulk}_{\Delta,J}$, and using the mapping \eqref{eq:mapping}. Since our mapping is linear, we can define also the inverse AdS-to-CFT mapping, by plugging the AdS definition of $C^{bulk}_{\Delta,J}$ in \eqref{eq:bulk_coeff} into the expansion of $\eta(P_1,P_2)$ in \eqref{eq:eta_decomp}. To do this, we first need to check that the resulting integral in \eqref{eq:bulk_coeff} is well-defined for every value of $d$, using our boundary conditions. 

Let us define $X=(x^i,z)$, then we need to check for convergence at large $|x|$ and at small $z$. At large $|x|$ the propagator has a power of $x^{-d}$ which cancels against the integral measure. We assume as above that $\Phi_J$ decays at large $|x|$ and then the integral converges. At small $z$ the measure has $z^{-d}$, the propagator has $z^{\Delta-J}$, the contraction of indices gives $z^{2J}$, and as discussed above, when we just have the principal series then $\Phi_J$ gives at most $z^{\frac{d}{2}-J}$. Thus, in this case we have no divergence as $z\to 0$ (on the principal series). But for $d<4$ and $J=0$ we had to add the contribution from the $\Delta=2\Delta_0=d-2$ pole, giving a scaling of $\Phi_0\sim z^{2\Delta_0}=z^{d-2}$. The total power of the integral is now negative, $z^{2\Delta_0-\frac{d}{2}}=z^{\frac{d}{2}-2}$, and the integral diverges.

For $d>4$, we conclude that \eqref{eq:bulk_coeff} is convergent, so we can immediately plug it into the decomposition of $\eta$ in \eqref{eq:eta_comp} to get an explicit AdS-to-CFT mapping
\begin{equation}\label{eq:ads_to_cft}
		\begin{split}
			\eta(P_1,P_2) & = 
			\sum_{J=0}^\infty 
			\int_{P.S.}\frac{d\Delta}{2\pi i} 
			\frac{1}{\alpha_J \, N_{\Delta,J} f_{\Delta,J}}
			\int\frac{dP}{\left(\frac{d-1}{2}\right)_{J}  \left(\frac{d}{2}-1\right)_J  J!^2}\\
			& \times 
			\int dX 
			\thpnorm{P_1}{P_2}{\Delta,J}{P,D_Z}
			G_{\tilde{\Delta},J}\left(X,P;K_W,Z\right)
			\Phi_{J}\left(X,W\right)\,.
		\end{split}
		\end{equation}
		
		For $d<4$, we can use the same formula for the $J>0$ terms on the right-hand side, but we need to replace the $J=0$ term by
\begin{equation} \label{eq:ads_to_cft_free}
		\begin{split}
			\int_{\gamma_0}\frac{d\Delta}{2\pi i} 
			\frac{1}{\alpha_0 N_{\Delta,0} \lambda_{\Delta,0} f_{\Delta,0}}
			\int dP \int dX  
			\thpnorm{P_1}{P_2}{\Delta,0}{P}
			G_{\tilde{\Delta},0}\left(X,P\right) \\
			\times 
			\left(\nabla_{X}^{2}-M_{d-2,0}^{2}\right)\left(\nabla_{X}^{2}-M_{d,0}^{2}\right)\Phi_{0}(X)\,,
		\end{split}
		\end{equation}		
as we will now show. The $J=0$ term must be modified because as discussed above the integral in \eqref{eq:bulk_coeff} is not convergent in this case. We can find a convergent integral by defining $\tilde \Phi_0$ using the $\tilde C_{\Delta,0}$ that we defined for this case in \eqref{eq:shadow_eta_coeff}, to get
		\begin{equation}
			\tilde\Phi_{0}\left(X\right)=\int_{P.S.}\frac{d\Delta}{2\pi i}\int dP\,
			f_{\Delta,0}
			\tilde C_{\Delta,0}\left(P\right)G_{\Delta,0}\left(X,P\right)\,. \label{eq:shadow_bulk_comp}
		\end{equation}
		Because the integration contour here is just the principal series, the small $z$ behavior of $\tilde\Phi_0(X)$ is at most $z^{\frac{d}{2}}$, and we can safely define the inverse relation
		\begin{equation}
			f_{\Delta,0} \, \tilde C_{\Delta,0}\left(P\right)=\frac{1}{\alpha_0 \, N_{\Delta,0}}
			\int dX \tilde \Phi_{0}\left(X\right)G_{\tilde{\Delta},0}\left(X,P\right)\,.\label{eq:shaodw_bulk_coeff}
		\end{equation}
		We can then relate $\Phi_0$ to $\tilde\Phi_0$ for the free theory as
\begin{equation}\label{phitophit}
			\tilde\Phi_0(X) = \left(\nabla_{X}^{2}-M_{d-2,0}^{2}\right)\left(\nabla_{X}^{2}-M_{d,0}^{2}\right) 
			\Phi_0(X)\,,
\end{equation}
which follows from \eqref{eq:bulk_comp}, \eqref{eq:shadow_bulk_comp}, \eqref{eq:coeff_shadow_relation}, and the identity
\begin{equation}
			\left(\nabla_{X}^{2}-M_{d+J-2,J}^{2}\right)\left(\nabla_{X}^{2}-M_{d+J,J}^{2}\right) G_{\Delta,J}\left(X,P;W,Z\right) = \lambda_{\Delta,J}G_{\Delta,J}\left(X,P;W,Z\right)\,,
			\label{eq:lambda_eigenvalue}
		\end{equation}
using \eqref{eq:free_lambda} and \eqref{eq:bulk_boundary_DE}. Finally, we can combine \eqref{eq:shaodw_bulk_coeff}, \eqref{phitophit}, and the definition of $C_{\Delta,0}$ in terms of $\tilde C_{\Delta,0}$ in \eqref{eq:coeff_shadow_relation}, to get a well-defined relation between $C_{\Delta,0}$ and $\Phi_0$:
		\begin{equation} \label{eq:bulk_coeff_scalar}
		\begin{split}
			f_{\Delta,0} \, C_{\Delta,0}\left(P\right)& =
			\int \frac{dX}{\alpha_0 \, N_{\Delta,0} \lambda_{\Delta,0}} G_{\tilde{\Delta},0}\left(X,P\right) 
			\left(\nabla_{X}^{2}-M_{d-2,0}^{2}\right)
			\left(\nabla_{X}^{2}-M_{d,0}^{2}\right)\Phi_{0}(X)\,,
		\end{split}
		\end{equation}
		which we can then plug into the decomposition of $\eta$ in \eqref{eq:eta_comp} for the $J=0$ term, to get the modified AdS-to-CFT map given in \eqref{eq:ads_to_cft_free}. Note that we can also plug \eqref{eq:bulk_coeff_scalar} into \eqref{eq:bulk_comp} to get the deformed version of the completeness relation \eqref{eq:bulk_comp_rel}:
		\begin{equation} \label{eq:bulk_comp_rel_scalar}
		\begin{split}
			\Phi_0(X) = \int dY 
			\left(\int_{\gamma_0}\frac{d\Delta}{2\pi i}
			\frac{1}{\lambda_{\Delta,0}}
			\Omega_{\Delta,0}\left(X,Y\right)\right)
			\left(\nabla_{Y}^{2}-M_{d-2,0}^{2}\right)
			\left(\nabla_{Y}^{2}-M_{d,0}^{2}\right)\Phi_{0}(Y)\,,
		\end{split}
		\end{equation}
		\ \linebreak
	which will be useful later.

		
		Equations \eqref{eq:ads_to_cft},\eqref{eq:ads_to_cft_free} are the direct mapping from the bulk fields $\Phi_J$ to the boundary bi-local field. Similarly to the inverse mapping (see section \ref{sec:cft_to_ads}), we understand the $\Delta=\frac{d}{2}+i s$ integral as having a large $s$ cutoff $|s|<s_\text{cutoff}$, and we take it to infinity only after performing all the integrals.
		We can think of the mapping \eqref{eq:ads_to_cft},\eqref{eq:ads_to_cft_free} as an explicit convolution
		\begin{equation}\label{eq:explicit_ads_to_cft}
					\eta(P_1,P_2)= \sum_{J=0} \int dX {\cal M}^{-1}_J\left(P_1,P_2 \mid X,K_W \right) \Phi_J(X,W).
		\end{equation}
		Again, to understand this equation we must perform different convolutions ${{\cal M}^{-1}_J}^{s_\text{cutoff}}$ as a function of the cutoff, and remove it only after performing the $X$ integral. Thus ${\cal M}^{-1}_J$ should be understood as a distribution in $X$.

		Showing the finiteness of \eqref{eq:ads_to_cft},\eqref{eq:ads_to_cft_free} is very similar to the discussion at the end of section \ref{sec:cft_to_ads}. At finite $s_\text{cutoff}$, the possible singularities of the $X$ integral were already taken care of at the beginning of this subsection (assuming the boundary condition for $\Phi_J$ discussed in section \ref{sec:bulk_bc}). The $P$ integrand behaves as $|P|^{-2d}$ at large $|P|$ and thus converges. At $P=P_1$ or $P=P_2$ the integrand has a power of $-\text{Re}(\Delta)=-\frac{d}{2}$, and thus also converges. With a finite $s_{\text{cutoff}}$, the integral over $s$ ($\Delta = \frac{d}{2} + i s$) is finite, as long as the inverse normalization $\frac{1}{f_{\Delta,J}}$ is non-singular (or integrable) on the principal series. 

		We still need to discuss what happens as we try to remove the cutoff, where divergences may arise because $\frac{1}{N_{\Delta,J}} \sim s^d$.
		To study this divergence we approximate the leading large $s$ behavior of the $P,X$ integrals using the stationary phase approximation. Denote by $x_1^i, x_2^i, y^i, (x^i,z)$ the coordinates of $P_1,P_2,P,X$, respectively. For any $d$ and $J$, the integrand of \eqref{eq:ads_to_cft},\eqref{eq:ads_to_cft_free} has the following dependence on $s$:
		\begin{equation}\label{eq:inv_s_depenence}
			\left( \frac{|x_1-x_2|}{|y-x_1||y-x_2|} \frac{(x-y)^2+z^2}{z}\right)^{i s}.
		\end{equation}
		The first term is from the three-point function, and the second is from the bulk-to-boundary propagator. We can find a saddle point for the integrals over $x^i$ and $y^i$, keeping the $z$ integral. The saddle point is $x^i=y^i=\frac{1}{2}(x^i_1+x^i_2)$, and the saddle-point value depends on $s$ as $\left| \frac{z}{x_1-x_2}\right|^{i s}$. The 1-loop determinant (we integrated over $2d$ variables) goes as $s^{-d}$ times an $s$-independent term. We end up with
		\begin{equation}
			\int dP dX \dots \underset{\text{large }s}{\sim} \int dz (\dots) \left| \frac{z}{x_1-x_2}\right|^{i s} \cdot s^{-d},
		\end{equation}
		where the $\dots$ in the brackets on the right-hand side denote a function of $z,x_1,x_2$ which is linear in $\Phi_J(\frac{x_1+x_2}{2},z)$.
		
		In the general case \eqref{eq:ads_to_cft}, we have another overall factor of $\frac{1}{N_{\Delta,J}f_{\Delta,J}} \sim s^d \frac{1}{f_{\Delta,J}}$ and so overall we have
		\begin{equation}
			\eta(x_1,x_2) \underset{\text{large }s}{\sim} \int_{-\infty}^{\infty} ds \frac{1}{f_{\Delta,J}} \int dz (\dots) \left| \frac{z}{x_1-x_2}\right|^{i s}.
		\end{equation}
		Thus, if we assume that at large $s$ $\frac{1}{f_{\Delta,J}}\sim s^a e^{i b s }$ for $a\in \mathbb{Z}$ and $b\in\mathbb{R}$, then the integral over $s$ converges when we remove the cutoff, localizing $z$ to a constant times $|x_1-x_2|$. As above, for $a \geq 0$ the resulting ${\cal M}_J^{-1}$ should be interpreted as a distribution. For $J=0, d<4$ we have outside the integral another factor of $\frac{1}{\lambda_{\Delta,J}} \sim s^{4\Delta_0-2d}=s^{-4}$, which only makes the integral more convergent.
		In section \ref{sec:cft_to_ads} we got the same constraint but on $f$ itself. Together, we find that our mappings are finite (as promised above) if we have
		\begin{equation}\label{eq:norm_integrability_cond}
			\lim_{s\rightarrow\pm\infty} |f_{\frac{d}{2}+i s,J}| \sim s^{a}
		\end{equation}
		for some $a\in \mathbb{Z}$.

		\subsection{The bulk dual of CFT single trace operators}
\label{sec:single_trace_dual}

We will now discuss a specific limit of our mapping, namely how singlet local operators in the CFT are mapped to the bulk. For clarity, we will use explicit coordinates instead of embedding space in this subsection. Recall that spin $J$ ``single trace'' singlet local operators $S^J_{i_1\cdots i_J}(x)$ in the $U(N)$ (or $O(N)$) free theory are defined in terms of the bi-local as 
\es{st}{
S_{i_1\cdots i_J}^J(x_1)\equiv \lim_{\eps\to 0}D_{i_1\cdots i_J}^{J,(x_1,x_2)}\eta(x_1,x_2)\vert_{x_2=x_1+\eps\hat e}\,,
}
where $\hat e$ is an arbitary unit vector, and the bi-local differential operator $D_{i_1\dots i_J}^{J,(x_1,x_2)}$ can be found in \cite{Craigie:1983fb} and is fixed such that $S_J(x)$ is a conformal primary normalized with two-point function
\es{2point2}{
\langle S_{i_1\dots i_J}^J(x_1)S^J_{j_1\dots j_J}(x_2)\rangle=&A_{J}\left(\frac{I_{j_1(i_1}(x_{12})\dotsb I_{i_J)j_J}(x_{12})}{x_{12}^{2(d-2+J)}}-\text{traces}\right)\,,\\
 I_{ij}\equiv\delta_{ij}-2\frac{x_i x_j}{x^2}\,,\quad A_{J}=&\frac{\pi ^{\frac{1}{2}-d}  \Gamma \left(\frac{d}{2}+J-1\right) \Gamma
   (d+J-3)}{2^{d+J}\Gamma (J+1) \Gamma \left(\frac{d-3}{2}+J\right)}\,.
}
For instance, for $J=0$ we have $D^0=1$ and we recover the coefficient $A_{0}=\frac{\Gamma(d/2-1)^2}{16\pi^d}$ in the bi-local 2-point \eqref{4pointfree}.

We will start by showing that $S^J(x)$ for general $J$ is related on-shell to the boundary limit of $\Phi_J(x,z)$ as
\es{toshow}{
\langle S_{i_1\dots i_J}^J(x)\cdots\rangle=\frac{1}{f_{d-2+J,J}}\lim_{\eps\to0}\eps^{2-d}\langle\Phi_{J,i_1\dots i_J}(x,\eps)\cdots\rangle\,,
}
where $(x,\eps)$ are the upper half space coordinates of the bulk position, and the AdS spin indices became boundary spin indices in this limit. The angle brackets and the $\cdots$ denote that this relation holds on-shell, i.e. with any other operator inserted inside an expectation value.

We start by acting with $D_{i_1\dots i_J}^{J,(x_1,x_2)}$ on the bi-local $\eta(x_1,x_1+\eps\hat e)$ in the $\eps\to0$ limit to get
\es{Sdecomp}{
	S^J_{i_1\dots i_J}(x_1) =&\lim_{\eps\to0}\sum_{J'=0}^\infty \int_{\gamma_{J'}}\frac{d\Delta}{2\pi i}\int d^dy C_{j_1\dots j_{J'}}^{\Delta,J'}(y)
D_{i_1\dots i_J}^{J,(x_1,x_2)}\langle   \mathcal{O}_{\Delta_0}(x_1)\hat{\mathcal{O}}_{\Delta_0}(x_2)  \mathcal{O}_{j_1\dots j_{J'}}^{\Delta,J'}(y) \rangle\vert_{x_2=x_1+\eps\hat e}\\
=&\lim_{\eps\to0}\sum_{J'=0}^\infty \int_{\gamma_{J'}}\frac{d\Delta}{2\pi i}\int d^dy C_{j_1\dots j_{J'}}^{\Delta,J'}(y)\frac{2\delta_{J,J'}\langle \mathcal{O}_{i_1,\dots i_{J'}}^{\Delta,J'}(y)\mathcal{O}_{j_1,\dots j_{J'}}^{\Delta,J'}(x_1)\rangle+(\text{$\hat e$-dependent})}{\eps^{2\Delta_0+J'-\Delta}} \\
=& \lim_{\eps\to0}\Big[\int_{\gamma_{J}}\frac{d\Delta}{2\pi i} \frac{2S^{(\tilde\Delta,J)}_{\Delta_0,\Delta_0}C_{i_1\dots i_{J}}^{\tilde\Delta,J}(x_1)}{\eps^{2\Delta_0+J-\Delta}}+ \sum_{J'=0}^\infty \int_{\gamma_{J'}}\frac{d\Delta}{2\pi i}\int d^dy C_{j_1\dots j_{J'}}^{\Delta,J'}(y)\frac{(\text{$\hat e$-dependent})_{i_1\dots i_J;j_i\dots j_{J'}}}{\eps^{2\Delta_0+J'-\Delta}}\Big]\\
}
where in the first equality we used the definition of $S_J$ \eqref{st} and the expansion of the bi-local in the $C_{\Delta,J'}$ basis \eqref{eq:eta_decomp}, in the second equality we used that when $D^J$ acts on the three-point function basis the only leading order in $\eps$ term that does not depend on the arbitrary direction $\hat e$ is given by the spin $J$ two-point function, and in the third equality we used the shadow transform \eqref{eq:basis_extension}. Note the factor of 2 in the second equality, which comes from the fact that in the $\eps\to0$ limit of the three-point, there are two terms related by shadow invariance, analogous to the boundary limit of the bulk-to-boundary propagator \eqref{eq:2_point_limit}. We can then compare this to the small $\eps$ expansion of $\Phi_{J,i_1\dots i_J}(x_1,\eps)$ in the $C^{bulk}_{\Delta,J}$ basis \eqref{eq:bulk_comp} to get
\es{Sdecomp2}{
\Phi_{J,i_1\dots i_J}(x_1,\eps)=&\int_{\gamma_J} \frac{d\Delta}{2\pi i} \int dy f_{\Delta,J}C_{j_1,\dots j_J}^{\Delta,J}(y)G_{i_1\dots i_J| j_1\dots j_J}^{\Delta,J}(x_1,\eps|y)\\
=&2\int_{\gamma_J} \frac{d\Delta}{2\pi i}   f_{\Delta,J}\eps^{\Delta-J} \left[S_{\Delta_0,\Delta_0}^{(\tilde\Delta,J)}C_{i_1,\dots i_J}^{\tilde\Delta,J}(x_1)+O(\eps)\right]\\
}
where in the first equality we used the mapping \eqref{eq:mapping} that holds so long as we change the principal series contour in \eqref{eq:bulk_comp} to $\gamma_J$ as discussed in section \ref{sec:cft_to_ads}, and in the second equality we used the boundary limit of $G_{\Delta,J}$ in \eqref{eq:2_point_limit} and the shadow transform \eqref{eq:basis_extension}. Note the factor of 2 since each term in \eqref{eq:2_point_limit} contributes equally under the $\Delta$ integral.

In general, it is difficult to perform the $\Delta$ integrals in \eqref{Sdecomp} and \eqref{Sdecomp2}, since we know very little about general $C_{\Delta,J}$. For instance, the contour $\gamma_{J'}$ for $d>4$ or $J'>0$ and $d\leq4$ is the principal series $\text{Re}(\Delta)=d/2$, so along this contour the leading term $\eps^{2-d/2-J'}$ diverges for the free theory. To get the finite answer for $S^J_{i_1\dots i_J}(x_1) $ that we expect, there must be complicated cancellations. When $S^J_{i_1\dots i_J}(x_1) $ and  $\Phi_{i_1\dots i_J}^J(x_1,\eps)$ appear in correlation functions, however, we expect to be able to close the contours in \eqref{Sdecomp} and \eqref{Sdecomp2} to get the expected discrete series of poles,\footnote{In computations in the CFT, these discrete poles arise when computing correlators of the $C_{\Delta,J}(y)$, as in the calculation of the bi-local propagator in Appendix \ref{sec:free_4_point}. We will see in the next section that this arises also directly in bulk computations, since for any reasonable choice of $f_{\Delta,J}$ the leading behavior of the bulk propagator of $\Phi_J$ when one of the points goes to the boundary gets a contribution just from a discrete pole at $\Delta=d-2+J$ which gives precisely the behavior in \eqref{toshow}.}
and the result should not depend on the direction $\hat e$ used to define $S^J_{i_1\dots i_J}(x_1) $. So on-shell, we can ignore the second term in \eqref{Sdecomp}, and evaluate both \eqref{Sdecomp} and \eqref{Sdecomp2} by taking the pole $\Delta=2\Delta_0+J=d-2+J$ that gives a finite result for $S^J_{i_1\dots i_J}(x_1) $ (other poles give contributions that vanish as $\eps \to 0$), which yields the relation \eqref{toshow}.

For $d<4$ and $J'=0$, recall that the contour $\gamma_0$ includes a deformation from the principal series to include the pole $\Delta=d-2<d/2$. Since the principal series contribution goes to zero as $\eps^{2-d/2}$ in this case, we know even off-shell 
that the only contribution as $\eps \to 0$ to the integrals in \eqref{Sdecomp} and \eqref{Sdecomp2} comes from the $\Delta=d-2$ pole, which is the only pole on the other side of the principal series. This yields the relation 
\es{toshow2}{
 S^0(x)\equiv\eta(x,x)=\frac{1}{f_{d-2,0}}\lim_{\eps\to0}\eps^{2-d}\Phi_0(x,\epsilon)\,,
}
which unlike the general equation \eqref{toshow} holds even off-shell, and in particular it continues to hold under deformations of the theory. This off-shell relation will prove useful when discussing ``double-trace'' deformations and the critical theory below.

\section{The bulk quadratic action}\label{sec:quadratic_action}

Now that we have defined an explicit mapping in the previous section between the bilocal $\eta(P_1,P_2)$ and the bulk fields $\Phi_J(X,W)$ for every spin $J$, we can use it to construct the bulk action by mapping the bilocal action. In this section, we focus on the quadratic bulk action, as obtained from the quadratic bilocal action in section \ref{quadFree}. For general bulk normalization $f_{\Delta,J}$, we will compute the bulk 2-point function using this action and analyze the particle spectrum. For a certain choice of $f_{\Delta,J}$, we will get a local quadratic action, and the 2-point function can then be simply written in terms of (modified) bulk-to-bulk propagators.

	\subsection{The quadratic action}
		We start from the bi-local quadratic action written in terms of the $C_{\Delta,J}$ in \eqref{eq:c_action} for the $U(N)$ theory, which we repeat here for clarity:
		\begin{equation}\label{eq:c_action_2}
		\begin{split}
			S^{(2)} & =  \sum_{J=0}^\infty \int_{\gamma_J} \frac{d\Delta}{2\pi i}
			\left(-1\right)^J \lambda_{\Delta,J} N_{\Delta,J}
			\frac{1}{J! \left(\frac{d}{2}-1\right)_J} \int dP \, C_{\Delta,J}(P,D_Z)C_{\tilde\Delta,J}(P,Z)\,.
		\end{split}
		\end{equation}
We can plug in the value of $C_{\Delta,J}$ in terms of the bulk fields, coming from \eqref{eq:bulk_coeff} and the mapping \eqref{eq:mapping}, to get
		\es{eq:bulk_S2}{
			S^{(2)}[\Phi_J] =& \sum_{J=0}^\infty  
			\int \frac{dX_1dX_2}{(\left(\frac{d-1}{2}\right)_{J}J!)^2}   \Phi_{J}\left(X_1,K_{W_1}\right) \Phi_{J}\left(X_2,K_{W_2}\right)\int_{\gamma_J} \frac{d\Delta}{2\pi i}
			\frac{\left(-1\right)^J \lambda_{\Delta,J}}{\alpha_J  f_{\Delta,J}f_{\tilde \Delta,J}}\Omega_{\Delta,J}(X_1,X_2;W_1,W_2)	 \\
	=& \sum_{J=0}^\infty  
			\int \frac{dX_1dX_2}{(\left(\frac{d-1}{2}\right)_{J}J!)^2} \int_{\gamma_J} \frac{d\Delta}{2\pi i} \Phi_{J}\left(X_1,K_{W_1}\right) \left(\nabla_{X_2}^{2}-M_{d+J-2,J}^{2}\right)\left(\nabla_{X_2}^{2}-M_{d+J,J}^{2}\right) \Phi_{J}\left(X_2,K_{W_2}\right) 
				 \\
			&  \qquad	\times\frac{\left(-1\right)^J}{\alpha_J f_{\Delta,J}f_{\tilde \Delta,J}}\Omega_{\Delta,J}(X_1,X_2;W_1,W_2)\,,
		}
		where in the second equality we used integration by parts and \eqref{eq:lambda_eigenvalue}. Note that for $J=0$ and $d<4$ we must use \eqref{eq:bulk_coeff_scalar} instead of  \eqref{eq:bulk_coeff}, so the $J=0$ term in \eqref{eq:bulk_S2} should be replaced by
		\begin{equation}\label{eq:bulk_S2_scalar}
		\begin{split}
			&\int dX_1 dX_2   \left(\nabla_{X_1}^{2}-M_{d-2,0}^{2}\right)
			\left(\nabla_{X_1}^{2}-M_{d,0}^{2}\right)\Phi_{0}(X_1) \\
			& \qquad \times \left(\nabla_{X_2}^{2}-M_{d-2,0}^{2}\right)
			\left(\nabla_{X_2}^{2}-M_{d,0}^{2}\right)\Phi_{0}(X_2) \int_{\gamma_0} \frac{d\Delta}{2\pi i} \frac{\Omega_{\Delta,0}(X_1,X_2)}{\alpha_0\lambda_{\Delta,0}f_{\Delta,0}f_{\tilde\Delta,0}}
	\,.
		\end{split}
		\end{equation}
		
For generic $f_{\Delta,J}$ this non-local quadratic action \eqref{eq:bulk_S2} is the best we can do. However, if we choose an $f^\text{local}_{\Delta,J}$ so that 
\es{floc}{
f^\text{local}_{\Delta,J} f^\text{local}_{\tilde\Delta,J} = (-1)^J\,,
}
then we can apply the bulk completeness relation \eqref{eq:bulk_comp_rel} to \eqref{eq:bulk_S2} to get the local quadratic action
\begin{equation}\label{eq:bulk_S2_local}
\begin{split}
	S^{(2)}_\text{local}[\Phi_J] = \sum_{J=0}^\infty &
	 \frac{1}{\alpha_J}
	\int \frac{dX}{\left(\frac{d-1}{2}\right)_{J}J!}  \Phi_{J}(X,K_W) \left(\nabla_{X}^{2}-M_{d+J-2,J}^{2}\right)\left(\nabla_{X}^{2}-M_{d+J,J}^{2}\right) \Phi_{J}(X,W)\,,
\end{split}
\end{equation}
where we used the fact that $\Phi_J(X,W)$ is transverse. So for this special choice of the normalization of our mapping we obtain a local bulk quadratic action\footnote{The same local bulk quadratic action was found in \cite{deMelloKoch:2018ivk}, using different considerations to determine the precise mapping.}. For the $O(N)$ theory, we get the same answer except with only even $J$ and an extra factor of $\frac12$ from the bilocal quadratic action. We will discuss our interpretation of the two masses that appear here below, when we compute the bulk 2-point function and analyze the resulting spectrum.
Note that for $J=0$ and $d<4$ we get a slightly different action \eqref{eq:bulk_S2_scalar}, which can naively be related by integration by parts to \eqref{eq:bulk_S2_local} (but this can lead to extra boundary terms in general). However, this modified action gives precisely the same bulk two-point function \eqref{eq:Omega_BB_props} as we get for $d>4$ below, and we can also obtain this bulk two-point function by directly mapping the CFT two-point function to the bulk (using the $d<4$ mapping).

Making this ``local'' choice for $f$, we can in fact solve the constraint \eqref{floc} for $f^\text{local}_{\Delta,J}$ using the shadow relation \eqref{eq:bulk_norm_shadow_cond} and the reality constraints \eqref{bulk_norm_reality_cond} on $f_{\Delta,J}$, to get the explicit form
		\begin{equation} \label{eq:local_f}
			f_{\Delta,J}^\text{local} = \left( (-1)^J \frac{S^{\tilde\Delta,J}_B}{S^{\tilde\Delta,J}_{\Delta_0,\Delta_0}}\right)^{\frac{1}{2}} 
			= \left( (-1)^J \frac{\Gamma(\Delta+J) \Gamma^2\left(\frac{\tilde\Delta+J}{2}\right)}{\Gamma(\tilde\Delta+J) \Gamma^2\left(\frac{\Delta+J}{2}\right)} \right)^{\frac{1}{2}}\,.
		\end{equation}
Notice that \eqref{eq:local_f} has a unit modulus $|f_{\Delta,J}^\text{local}|=1$ everywhere on the principal series. Thus it also satisfies the condition \eqref{eq:norm_integrability_cond} for a finite mapping. Unfortunately, this form of $f_{\Delta,J}$ is not analytic in $\Delta$ (though it is analytic near the principal series), so that we have to be careful when continuing expressions involving it away from the principal series. The function \eqref{eq:local_f} has branch points at $\Delta=d+J+n$ and $\Delta=-J-n$ for $n=0,1,\cdots$, and we can choose it to have branch cuts on the intervals $[d+J+n,d+J+n+1]$ and $[-J-n-1,-J-n]$ for even non-negative integers $n$. With this choice $f_{\Delta,J}^\text{local}$ is analytic between $-J<\Delta<d+J$, and specifically in the region between the principal series $\text{Re}(\Delta)=\frac{d}{2}$ and the physical dimension $\Delta=d-2+J$ (for $d>2)$, which is what we need for our analytic continuations.

There are other choices of $f$ which are more complicated but which still give local quadratic terms. All these choices (if we do not introduce additional singularities in $f$ beyond those of \eqref{eq:local_f}) still have a pole of the bulk propagator at $M^2_{d+J-2,J}$, while the pole at $M_{d+J,J}$ may be shifted to other locations.

\subsection{Bulk-to-Bulk propagators}

We can now use the quadratic bulk action to compute bulk correlation functions, which will be expressed in terms of bulk-to-bulk propagators\footnote{We use this term to refer to the propagator of bulk fields with a canonical kinetic term, it should not be confused with the two-point function of our bulk fields.}. Since our bulk fields $\Phi_J(X,W)$ are transverse, the bulk-to-bulk propagators that naturally arise for us differ slightly from the standard bulk-to-bulk propagator discussed e.g. in \cite{Costa:2014kfa}, which naturally applies to bulk fields that are not transverse. We will first review the standard propagators, and then discuss our modification.

The massive bulk-to-bulk propagator $\Pi_{\Delta,J}(X_1,X_2;W_1,W_2)$ is defined in \cite{Costa:2014kfa} as the solution to the differential equation 
\es{PbulkODE}{
(\nabla^2_{1}-M^2_{\Delta,J})\Pi_{\Delta,J}(X_1,X_2;W_1,W_2)=-\delta(X_1,X_2)W_{12}^J+\cdots\,,
}
subject to the modified transversality condition
\es{modtrans}{
\nabla_1\cdot K_{W_1} \Pi_{\Delta,J}(X_1,X_2;W_1,W_2)=\cdots\,,
}
and to the boundary condition relation to the bulk-to-boundary propagator 
\begin{equation}
	\lim_{z\rightarrow 0}\Pi_{\Delta,J}(X,\frac{1}{z}P;W,\frac{1}{z}Z) = z^{\Delta-J} {\cal C}_{\Delta,J} G_{\Delta,J}(X,P;W,Z) +O(z^{\Delta-J+1})\,,  \label{eq:BB_limitP}
\end{equation}
with\footnote{Our normalization for $\Pi_{\Delta,J}(X_1,X_2;W_1,W_2)$ is the same as \cite{Costa:2014kfa}, but our $G_{\Delta,J}(X,P;W,Z)$ differs by the constant ${\cal C}_{\Delta,J} $, as mentioned before.}
\begin{equation}
	{\cal C}_{\Delta,J} \equiv \frac{(J+\Delta-1)\Gamma(\Delta-1)}{2\pi^{\frac{d}{2}} \Gamma(\Delta+1-\frac{d}{2})}\,. \label{eq:bb_limit_coeff}
\end{equation}
The $\cdots$ in \eqref{PbulkODE} and \eqref{modtrans} signify that these equations hold up to nonzero contact terms, so that $\Pi_{\Delta,J}(X_1,X_2;W_1,W_2)$ is not in fact transverse. The explicit $\Pi_{\Delta,J}(X_1,X_2;W_1,W_2)$ was constructed including these contact terms in \cite{Costa:2014kfa}, using the split presentation
\es{massprop}{
&	\Pi_{\Delta,J}(X_1,X_2;W_1,W_2) =\sum_{l=0}^J \int_{\gamma_J} \frac{d \Delta^\prime}{2\pi i} a_l(\Delta^\prime,\Delta)\frac{((W_1\cdot \nabla_1)(W_2\cdot \nabla_2))^{l}}{M^2_{\Delta,J}-M^2_{\Delta^\prime,J}} \Omega_{\Delta',J-l}(X_1,X_2;W_1,W_2)
\,,
	}
	where $\gamma_J$ is the contour used in previous sections. The coefficients $a_l(\Delta^\prime,\Delta)$ are given by a recursion formula, of which we will only use the fact that $a_0(\Delta^\prime,\Delta)=1$, and that the poles in $\Delta$ for $\Pi_{\Delta,J}$ are then given by
	\es{polesBB}{
\Pi_{\Delta,J}\quad \text{poles}:\quad \bold{P}_J=d-1,d,\dots d+J-2 \quad\text{and}\quad \tilde{\bold{P}}_J=1,0,\dots, (2-J)\,.
}
Note that $\tilde{\bold{P}}_J$ is the shadow of $\bold{P}_J$, and that if we only had the $l=0$ term in \eqref{massprop}, then $\Pi_{\Delta,J}$ would have been transverse since $G_{\Delta,J}$ is transverse \eqref{eq:bulk_boundary_lorentz}. The AdS harmonic function was shown in \cite{Costa:2014kfa} to equal
	\es{omega}{
\Omega_{\Delta,J}(X_1,X_2;W_1,W_2)=\left(\Delta-\frac d2\right)\left(\Pi_{\Delta,J}(X_1,X_2;W_1,W_2) -\Pi_{\tilde\Delta,J}(X_1,X_2;W_1,W_2)\right) \,,
	}
which we can then plug back into \eqref{massprop} and use the fact that the contour, the denominator, and the $a_l(\Delta^\prime,\Delta)$ are invariant under $\Delta'\leftrightarrow \tilde\Delta'$ to get 
\es{massprop2}{
	\Pi_{\Delta,J}(X_1,X_2;W_1,W_2) =\sum_{l=0}^J \int_{\gamma_J} \frac{d \Delta^\prime}{2\pi i} a_l(\Delta^\prime,\Delta)((W_1\cdot \nabla_1)(W_2\cdot \nabla_2))^{l}\frac{(2\Delta'-d)\Pi_{\Delta^\prime,J-l}(X_1,X_2;W_1,W_2)}{M^2_{\Delta,J}-M^2_{\Delta^\prime,J}}\,.
	}
We can then perform the $\Delta$ integral by closing the contour to the right, where for each $l$ we will pick up both the single pole $\Delta'=\Delta$ from $(M^2_{\Delta,J}-M^2_{\Delta^\prime,J})^{-1}$, as well as all the poles $\bold{P}_{J-l}$ in \eqref{polesBB} from $\Pi_{\Delta^\prime,J-l}$. We see that the entire right-hand side in the expression for $\Pi_{\Delta,J}$ is in fact given by the $\Delta'=\Delta$ pole for $l=0$, and that the role of the $l>0$ terms is just to cancel the spurious poles $\bold{P}_{J-l}$ for each $l$. This is in fact one way of fixing the coefficients $a_l(\Delta^\prime,\Delta)$.

The preceding discussion was for the massive propagator with $\Delta>d-2+J$. In the massless limit $\Delta\to d-2+J$, we see from \eqref{polesBB} that the propagator diverges, which is the standard story for gauge fields that one needs to fix a gauge to find a finite propagator. Several such gauges are discussed in \cite{Bekaert:2014cea}, and their ultimate effect is to modify the explicit form of the $a_l(\Delta^\prime,\Delta)$ in \eqref{massprop} for $l>0$. The $l=0$ term is gauge-invariant, however, and corresponds to the physical transverse propagating modes. One can define a bulk-to-bulk propagator $\Pi^{TT}$ that only includes this physical mode as
\es{usprop}{
&	\Pi^{TT}_{\Delta,J}(X_1,X_2;W_1,W_2)=  \int_{\gamma_J} \frac{d \Delta^\prime}{2\pi i}\frac{\Omega_{\Delta^\prime,J}(X_1,X_2;W_1,W_2)}{M^2_{\Delta,J}-M^2_{\Delta^\prime,J}}\,,
	}
	where we took just the $l=0$ term from the right-hand side of \eqref{massprop}. The superscript $TT$ refers to the fact that since we only include the $l=0$ term, this propagator is both traceless and transverse
	\es{trans}{
\nabla_1\cdot K_{W_1} \Pi^{TT}_{\Delta,J}(X_1,X_2;W_1,W_2)=0\,.
}
These properties are in fact exactly what we need for our transverse bulk field. We can then evaluate the $\Delta$ integral just as we did above by replacing $\Omega_{\Delta,J}$ by $\Pi_{\Delta,J}$ using \eqref{omega} and the shadow invariance of the contour, then closing the contour to the right and collecting poles to get
	\es{usprop2}{
	\Pi^\text{TT}_{\Delta,J}(X_1,X_2;W_1,W_2)& =\Pi_{\Delta,J}(X_1,X_2;W_1,W_2)-\sum_{p=d-1}^{d+J-2}  \frac{(2p-d)}{M^2_{\Delta,J}-M^2_{p,J}}\text{Res}\left[  \Pi_{\Delta^\prime,J}(X_1,X_2;W_1,W_2)   \right]_{\Delta^\prime=p}\,,
	}
	where the first term came from the $\Delta'=\Delta$ pole from $(M^2_{\Delta,J}-M^2_{\Delta^\prime,J})^{-1}$, and the second term came from the poles $\bold{P}_J$ of $\Pi_{\Delta^\prime,J}$. In the massless limit $\Delta\to d-2+J$, the $\Delta'=d-2+J$ pole becomes a double pole and we get the finite result
	 \es{usprop3}{
	&\Pi^\text{TT}_{d-2+J,J}(X_1,X_2;W_1,W_2) =\\
	&\partial_\Delta\Big[(\Delta-d-J+2)\Pi_{\Delta,J}(X_1,X_2;W_1,W_2)\Big]_{\Delta=d-2+J}- \frac{\text{Res}   \left[   \Pi_{\Delta^\prime,J}(X_1,X_2;W_1,W_2) \right]_{\Delta^\prime=d-(2-J)}}{4-d-2J} \\
	&-\sum_{p=d-1}^{d+J-3}  \frac{(2p-d)}{M^2_{\Delta,J}-M^2_{p,J}}\text{Res}\left[  \Pi_{\Delta^\prime,J}(X_1,X_2;W_1,W_2)   \right]_{\Delta^\prime=p}\,.
	}
	
	The transverse bulk-to-bulk propagator satisfies many of the features of the standard bulk-to-bulk propagator, with some slight modifications. We find that \eqref{omega} applies identically to $\Pi^{TT}_{\tilde\Delta,J}$ as
	\es{omega2}{
\Omega_{\Delta,J}(X_1,X_2;W_1,W_2)=\left(\Delta-\frac d2\right)\left(\Pi^{TT}_{\Delta,J}(X_1,X_2;W_1,W_2) -\Pi^{TT}_{\tilde\Delta,J}(X_1,X_2;W_1,W_2)\right) \,,
	}
	since the difference between $\Pi^{TT}_{\Delta,J}$ and $\Pi_{\Delta,J}$ is the second term in \eqref{usprop2}, which is shadow invariant. We also find that $\Pi^{TT}_{\Delta,J}$ satisfies a differential equation similar to \eqref{PbulkODE}, which takes the form
	\begin{equation}\label{eq:bulk_bulk_DE_2}
\begin{split}
 	\left( \nabla^2_1 - M^2_{\Delta,J}\right) &\Pi^\text{TT}_{\Delta,J}(X_1,X_2;W_1,W_2)   = -\delta^{TT}\left(X_{1},X_{2}\right)\left(W_{12}\right)^{J} \,.
\end{split}
\end{equation}
This follows from writing the right-hand side of \eqref{usprop} in terms of $G_{\Delta,J}$'s using \eqref{omega}, applying the differential equation on $G_{\Delta,J}$ in \eqref{eq:bulk_boundary_DE}, and then using the bulk completeness relation \eqref{eq:bulk_comp_rel}, where recall that $\delta^{TT}\left(X_{1},X_{2}\right)$ acts as a delta function on traceless transverse functions. Lastly, from \eqref{usprop2} we see that the small $z$ limit is
\begin{equation} \label{eq:BB_limit}
	\lim_{z\rightarrow 0}\Pi^\text{TT}_{\Delta,J}(X,\frac{1}{z}P;W,\frac{1}{z}Z) =
	\begin{cases}
z^{\Delta-J} {\cal C}_{\Delta,J} G_{\Delta,J}(X,P;W,Z) +O(z^{\Delta-J+1}) \qquad \Delta<d-J+4\\
O(z^{d-2J+4})\qquad\qquad\qquad\qquad\qquad\qquad\qquad\;\;\;\, \Delta\geq d-J+4 \\
\end{cases}
\end{equation}
where the $\Delta<d-J+4$ case comes from the $\Pi_{\Delta,J}$ term in \eqref{usprop2} and the standard small $z$ limit \eqref{eq:BB_limitP}, while the scaling for the $\Delta\geq d-J+4$ case comes from the $\text{Res}\left[  \Pi_{\Delta^\prime,J}  \right]_{\Delta^\prime=d-1}$ term in \eqref{usprop2}, whose small $z$ limit can be computed from the explicit results in \cite{Costa:2014kfa} and takes a complicated form that we will not use (aside from its scaling in $z$).

	\subsection{The tree-level bulk two-point function}
	
	We are now ready to use the quadratic bulk action to compute the bulk two-point function, by inverting the operator that appears in our quadratic term. In our case, we see from the first line of \eqref{eq:bulk_S2} that this operator includes the AdS harmonic function $\Omega_{\Delta,J}(X_1,X_2;W_1,W_2)$, which on the principal series satisfies the orthogonality property \cite{Costa:2014kfa}
	\es{orthoOmega}{
	\int dX \Omega_{ \Delta,J}(X_1,X;W_1,K_W)\Omega_{\Delta',J}(X,X_2;W,W_2)=\frac{2\pi i}{2}(\delta(\Delta-\Delta')+\delta(\Delta-\tilde\Delta'))\Omega_{\Delta,J}(X_1,X_2;W_1,W_2)\,.
	}
	We can use this identity and the bulk completeness relation \eqref{eq:bulk_comp_rel} (as applied to transverse functions) to compute the two-point function in the bulk dual of the $U(N)$ theory
		\es{eq:Omega_BB_props}{
			\left< \Phi_J(X_1,W_1) \Phi_J(X_2,W_2)\right> =&\frac{\alpha_J}{2}\int_{\gamma_J} \frac{d\Delta}{2\pi i} \frac{f_{\Delta,J}f_{\tilde \Delta,J}}{(-1)^J \lambda_{\Delta,J}} \Omega_{\Delta,J}(X_1,X_2;W_1,W_2)\\
			=&\frac{\alpha_J}{2} \int_{\gamma_J} \frac{d\Delta}{2\pi i} 
			\frac{f_{\Delta,J}f_{\tilde \Delta,J}\left(2\Delta-d\right)}
			{(-1)^J \lambda_{\Delta,J}} \Pi^\text{TT}_{\Delta,J}(X_1,X_2;W_1,W_2)\,,
}
where in the second equality we used \eqref{omega2}. For the bulk dual of the $O(N)$ theory, we would get an extra factor of 2 since the bulk quadratic action had an extra factor of $\frac12$. For the local $f^\text{local}_{\Delta,J}$ defined in \eqref{eq:local_f}, we can further simplify this expression as
\begin{equation}\label{eq:2p_local}
		\begin{split}
			\langle & \Phi_J (X_1,W_1) \Phi_J(X_2,W_2)\rangle = 
			\frac{\alpha_J}{2} \int_{\gamma_J} \frac{d\Delta}{2\pi i} 
			\frac{\left(2\Delta-d\right)}
			{\lambda_{\Delta,J}} \Pi^\text{TT}_{\Delta,J}(X_1,X_2;W_1,W_2)\\
			& = \frac{\alpha_J/2}{M^2_{d+J,J}-M^2_{d+J-2,J}} 
			\left(
				\Pi^\text{TT}_{d-2+J,J}(X_1,X_2;W_1,W_2)-\Pi^\text{TT}_{d+J,J}(X_1,X_2;W_1,W_2)
			\right)\,,
		\end{split}
		\end{equation}
		where in the second line we closed the contour to the right and picked up the poles of $1/\lambda_{\Delta,J}$ at $\Delta=d-2+J$ and $\Delta=d+J$. For $f^\text{local}_{\Delta,J}$, we see that we have in the bulk a physical particle with spin $J$ and mass related to $\Delta=d-2+J$ (for $J>0$ this is a massless particle), as well as a non-physical particle with spin $J$ and mass related to $\Delta=d+J$ with a wrong-sign propagator, which is sub-leading in the boundary limit according to \eqref{eq:BB_limit}. Note that the existence of a physical particle at $\Delta=d-2+J$ is expected since it should be dual to the local operators in our theory, so we will only discuss choices of $f_{\Delta,J}$ for which it appears: this imposes the mild limitation on $f_{\Delta,J}$ that it should be analytic between the principal series and $\Delta=d-2+J$, so that the leading pole at $\Delta=d-2+J$ in \eqref{eq:Omega_BB_props} will give in the boundary limit \eqref{eq:BB_limit}:
		\begin{equation}\label{eq:2p_localf}
		\begin{split}
			\lim_{z_1\to0}\langle & \Phi_J ({z^{-1}}P,{z^{-1}}Z) \Phi_J(X,W)\rangle
			 =\lim_{z_1\to0} \frac{(-1)^J\alpha_Jf_{d-2+J,J}f_{2-J,J}\mathcal{C}_{d-2+J,J}}{2(M^2_{d+J,J}-M^2_{d+J-2,J})} z^{d-2}G_{d-2+J,J}(P,X;Z,W)\,,
		\end{split}
		\end{equation}
		where any poles $\Delta>d-2+J$ will be subleading in $z$.
		

	For the local choice of $f$, we found that our bulk propagator is the difference of two propagators of on-shell symmetric traceless transverse spin $J$ fields, a field with positive propagator and mass $M_{d+J-2,J}$, and a field with negative propagator and mass $M_{d+J,J}$. We identify the on-shell fields of the first type with the physical operators in our theory; note that for $J>0$ these are massless spin $J$ fields in the bulk, dual to conserved currents (and, in particular, for $J=2$ we obtain the standard physical part of the propagator for a graviton in a fixed AdS space-time). The particle spectrum that we find, in the $U(N)$ case, is exactly the same as the particle spectrum of Vasiliev's theory, including massless spin $J$ fields with all $J$. The canonical quadratic action for such a massless spin $J$ field (which is symmetric and double-traceless, see \cite{Gaberdiel:2010ar,Gaberdiel:2010xv,Gupta:2012he}), has (for $J>0$) a gauge freedom, which we can gauge-fix to obtain a symmetric traceless transverse field (the details of this gauge-fixing may be found in \cite{Gaberdiel:2010ar,Gaberdiel:2010xv,Gupta:2012he}), and in this gauge we obtain precisely the spin $J$ fields that we found in our theory (with the same propagator as for our physical particles, up to a constant). The gauge-fixing leads to an extra ghost field in the bulk, coming from the Faddeev-Popov determinant, with spin $(J-1)$ and mass $M_{d+J-1,J-1}$, so we can identify the extra negative-propagator modes that we found above with these ghosts (see \cite{Giombi:2016ejx} and references therein). More precisely, for $J>2$, the original spin $J$ field contains also a trace part which is a symmetric spin $(J-2)$ field, but on-shell one can use extra gauge transformations to set this trace part to zero, consistent with the fact that we do not see the corresponding on-shell modes in our approach; moreover, this extra gauge-fixing precisely makes the ghosts into symmetric traceless transverse tensors, which is what we obtain in our formalism.

	While this identification with the expectations from Vasiliev's theory works very nicely for the $U(N)$ case, it actually fails for the $O(N)$ case. This is because in our spectrum we find physical and unphysical fields just for even spins, while from the minimal Vasiliev action that would start from massless gauge fields with even spins $J$, we would get physical fields of even spin but unphysical ``ghosts'' with odd spin. It would be interesting to understand the reason for this discrepancy. In any case, our bulk action by construction provides a consistent dual for the $U(N)$ and $O(N)$ theories, whether or not it is related to the standard Vasiliev theory in a simple way.

	Note that even though on-shell we found (for $U(N)$) precisely the same spectrum as in Vasiliev's theory, the quadratic terms we find are very different from the naive quadratic terms one would associate with the fields of that theory. Of course in any case Vasiliev's theory is only known classically (on-shell), so we claim that our action is the correct way to define it off-shell (in perturbation theory in $1/N$), in the specific gauge-choice described above (more precisely, we should probably work with the formalism where this theory is written on a fixed AdS background, as in \cite{Neiman:2015wma}). In our off-shell formulation, our analysis guarantees that at all loop orders this theory (with the specific counter-terms we found) will give the correlation functions of the free $U(N)$ theory. Note that, as discussed above, for finite $N$ the bulk fields that we find obey complicated constraints, such that even empty AdS space ($\Phi_J=0$) is not a consistent configuration in the finite $N$ path integral; but at any order in $1/N$ we can ignore this.

\section{The higher order bulk action} \label{sec:bulk_theory}
In the previous section, we only discussed the quadratic term in the bulk higher spin theory, which we constructed by mapping the corresponding term in the bi-local action. We will now construct all the higher order bulk interaction and counter terms from the CFT, which will be non-local for any choice of $f_{\Delta,J}$.  We then derive the bulk Feynman rules, and apply them to show that 1-loop terms cancel in the bulk, just as they did in the CFT. For simplicity, we discuss here the case $d>4$. For $d<4$ the $J=0$ fields need to be handled differently, but at the end of the day they lead to the same bulk action (see appendix \ref{app:d4_mod}). 


\subsection{Map identities and diagrammatics}

In the previous section, we derived the bulk dual of the bi-local quadratic action in \eqref{eq:free_pert_bilocal_action} by plugging in the bulk definition of $C_{\Delta,J}$ in \eqref{eq:bulk_coeff} and \eqref{eq:mapping} into the bi-local quadratic action as written in terms of $C_{\Delta,J}$ in \eqref{eq:c_action}. To derive the bulk dual of the higher order terms in \eqref{eq:free_pert_bilocal_action}, we will use a conceptually identical but technically simpler approach, where we directly apply the CFT-to-AdS map \eqref{eq:cft_to_ads_1} to each term in \eqref{eq:free_pert_bilocal_action}. To apply this map, we will find it convenient to first write the map in terms of various bulk bi-local functions, which satisfy some key identities. 

Our first step is to write the maps purely using bulk variables. The original definition of the CFT-to-AdS map and its inverse in \eqref{eq:explicit_cft_to_ads} and \eqref{eq:explicit_ads_to_cft} were in terms of the functions $\mathcal{M}_J(X,W|P_1,P_2)$ and $\mathcal{M}_J^{-1}(P_1,P_2|X,W)$, respectively, each of which were expressed as an integral over a boundary point $P$ in \eqref{eq:cft_to_ads_1} and \eqref{eq:ads_to_cft}, respectively. We can exchange the boundary integral in
\begin{equation}\label{eq:ads_to_cft_theta1}
	{\cal M}^{-1}_J \left(P_1,P_2 | X,W\right) = \int_{\gamma_J}\frac{d\Delta}{2\pi i} 
	\int  \frac{dP}{J! \left(\frac{d}{2}-1\right)_J}
	\thpnorm{P_1}{P_2}{\Delta,J}{P,D_Z}
	 \frac{G_{\tilde{\Delta},J}\left(X,P;W,Z\right) }{\alpha_J f_{\Delta,J} N_{\Delta,J}}  
\end{equation}
for a bulk integral by using the expression for a CFT 3-point function in terms of its Witten diagram in the bulk \cite{Costa:2014kfa,Freedman:1998tz}
\begin{equation} \label{eq:penedones_3point}
\begin{split}
	\thp{\Delta_1}{P_1}{\Delta_2}{P_2}{\Delta,J}{P,Z}& =\frac{1}{  b^{(\Delta,J)}_{\Delta_1,\Delta_2} } \int \frac{dX'}{J! \left(\frac{d-1}{2}\right)_J} G_{\Delta_2,0}(X',P_2)  \\
	& \quad \times G_{\Delta,J}(X',P; K_{W'},Z) (W'\cdot \nabla)^J G_{\Delta_1,0}(X',P_1)\,,
\end{split}
\end{equation}
with
\begin{equation}
	b^{(\Delta,J) }_{\Delta_1,\Delta_2}= 
	\frac{
	\Gamma\left(\frac{\Delta_1 + \Delta_2 + \Delta-d + J}{2}\right)
	\Gamma\left(\frac{\Delta_1 + \Delta_2 - \Delta   + J}{2}\right)
	\Gamma\left(\frac{\Delta   + \Delta_1 - \Delta_2 + J}{2}\right)
	\Gamma\left(\frac{\Delta_2 + \Delta   - \Delta_1 + J}{2}\right)
	}{2^{1-J}\pi^{-\frac{d}{2}}\Gamma(\Delta_1)\Gamma(\Delta_2)\Gamma(\Delta+J)}\,,
\end{equation}
along with the definition of $\Omega_{\Delta,J}(X',X;W',W)$ in \eqref{eq:omega_def}. This leads to
\begin{equation}\label{eq:ads_to_cft_theta}
	{\cal M}^{-1}_J \left(P_1,P_2 | X,W\right) =
	\int  \frac{dX' }{J! \left(\frac{d-1}{2}\right)_J} (K_{W'} \cdot \nabla_{X'})^J G_{\Delta_0,0}(X',P_1) G_{\Delta_0,0}(X',P_2) 
	\theta_J(X',X;W',W)\,,
\end{equation}
where we define the bulk bi-local function 
\begin{equation}
	\theta_J(X',X;W',W) = \int_{\gamma_J}\frac{d\Delta}{2\pi i} \frac{1}{b^{(\Delta,J)}_{\Delta_0,\Delta_0} f_{\Delta,J}} \Omega_{\Delta,J}(X',X;W',W)\,. \label{eq:theta_def}
\end{equation}
We can similarly rewrite $\mathcal{M}_J(X,W|P_1,P_2)$ as
\begin{equation}\label{eq:cft_to_ads_theta}
\begin{split}
	{\cal M}_J\left(X,W | P_1,P_2 \right)
	&=\frac12\int_{\gamma_J}\frac{d\Delta}{2\pi i} \frac{f_{\Delta,J}}{N_{\Delta,J}}
	\int \frac{dP}{J!\left(\frac{d}{2}-1\right)_{J}} 
 	G_{\Delta,J}\left(X,P;W,D_Z\right)
 	\thptild{P_1}{P_2}{\tilde\Delta,J}{P,Z}\\
	& 	=\frac{1}{J! \left(\frac{d-1}{2}\right)_J}
	\int dX' (K_{W'}\cdot \nabla_{X'})^J G_{\tilde\Delta_0,0}(X',P_1) G_{\tilde\Delta_0,0}(X',P_2)
	{\tilde \theta}_J(X',X;W',W)\,,
\end{split}
\end{equation}
where we define the bulk bi-local function ${\tilde \theta}_J$ as
\begin{equation}
	{\tilde \theta}_J(Y,X;Q,W) = \int_{\gamma_J}\frac{d\Delta}{2\pi i} \frac{\alpha_J f_{\Delta,J}}{2 b^{(\tilde \Delta,J)}_{\tilde\Delta_0,\tilde\Delta_0}}\Omega_{\Delta,J}(X',X;W',W)\,. \label{eq:tilde_theta_def}
\end{equation}
It is useful to define the ``multiplication" of $\theta_J,\tilde \theta_J$ as\footnote{Adding back the cutoffs on the two $\Delta$ integrals, the right-hand side involves an integral over the region $|s|<\text{min}(s_\text{cutoff}^1,s_\text{cutoff}^2)$.}
\begin{equation} \label{eq:Theta_def}
\begin{split}
	\Theta_J(X_1,X_2;W_1,W_2) & = \int  \frac{dX}{J! \left(\frac{d-1}{2}\right)_J} \theta_J(X_1,X;W_1,K_{W}) {\tilde \theta}_J(X,X_2;W,W_2)\\
	& = \int \frac{d\Delta}{2\pi i} \frac{\alpha_J \Omega_{\Delta,J}(X_1,X_2;W_1,W_2)}{2 b^{(\Delta,J)}_{\Delta_0,\Delta_0} b^{(\tilde \Delta,J)}_{\tilde \Delta_0,\tilde\Delta_0}}\,,
\end{split}
\end{equation}
where the second line follows from \eqref{orthoOmega}. Note that $\Theta_J$ is independent of $f_{\Delta,J}$, unlike $\mathcal{M}_J$ and $\mathcal{M}^{-1}_J$ which depend on $f_{\Delta,J}$ via $\tilde\theta_J$ and $\theta_J$, respectively. Another useful property of $\theta_J,\tilde \theta_J$ is the identity
\begin{equation}\begin{split}
	\delta^{TT} & (X_1,X_2) W_{12}^{J_1} \delta_{J_1,J_2} = \int dP_1 \int dP_2 \ {\cal M}_{J_1}\left(X_1,W_1 | P_1,P_2 \right) 
	{\cal M}_{J_2}^{-1}\left(P_1,P_2 | X_2,W_2\right)\\
	& = 
	\int  \frac{dX'_1 dX'_2  }{J_1! J_2! \left(\frac{d-1}{2}\right)_{J_1}\left(\frac{d-1}{2}\right)_{J_2}}
	{\tilde \theta}_{J_1}(X'_1,X_1;W'_1,W_1){\theta}_{J_2}(X'_2,X_2;W'_2,W_2) \\
	& \quad \times  \left(K_{W'_1} \cdot \nabla_{X'_1}\right)^{J_1} \Omega_0(X'_1,X'_2)\left(K_{W'_2} \cdot \nabla_{X'_2}\right)^{J_2} \Omega_0(X'_2,X'_1)\,, \label{eq:Ot_Omega2_O}
\end{split}\end{equation}
where the first equality follows from the CFT orthogonality relation \eqref{eq:orthogonality} and the bulk completeness relation \eqref{eq:bulk_comp_rel} as applied to the original definition of the maps in the first lines of \eqref{eq:cft_to_ads_theta} and \eqref{eq:ads_to_cft_theta}. The second equality follows from the definition of the maps in terms of $\theta_J,\tilde\theta_J$ in the second lines of \eqref{eq:cft_to_ads_theta} and \eqref{eq:ads_to_cft_theta}, and we have defined the new bulk bi-local function
\begin{equation} \label{eq:omega0_def}
	\Omega_0(X_1,X_2) = \int dP G_{\Delta_0,0}(X_1,P) G_{\tilde\Delta_0,0}(X_2,P)\\
	 = \lim_{\Delta\rightarrow \Delta_0} \alpha_0 N_{\Delta,0} \Omega_{\Delta,0}(X_1,X_2)\,,
\end{equation}
where the pole in $N_{\Delta_0,0}$ \eqref{eq:CFT_norm} cancels the zero in $\Omega_{\Delta_0,0}$ in the explicit expression in \cite{Costa:2014kfa}.

\begin{table}[!ht]
\centering
	\includegraphics{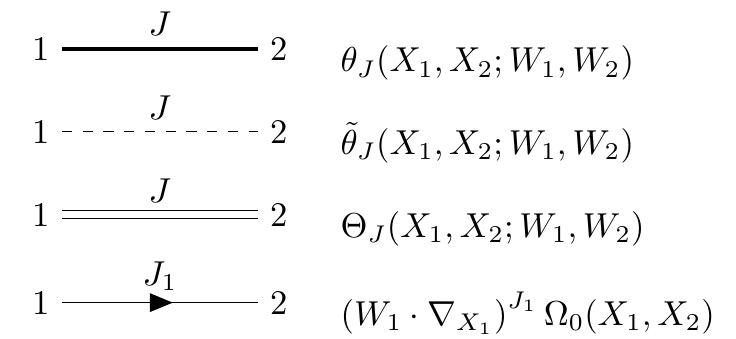}
	\caption{The line notations for the different bulk bi-local functions.}
	\label{tb:feynman_lines}
\end{table}

In order to use these identities in a more transparent way, we will use the diagrammatic notation defined in Table \ref{tb:feynman_lines}. Every point has its own position and usually spin. Note that the first three lines are function of the two spins indices at each end, and both ends are in the same representation. The arrow line on the other hand depends only on the spin index at the origin of the arrow (in the drawing, $X_1,W_1,J_1$). When we draw diagrams with these lines, every internal dot is implicitly integrated over, and the spin indices are contracted in the standard embedding space manner. If the spin representation of an internal edge or point is not defined by the external legs, we implicitly sum over all possible values (all integers for $U(N)$ and even integers only for $O(N)$). For example, the definition of $\Theta_J$ in \eqref{eq:Theta_def} can be drawn as
\begin{equation}\label{eq:Theta_def_diag}
	\vcenter{\hbox{\includegraphics{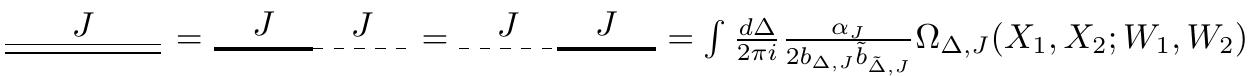}}}
\end{equation}
while the invertibility equation \eqref{eq:Ot_Omega2_O} is
\begin{equation}
	\label{eq:invertibility}
	\vcenter{\hbox{\includegraphics{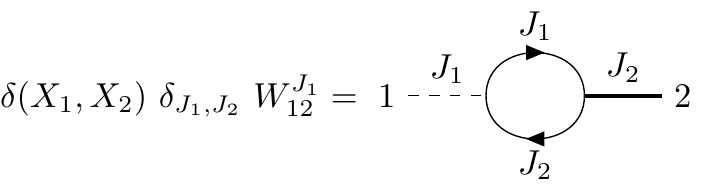}}}
	\,.
\end{equation}
Two important identities that we will use in the following are 
\begin{equation} \label{eq:Theta_Omega4_diag}
\vcenter{\hbox{\includegraphics{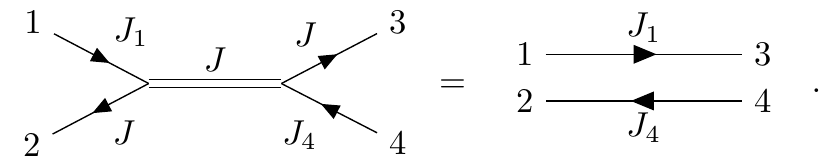}}}\,,
\end{equation}
and
\begin{equation}
	\vcenter{\hbox{\includegraphics{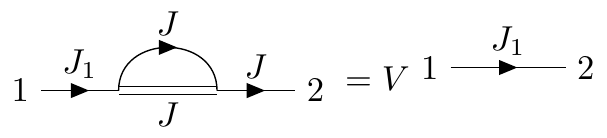}}}\,,
	\label{eq:Theta_loop}
\end{equation}
where the first identity allows us to eliminate factors of $\Theta_J$ for general diagrams, while the second identity eliminates a $\Theta_J$ and introduces the regularization factor $V$ given in section \ref{sec:local_to_bilocal}, which will cancel with contributions from counterterm diagrams. We prove both of these identities in Appendix \ref{bulkIdentities}.

\subsection{Feynman rules}
Recall that in the bi-local formalism the free theory has a non-local action with an explicit $1/N$ expansion \eqref{eq:free_pert_bilocal_action}. We will now use the CFT-to-AdS map as expressed using the bulk bi-local functions described above to construct the bulk action to all orders in $1/N$, which defines the Feynman rules for the bulk theory. 

The $n$-th interaction term in \eqref{eq:free_pert_bilocal_action} is proportional to 
\begin{equation}\begin{split}
 	 \text{Tr}\( \( G_0^{-1} \eta \) ^n \)  = \int \prod_{i=1}^n dP_i \ \square_{P_1} \eta(P_1,P_2) ... \square_{P_n} \eta(P_n,P_1) \label{eq:S_n}\,.
\end{split}\end{equation}
We can map each term in this product to the bulk using \eqref{eq:ads_to_cft_theta} to get
\begin{equation}\begin{split}
 	 \text{Tr}\( \( G_0^{-1} \eta \) ^n \)  = &\prod_{m=1}^n\sum_{J_m} a_{\Delta_0}
	 \int \frac{dX_m dX'_m }{J_m! \left(\frac{d-1}{2}\right)_{J_m}} (K_{W'_m} \cdot \nabla_{X'_m})^{J_m} \Omega_0(X'_{m-1},X'_m)\\
	&\qquad\qquad\qquad\qquad\times \theta_{J_m}(X'_m,X_m;W'_m,K_{W_m}) \Phi_{J_m}(X_m,W_m)\,, \label{interId}\\
\end{split}\end{equation}
where $X'_{0}\equiv X'_n$, the boundary Laplacian acts as
\begin{equation}\label{eq:a_def}
\square_{P} G_{\Delta_0,0} (X,P) = a_{\Delta_0} G_{{\tilde \Delta_0,0}}(X,P)\,,\qquad	a_{\Delta_0}=-\frac{4\pi^{\frac{d}{2}} \Gamma(\frac{d}{2}-\Delta_0)}{\Gamma(\Delta_0) S_B^{\tilde \Delta_0,0}}\,,
\end{equation}
with $a_{\Delta_0}=d(2-d)$ for the free theory value $\Delta_0=\frac{d-2}{2}$, and we performed the $P_m$ integrals using the definition of $\Omega_0$ in \eqref{eq:omega0_def}. For instance, the quadratic term in \eqref{eq:free_pert_bilocal_action} is mapped to 
\begin{equation}\begin{split}
 	S_2 = &\sum_{J_1,J_2} \frac{a^2_{\Delta_0}}{2}
	 \int \frac{dX_1 dX'_1dX_2 dX'_2 }{J_1! \left(\frac{d-1}{2}\right)_{J_1}J_2! \left(\frac{d-1}{2}\right)_{J_2}} (K_{W'_1} \cdot \nabla_{X'_1})^{J_1} \Omega_0(X'_{2},X'_1) (K_{W'_2} \cdot \nabla_{X'_2})^{J_2} \Omega_0(X'_{1},X'_2)\\
	&\times \theta_{J_1}(X'_1,X_1;W'_1,K_{W_1}) \theta_{J_2}(X'_2,X_2;W'_2,K_{W_2}) \Phi_{J_1}(X_1,W_1) \Phi_{J_2}(X_2,W_2)\\
	=& \frac{a_{\Delta_0}^2}{2}
		\vcenter{\hbox{\includegraphics{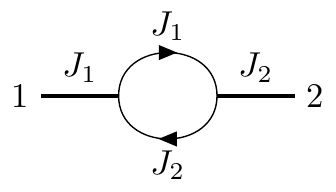}}}
	\,, \label{eq:S2_bulk_theta}
\end{split}\end{equation}
which is a rewriting of \eqref{eq:bulk_S2} in the bulk bi-local function language. We can then recompute the bulk 2-point function in this formalism by inverting the operator that acts on the quadratic term in $\Phi_J$, using the identity 
\begin{equation}\begin{split}
&	\vcenter{\hbox{\includegraphics{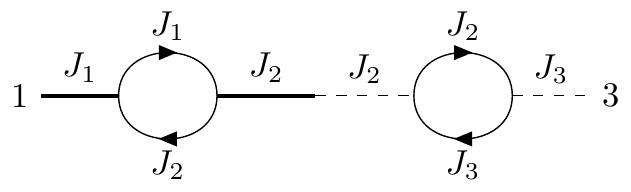}}} 
=
	\vcenter{\hbox{\includegraphics{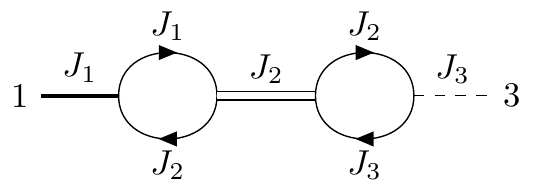}}}\\
	&\qquad\qquad\qquad\qquad\qquad\qquad = \vcenter{\hbox{\includegraphics{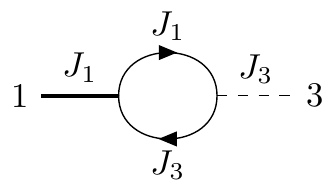}}} \\
	& \qquad\qquad\qquad\qquad\qquad\qquad= \delta_{J_1,J_3} \delta^{TT}(X_1,X_3) W_{13}^{J_1}\,,
\end{split}\end{equation}
where in the first equality we used the definition of $\Theta_J$ in \eqref{eq:Theta_def_diag}, in the second equality we used the four-Omega identity \eqref{eq:Theta_Omega4_diag}, and in the last equality we used the invertibility identity \eqref{eq:Ot_Omega2_O}. The bulk propagator is then 
\begin{equation} \label{eq:bulk_prop}
			\langle\Phi_{J}(X_1,W_1)\Phi_{J}(X_2,W_2)\rangle = \frac{1}{a_{\Delta_0}^2}\ 
			 \vcenter{\hbox{\includegraphics{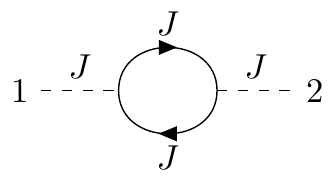}}}\,,
		\end{equation}
which is a rewriting of \eqref{eq:Omega_BB_props} in the bulk bi-local function language.

We can similarly map the interaction terms in \eqref{eq:free_pert_bilocal_action} to get
\begin{equation} \label{eq:bulk_vertex}
	n\geq3:\qquad\qquad	\frac{\left(-1\right)^{n+1}}{n}N^{1-\frac{n}{2}}
		a_{\Delta_0}^n
		\vcenter{\hbox{\includegraphics{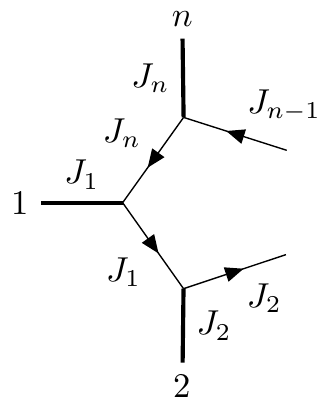}}}
		\vdots\quad \,,
	\end{equation}
as well as the counterterms in \eqref{eq:free_pert_bilocal_action} to get 
\begin{equation} \label{eq:bulk_couterterm}
	n\geq1:\qquad\qquad	\frac{\left(-1\right)^{n}}{n}VN^{-\frac{n}{2}} a_{\Delta_0}^n
		\vcenter{\hbox{\includegraphics{S_n.pdf}}}
		\vdots\quad \,.
	\end{equation}
The Feynman rules for the bulk dual of the $U(N)$ theory are then given by these vertices along with the propagator \eqref{eq:bulk_prop}, and a symmetry factor that should be taken with regard to the ordering in the loop up to cyclic transformations. For the bulk dual of the $O(N)$ theory, the vertices have an extra factor of $\frac12$, the propagator an extra factor of $2$, and we sum only over even spins.

\subsection{Properties and cancellation of loop diagrams}

We will now use the Feynman rules and identities discussed above to show that all 1-loop corrections vanish in the bulk, just as we showed for the CFT in section \ref{sec:bilocal_1_loop}. The general strategy will be to use the four-Omega identity \eqref{eq:Theta_Omega4_diag} to systematically eliminate $\Theta_J$ double lines from complicated loop diagrams, until the diagram identically cancels a suitable counterterm diagram, just as we observed in the CFT 1-loop calculation. We first show these cancellations ignoring the need to regularize the Feynman diagrams (beyond assuming that the regularization leads to \eqref{eq:Theta_loop} for some $V$), and we then discuss the effects of regularization.

As an example, we first consider the connected tree level $n$-point correlators, which we do not expect to vanish. These correlators get a contribution from the bulk $n$-point vertex, which takes the form
\begin{equation}\begin{split}
	\langle \Phi_{J_1}(X_1,W_1)&...\Phi_{J_n}(X_n,W_n)\rangle_\text{tree, connected n-point vertex} 
	\\ & = \frac{\left(-1\right)^{n+1}}{n}N^{1-\frac{n}{2}}
		\left(a_{\Delta_0} \cdot \frac{1}{a_{\Delta_0}^2}\right)^n 
		\cdot
		\vcenter{\hbox{\includegraphics{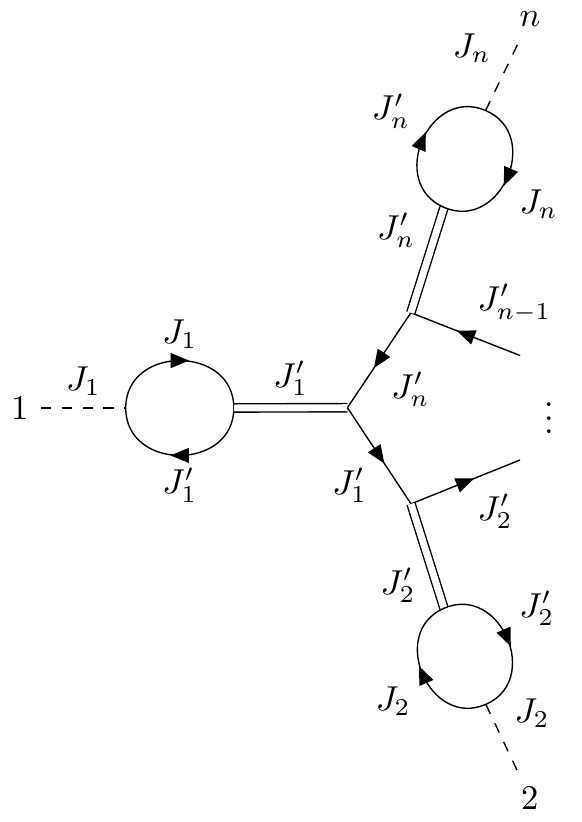}}}
		 + \text{permutations}
	\\ & = \frac{\left(-1\right)^{n+1}}{n}N^{1-\frac{n}{2}}
		a_{\Delta_0}^{-n} 
		\cdot
		\vcenter{\hbox{\includegraphics{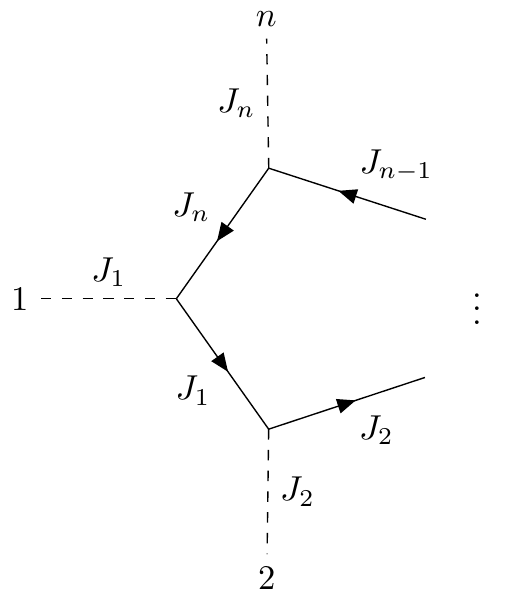}}}
		+ \text{permutations}\,,
\end{split}\end{equation}
where the second equality came from repeatedly applying the identity \eqref{eq:Theta_Omega4_diag} to eliminate all the $\Theta_J$ double lines. The result is a convolution of $n$ $\tilde \theta$ operators with a bilocal function that is just multiplications of $n$ $\partial \Omega_0$s. This latter function is independent of our normalization, so the entire dependence on $f_{\Delta,J}$ is through the external legs. We will find that all the diagrams have this structure. This is expected, as each diagram can also be written as the CFT-to-AdS mapping of the corresponding bi-local diagram.

We now turn to 1-loop calculations, where for simplicity we follow the CFT calculation in section \ref{sec:bilocal_1_loop} and consider the 1-loop contribution to the one-point function coming from the $3$-point vertex. Using the Feynman rules above we get
\begin{equation} \label{eq:bulk_VEV_1loop_steps}
\begin{split}
	\langle \Phi_J(X,W)\rangle_{S_3, \text{1-loop}} & = 3 \cdot
	\frac{N^{-\frac{1}{2}}}{3} 
	\frac{1}{a_{\Delta_0}}
	\vcenter{\hbox{\includegraphics{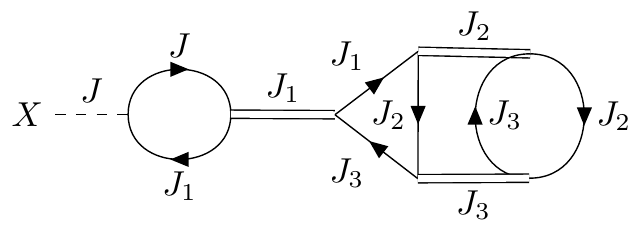}}}
	\\
	& = N^{-\frac{1}{2}}
		\frac{1}{a_{\Delta_0}}
		\vcenter{\hbox{\includegraphics{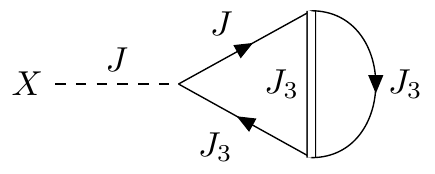}}}
		\\
	& = V N^{-\frac{1}{2}} \frac{1}{a_{\Delta_0}}
		\vcenter{\hbox{\includegraphics{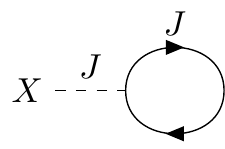}}}
		\,,
\end{split}
\end{equation}
where in the second equality we used \eqref{eq:Theta_Omega4_diag} twice, and in the third equality we used the identity \eqref{eq:Theta_loop} to get the factor of $V$. The result is exactly minus the contribution from the 1-vertex counter-term:
\begin{equation}\begin{split}
	\langle \Phi_J(X,W)\rangle_\text{counter-term} & = 
	(-1) V N^{-\frac{1}{2}} a_{\Delta_0} \cdot \frac{1}{a_{\Delta_0}^2}\cdot
	\vcenter{\hbox{\includegraphics{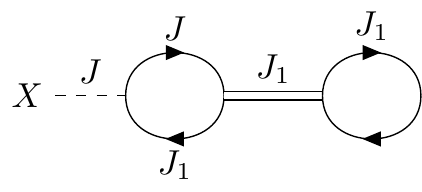}}}
	\\
	& = - V N^{-\frac{1}{2}} \frac{1}{a_{\Delta_0}} \cdot 
	\vcenter{\hbox{\includegraphics{VEV_calc_3.pdf}}}
	\,,
\end{split}\end{equation}
so we have exact cancellation of the 1-loop contribution, just like in section \ref{sec:bilocal_1_loop}.

As we saw in the bilocal theory, up to combinatorics, the cancellation of loops by the counter-terms is basically because the diagrams themselves are equal to their tree level counterparts, multiplying some power of a specific divergence we called $V$. We can list the general mechanism as follows:
\begin{enumerate}
	\item Every loop is made of contractions between vertices. As each contraction has $\tilde \theta_J$ at each end \eqref{eq:bulk_prop}, and every vertex in \eqref{eq:bulk_vertex} and \eqref{eq:bulk_couterterm} has a $\theta_J$ at each end, we can always write the loop using $\Omega_0$ and $ \Theta$ alone. As an example, consider a loop made of three contractions. There are two possible options for this loop, depending on the relative orientation of the vertices:
	\begin{equation} \label{eq:3_vertices}
		\vcenter{\hbox{\includegraphics{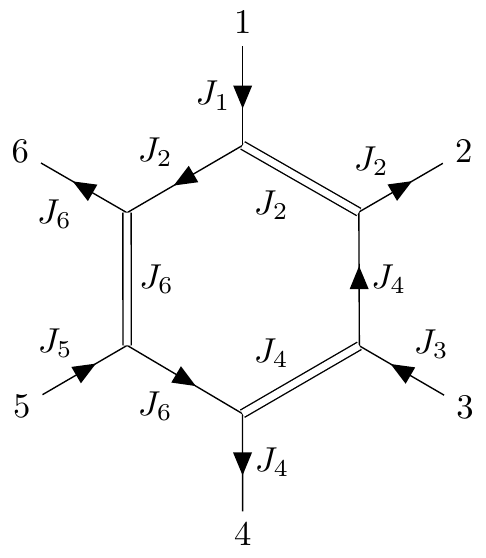}}}
		, \quad\quad
		\vcenter{\hbox{\includegraphics{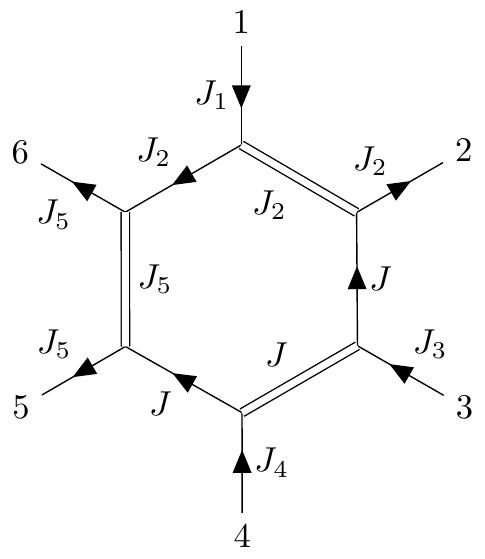}}}
	\end{equation}
	Note the ``external legs" in the drawing are actually part of the full expression of the vertex and are not the actual external legs of the complete diagram (see \eqref{eq:bulk_vertex}).
	For the same reasons, external contractions keep their $\tilde \theta_J$ legs. Eventually the entire diagram is independent of $f_{\Delta,J}$ except via the external $\tilde \theta_J$'s. As mentioned above, this is expected from the linear nature of our mapping.
	\item Using both \eqref{eq:Theta_Omega4_diag} and \eqref{eq:Theta_loop}, we can now `disentangle' the diagrams. The result may or may not have closed loops, which correspond to factors of $V$, depending on the orientation of each vertex (and eventually the total topology of the diagram). For example, the disentangling for the 3-vertices \eqref{eq:3_vertices} gives:
	\begin{equation}
		\vcenter{\hbox{\includegraphics{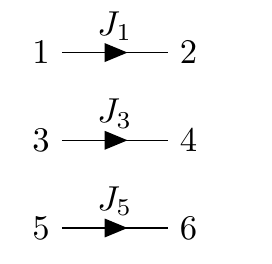}}}
		\times \quad V\,, \quad\quad
		\vcenter{\hbox{\includegraphics{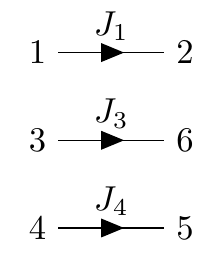}}}
	\end{equation}
	thus reducing the number of loops in the complete diagram by one (times a possible factor of $V$).
	\item Eventually we can reduce any diagram (including those that include counterterm vertices) to a power of $V$ multiplied by a tree-level diagram.
\end{enumerate}
The algebraic relation between each diagram to the tree level diagram has the same form as in the bi-local theory. We expect that the combinatorics of the bi-local diagrams that cancel all loops in section \ref{sec:bilocal_1_loop} should therefore carry over to the same combinatorics here.

The considerations above are somewhat formal since all the loop diagrams need to be regularized. As discussed in appendix \ref{app:bulk_dive}, one way to regularize is to put a cutoff $z > \varepsilon$ on the radial direction of AdS space, and to put an IR cutoff $|x|<L$ on the space-time coordinates. With this regularization (and with a finite cutoff on the integrals over $\Delta$, as we have throughout the paper) we claim that all diagrams are finite. Furthermore, we claim that when we remove the regulators by taking $\varepsilon \to 0$ and $L \to \infty$, all the divergences arise in subdiagrams of the form \eqref{eq:Theta_loop}, with $V \propto L^d / \varepsilon^d$. These divergences then cancel in all loop computations, as described above. In the free theory, in fact, all loops cancel exactly (not just their divergent terms), but this will no longer be true in the critical theory that we discuss below.

%

\section{The mass deformation}
\label{sec:massDef}

In this section we discuss the deformation of the bulk theory by a finite scalar mass $m$ and its effect on the vacuum expectation value (VEV) of $\Phi_0$, which we would like to derive from $\langle \eta(x_1,x_2)\rangle$ (computed in the massive theory) using our CFT-to-AdS map. In the CFT, we add the standard mass term $m^2 N G(x,x)$, which up to an irrelevant constant corresponds to adding $m^2\sqrt{N}\eta(x,x)$. The VEV for $\eta(x_1,x_2)$ under this deformation is then
\es{bilocalmass}{
	\langle \eta(x_1,x_2) \rangle_{m} =  \sqrt{N} \Big[\frac{m^{d/2-1}K_{d/2-1}(m|x_{12}|)}{(2\pi)^{d/2}|x_{12}|^{d/2-1}}-\frac{\Gamma(d/2-1)}{4\pi^{d/2}|x_{12}|^{d-2}}\Big]\,,
}
where in $d=3$ the massive propagator simplifies to give $\langle \eta(x_1,x_2) \rangle_{m} =\sqrt{N}\frac{e^{-m|x_{12}|}-1}{4\pi |x_{12}|}$.
Since this configuration does not decay at $x_1,x_2\to\infty$, we cannot expand it in the conformal basis of $C_{\Delta,J}(y)$ in section \ref{sec:bi_local_to_conformal}, and so cannot apply the CFT-to-AdS map derived in section \ref{sec:mapping_to_the_bulk}. Instead, we will derive a new expansion of non-decaying bi-locals $\eta_{ND}(|x_{12}|)$ in terms of a basis $d_\Delta$, which we will then identify with a similar expansion of non-decaying bulk fields $\Phi_{ND}(z)$ to derive a non-decaying CFT-to-AdS map (and its inverse). This non-decaying map will correspond to the limit of the standard decaying map, for configurations that only depend on $|x_{12}|$ in the CFT, and that only depend on $z$ in the bulk. We will then use this non-decaying map to compute the bulk dual $\langle\Phi_{ND}(z)\rangle_m$ of \eqref{bilocalmass}. In this section, we do not use embedding space for clarity.

\subsection{Mapping the small mass deformation}
\label{sec:smallmass}

Before we address the more difficult finite mass case, we will first discuss the simpler case where $m^2\sim1/\sqrt{N}$ so that adding $m^2\sqrt{N}\eta(x,x)$ can be treated as a small deformation to the bi-local action. In addition, we consider
a slightly more general position-dependent mass term $m(x)$, which will enable us (for $m(x)$ that decays at infinity) to use the standard CFT-to-AdS map for decaying configurations that we derived above. The CFT path integral is 
\begin{equation}\label{eq:mass_term}
\begin{split}
	Z_{m(x)} = \int D\eta(x,x')\text{exp}\left(-S[\eta]-\int d^dx\, m^2(x)\sqrt{N} \eta(x,x)\right)\,. 
\end{split}
\end{equation}
For $d<4$ we can map this to the bulk using the off-shell mapping \eqref{toshow2} to get
\begin{equation}\label{eq:mass_term2}
\begin{split}
	Z_{m(x)}=\int D\Phi_J(x,z)\text{exp}\left(-S[\Phi]-\frac{1}{f_{d-2,0}}\lim_{\eps\rightarrow 0} \eps^{-d+2} \int d^dx m^2(x)\sqrt{N} \Phi_0(x,\eps)\right)\,,
\end{split}
\end{equation}
where the bulk action $S[\Phi]$ was described in section \ref{sec:bulk_theory}, and note that this deformation only affects the boundary of AdS. We can then compute the VEV of $\Phi_J$ under this deformation at tree level, to get the leading order in $1/N$ result
\es{vev}{
	\langle \Phi_J(x,z)\rangle_{m(x)} 
	& = \frac{1}{f_{d-2,0}} \int d^dy \,m^2(y)\sqrt{N} 
	\left( \lim_{\eps\rightarrow 0} \eps^{-d+2} \langle \Phi_J(x,z) \Phi_0(y,\eps)\rangle \right) +O(N^{-1})\\
	& =\delta_{J,0} f_{2,0}  \frac{\Gamma(d-2)}{16\pi^d}
	\int d^dy\, m^2(y)\sqrt{N} 
	G_{d-2,0}(x,z|y)+O(N^{-1}),\\
}
where in the second equality we used the boundary limit of the bulk two-point function in \eqref{eq:2p_localf}, which is only nonzero when $J=0$. Since this calculation was done on-shell, we in fact only needed to use the on-shell mapping \eqref{toshow}, so the computation holds for any $d$ \footnote{For $d>4$, \eqref{vev} does not obey the original boundary conditions for $\Phi_0$ at small $z$ (which had the leading term going as $z^{d/2}$); it turns out that in this case we need to modify the boundary conditions when deforming the action by \eqref{eq:mass_term}, related to the fact that $\eta(x,y)$ is singular at $x\to y$ after the deformation, so that the first line of \eqref{eq:bc_eta} is no longer obeyed. The modified boundary condition for the $z^2$ term in the expansion is exactly the usual one that would be associated to this deformation by the AdS/CFT correspondence.}.

If the mass $m$ is independent of the position\footnote{Strictly speaking, we need to use in this case the mapping for non-decaying field configurations, but we will see that it gives the same answer as the naive application of the equations above.}, we can perform the integral over the bulk-to-boundary propagator in \eqref{vev} to get
\es{vev2}{
	\langle \Phi_J(x,z)\rangle_{m}  =\delta_{J,0} m^2\sqrt{N}z^2\frac{f_{2,0}}{16\pi ^{d/2}}  \Gamma \left(\frac{d}{2}-2\right)+O(N^{-1})\,.\\
}
Note that the $z$ dependence corresponds to the shadow of the mass operator scaling dimension $\Delta=d-2$, and that as expected this bulk VEV only depends on the dimensionless product $mz$, and not on the $x$ direction. Both of these properties come from the fact that the $y$ integral in \eqref{vev} picks out the second term in the expansion of the bulk-to-boundary propagator at small $z$ \eqref{eq:2_point_limit}. For the finite mass deformation that we discuss next, we will similarly find that the bulk VEV depends only on $mz$, though it will include more than just a quadratic term.

\subsection{The conformal basis for non-decaying bi-locals}
\label{sec:NDbasisCFT}

Before we can discuss the mapping of the finite mass configuration in \eqref{bilocalmass}, we first need an CFT-to-AdS map that applies to non-decaying bi-locals. We start on the CFT side by defining the non-decaying $\eta_{ND}(|x_{12}|)$ and decaying $\eta_{D}(x_1,x_2)$ parts of $\eta(x_1,x_2)$ as
\es{decay}{
	\eta_{ND}(|x_{12}|)=\frac{1}{v}\int d^dy\eta(y,y+|x_{12}|)\,,\qquad \eta_{D}(x_1,x_2)=\eta(x_1,x_2)-\eta_{ND}(|x_{12}|)\,,
	}
	where $v\equiv \int d^dy$ is the volume of spacetime, and for simplicity we assume that the part of $\eta(x_1,x_2)$ that does not decay at infinity is just a function of $|x_1-x_2|$ \footnote{The non-decaying map discussed in this section could be easily generalized to non-rotationally symmetric configurations, but we will not consider them in this work.}. We use $v$ as an IR regulator, where at intermediate steps we consider large but finite $v$, and take $v\to\infty$ at the end. The definition \eqref{decay} implies the constraint
	\es{etaconstraint}{
	\frac1v\int d^dx\eta_D(x,x+\epsilon)=0\,
	}
	for any $\epsilon\in\mathbb{R}^d$.
	
	In section \ref{sec:bi_local_to_conformal}, we showed that decaying configurations $\eta(x_1,x_2)$ can be expanded in the basis of 3-point functions, with coefficients $C_{\Delta,J}(y)$. We can apply the same expansion to $\eta_D(x_1,x_2)$, except that we should impose the constraint \eqref{etaconstraint} in the language of the $C_{\Delta,J}(y)$ by requiring that
		\es{Cconstraint}{
	\frac1v\int d^dy C_{\Delta,J}(y)=0\,,
	}
	for any $\Delta,J$. 
	
	Since $\eta_{ND}(|x_{12}|)$ just depends on a single degree of freedom $|x_{12}|$, we can expand it in a simpler basis of scalar 2-point functions $|x_{12}|^{-2\Delta}$. This expansion is just the standard Mellin transform of a single variable function with a specific choice of contour, leading to the completeness relation
		\es{NDcomplete}{
	\int_{\gamma_0}\frac{d\Delta}{2\pi i}(|x|/|y|)^{\Delta-\frac d2}=|x|\delta(|x|-|y|)\,,
	}
and to the orthogonality relation
	\es{NDortho}{
	\int_0^\infty \frac{d|x_{12}|}{|x_{12}|} x_{12}^{\Delta-\Delta'}=2\pi i\delta(\Delta-\Delta')\,\,,
	}
where we have assumed that $\Delta,\Delta'\in\gamma_0$. We can then decompose $\eta_{ND}(|x_{12}|)$ in this basis using the completeness relation to get
\es{eq:eta_decompND}{
	\eta_{ND}(x_1,x_2) &=	 \int_{\gamma_0}\frac{d\Delta}{2\pi i} d_\Delta x_{12}^{2-\Delta}\,,
}
where $d_\Delta$ is defined by \eqref{NDcomplete} to be
\es{eq:coeffND}{
	d_{\Delta}=  \int_0^\infty d|x_{12}|\, \eta_{ND}(|x_{12}|)x_{12}^{\Delta-3}\,,
	}
	and is well-defined so long as this integral converges for $\Delta\in\gamma_0$. Note that \eqref{NDcomplete} and \eqref{NDortho} are the standard representations of delta functions in Mellin space.\footnote{These relations also hold for $2<d<4$, when our contour $\gamma_0$ requires that we remove the pole at $\Delta=2$ and include the one at $\Delta=d-2$ \cite{Caron-Huot:2017vep}.} The choice of contour we used comes from taking the limit of the analogous formulae in section \ref{sec:bi_local_to_conformal}, for the case where $\eta(x_1,x_2)$ only depends on $|x_{12}|$ and $C_{\Delta,J}(y)$ no longer depends on $y$ (i.e. $P$ in embedding space), and is nonzero only for $J=0$. For instance, we can compute the $P$ integral in \eqref{eq:eta_decomp} in this case to get \eqref{eq:eta_decompND} if we identify
	\es{Ctod}{
	C_{\Delta,0}\to  \frac{d_{\Delta}}{S_{\Delta_0,\Delta_0}^{(\Delta_0,0)}}\,.
	}
	
The $d_\Delta$ basis that we use for non-decaying functions is simpler than the $C_{\Delta,J}(y)$ basis that we use for decaying functions in several ways. First, recall that the $C_{\Delta,J}(y)$ were doubly complete due to the shadow transform that relates $C_{\Delta,J}(y)$ to $C_{\tilde\Delta,J}(y)$, while the $d_\Delta$ are complete without any redundancies. Second, the expression for $C_{\Delta,J}(y)$ in \eqref{eq:coeff} did not converge for the free theory in $2<d<4$, so we had to define a more complicated expansion of $\eta_D(x_1,x_2)$ in terms of $C_{\tilde\Delta,J}(y)$. The expression for $d_\Delta$ in \eqref{eq:coeffND}, however, has the same convergence properties for any $d$.

We can now write the bi-local action in \eqref{eq:free_pert_bilocal_action} in terms of the conformal basis of $C_{\Delta,J}(y)$ and $d_\Delta$ by first writing $\eta(x_1,x_2)$ in terms of $\eta_D(x_1,x_2)$ and $\eta_{ND}(x_1,x_2)$ as in \eqref{decay}, and then expanding the $\eta_D(x_1,x_2)$ in terms of the $C_{\Delta,J}(y)$ as discussed in section \ref{sec:bi_local_to_conformal} and the $\eta_{ND}(x_1,x_2)$ in terms of $d_\Delta$ as in \eqref{eq:eta_decompND}. For instance, at quadratic order, the decaying action in terms of the $C_{\Delta,J}(y)$ was written in \eqref{eq:c_action}, and can be used to compute tree level correlators as discussed before. The mixed term with both $d_\Delta$ and $C_{\Delta,J}(y)$ vanishes due to the constraint \eqref{Cconstraint}. Finally, the quadratic term in terms of just the $d_\Delta$ can be computed for the $U(N)$ theory as 
\es{quad}{
S^{(2)}(d_\Delta)&=\frac{1}{{2}}\int d^dx_1d^dx_2  \nabla^2_1\eta_D(x_1,x_2)\nabla^2_2\eta_D(x_1,x_2)\\
&=\frac12\int d^dx_1d^d x_2\int_{\gamma_0}\frac{d\Delta d\Delta'}{(2\pi i)^2} \frac{d_\Delta d_{\Delta'}}{|x_{12}|^{\Delta+\Delta'}}  (\Delta-2)(\Delta'-2)(\Delta-d)(\Delta'-d)\\
&= \frac{v}{\alpha_0}\int_{\gamma_0}\frac{d\Delta }{2\pi i} d_\Delta d_{\tilde\Delta} \lambda_{\Delta,0}\,,
	}
	where in the last equality we used rotational symmetry and then the orthogonality relation \eqref{NDortho} to do the $x_1$ integral, while the $x_2$ integral gave the factor of $v$. Since $v\to\infty$ when we remove the IR cutoff, this factor ensures that the non-decaying modes are non-dynamical. In the $O(N)$ theory, we would get the same result except with an extra factor of $\frac12$. We can similarly write the interaction terms in \eqref{eq:free_pert_bilocal_action} in this conformal basis, where in general there will be nonvanishing cross terms with both $C_{\Delta,J}(y)$ and $d_\Delta$. 

\subsection{The non-decaying AdS/CFT map}
\label{sec:NDbasisBulk}

In the bulk, we can similarly define the non-decaying $\Phi_{ND}(z)$ and decaying $\Phi_{D}(x,z)$ parts of the scalar bulk field $\Phi_{0}(x,z)$ as
\es{decayBulk}{
	\Phi_{ND}(z)=\frac 1v\int d^dy\Phi_0(y,z)\,,\qquad \Phi_{D}(x,z)=\Phi_0(x,z)-\Phi_{ND}(z)\,,
	}
where again we only consider non-decaying configurations where only the scalar is non-zero, and approaches a constant at large distances. This definition implies the constraint
\es{phiconstraint}{
	\frac1v\int d^dx\Phi_D(x,z)=0\,.
	}
The decaying $\Phi_{D}(x,z)$ are expanded in the same bulk-to-boundary basis as in section \ref{sec:mapping_to_the_bulk}, except with $C^{bulk}_{\Delta,J}(y)$ subject to the constraint \eqref{Cconstraint}. The non-decaying $\Phi_{ND}(z)$ can be expanded using the same Mellin transform as $\eta_{ND}(|x_{12}|)$, except with the shifted contour
\es{eq:eta_decompNDB}{
	\Phi_{ND}(z) &=	 \int_{\gamma_0}\frac{d\Delta}{2\pi i} d^{bulk}_\Delta z^{d-\Delta}\,,
}
where $d^{bulk}_\Delta$ is defined by \eqref{NDcomplete} to be
\es{eq:coeffNDB}{
	d^{bulk}_{\Delta}=  \int_0^\infty dz\, \Phi_{ND}(z)z^{\Delta-d-1}\,.
	}
	As in the non-decaying CFT case, the choice of contour comes from taking the limit of the analogous formulae in section \ref{sec:mapping_to_the_bulk}, where $\Phi_0(x,z)$ only depends on $z$, and $C^{bulk}_{\Delta,J}(y)$ no longer depends on $y$ and is nonzero only for $J=0$. For instance, we can compute the $P$ integral in \eqref{eq:bulk_comp} in this case to get \eqref{eq:eta_decompNDB} if we identify
	\es{CtodB}{
	C^{bulk}_{\Delta,0}\to \frac{d^{bulk}_{\Delta}}{S_B^{\Delta,0}}\,.
	}
	
	We can then derive the non-decaying AdS-to-CFT map by identifying $d^{bulk}_{\Delta}$ with $d_{\Delta}$ as
	\es{g}{
	d^{bulk}_{\Delta}=f_{\tilde\Delta,0}d_{\Delta}\,,
	}
	where the normalization factor $f_{\tilde\Delta,0}$ was fixed from the limit of the decaying map in \eqref{Ctod} and \eqref{CtodB} as well as the consistency condition \eqref{eq:bulk_norm_shadow_cond}. The non-decaying AdS-to-CFT map is then
\es{adstocftND}{
\eta_{ND}(|x_{12}|) &=	 \int_{\gamma_0}\frac{d\Delta}{2\pi i} |x_{12}|^{2-\Delta}\frac{1}{f_{\tilde\Delta,0}} \int_0^\infty dz \Phi_{ND}(z)z^{\Delta-d-1}\,,
}
while the non-decaying CFT-to-AdS map is
\es{cfttoadsND}{
\Phi_{ND}(z) &=	 \int_{\gamma_0}\frac{d\Delta}{2\pi i} f_{\tilde\Delta,0} z^{d-\Delta} \int_0^\infty d|x_{12}|\, \eta_{ND}(|x_{12}|)|x_{12}|^{\Delta-3}\,.
}
Note that since we used the same Mellin transform (up to a shift in contour) for both the bulk and CFT expansions, this map would have simply identified $|x_{12}|=z$, if we did not have the nontrivial factor $f_{\tilde\Delta,0}$.

We can now map the bi-local action in \eqref{eq:free_pert_bilocal_action} as written in terms of $\eta_D(x_1,x_2)$ and $\eta_{ND}(x_1,x_2)$ into the bulk using the decaying AdS-to-CFT map in section \ref{sec:mapping_to_the_bulk} for $\eta_D(x_1,x_2)$, and the non-decaying map in this section for $\eta_{ND}(x_1,x_2)$. For instance, at quadratic order for the $U(N)$ theory, we can find the non-decaying bulk term by expanding \eqref{quad} using \eqref{eq:coeffNDB} and \eqref{g} as
	\es{quadBulkD}{
S^{(2)}[\Phi_{ND}]&=	\frac{v}{\alpha_0}\int_{\gamma_0}\frac{d\Delta }{2\pi i} d_\Delta d_{\tilde\Delta} \lambda_{\Delta,0}\\
&=\frac{v}{\alpha_0}\int_0^\infty \frac{dz dz'}{z z'} \Phi_{ND}(z)\Phi_{ND}(z')\int_{\gamma_0} \frac{d\Delta }{2\pi i} \frac{z^{-\tilde\Delta}z'^{-\Delta}}{f_{\Delta,0}f_{\tilde\Delta,0}}\lambda_{\Delta,0}\\
&= \frac{v}{\alpha_0}\int_0^\infty \frac{dz dz'}{zz'} \Phi_{ND}(z')(\nabla^2_z-M^2_{d-2,0})(\nabla^2_z-M^2_{d,0})\Phi_{ND}(z) \int_{\gamma_0}\frac{d\Delta }{2\pi i}\frac{z^{-\tilde\Delta}z'^{-\Delta}}{f_{\Delta,0}f_{\tilde\Delta,0}}\,, \\
	}
	where in the third equality we used integration by parts and the action of the AdS Laplacian in the upper half space when acting on functions of only $z$:
	\es{ufsLap}{
	\nabla_z^2=z^2(\partial^2_z-(d-1)z^{-1}\partial_z)\,.
	}
As before, the factor of $v$ ensures that the non-decaying modes are non-dynamical. This non-decaying quadratic term could equivalently be found by taking the limit of the decaying scalar quadratic term in \eqref{eq:bulk_S2}, if we assume that $\Phi_0(x,z)$ does not depend on $x$. 	The non-decaying bulk propagator is then
\es{bulkPropz}{
&	\langle \Phi_{ND}(z_1) \Phi_{ND}(z_2)\rangle=  \frac{\alpha_0}{2v} \int_{\gamma_0} \frac{d\Delta }{2\pi i} \frac{f_{\Delta,0}f_{\tilde\Delta,0}}{\lambda_{\Delta,0}}z_1^{\tilde\Delta}z_2^{\Delta}\,,
	}
which is the inverse of the second line of \eqref{quadBulkD} due to the completeness relation \eqref{NDcomplete}.  

As in the decaying case, if we choose $f^\text{local}_{\Delta,0}$ then we get a local quadratic action
		\es{quadBulkDlocal}{
S^{(2)}[\Phi_{ND}]&=  \frac{v}{\alpha_0}\int_0^\infty \frac{dz }{z^{d+1}} \Phi_{ND}(z)(\nabla^2_z-M^2_{d-2,0})(\nabla^2_z-M^2_{d,0})\Phi_{ND}(z)\,,
	}
	where we used the fact that $f^\text{local}_{\Delta,0}f^\text{local}_{\tilde\Delta,0}=1$ as well as the completeness relation \eqref{NDcomplete}. In the $O(N)$ theory, we would get the same results except with an extra factor of $\frac12$. The non-decaying bulk propagator can be explicitly computed in this case to get 
	\es{bulkPropz2}{
	\langle \Phi_{ND}(z_1) \Phi_{ND}(z_2)\rangle=\frac{\alpha_0}{2v}\Big[\frac{\text{min}(z_1,z_2)^d}{2 (d-2) d}-\frac{\text{max}(z_1,z_2)^2 \text{min}(z_1,z_2)^{d-2}}{2 (d-4) (d-2)}\Big]\,,
}
where we close the contour in \eqref{bulkPropz} to either the left or the right, depending on whether $z_1>z_2$.
	
	We could similarly map the interaction terms in \eqref{eq:free_pert_bilocal_action} to the bulk, where in general there will be nonvanishing cross terms with both $\Phi_{D}(x,z)$ and $\Phi_{ND}(z)$. Note that since the decaying and non-decaying AdS-to-CFT maps are linear, all the $\Phi_{D}(x,z)$ and $\Phi_{ND}(z)$ can be recombined into $\Phi_0(x,z)$ to get the terms discussed in section \ref{sec:bulk_theory}, just as we know that in the bi-local language $\eta_D(x_1,x_2)$ and $\eta_{ND}(x_1,x_2)$ recombine into $\eta(x_1,x_2)$ by definition to get \eqref{eq:free_pert_bilocal_action}. So from the perspective of the bulk action and computing loops, nothing actually changes by introducing the non-decaying map. This is not surprising, as the non-decaying modes are all non-dynamical due to the factors of $v$. On the other hand, the non-decaying map is necessary to map the finite mass deformation, as we discuss next, and for similar computations that modify the theory everywhere in space-time.
		
	\subsection{Mapping the finite mass deformation}
\label{sec:finmass}

We will now use the non-decaying CFT-to-AdS map \eqref{cfttoadsND} to find the bulk VEV dual to the mass deformed CFT VEV \eqref{bilocalmass}, where for simplicity we will work in $d=3$.\footnote{We could also find this VEV directly in the bulk as a solution to the full non-linear bulk equations of motion in the presence of the mass, but we find it easier to just map it from the CFT.} An immediate difficulty with applying this map is that the expansion in the $d_\Delta$ basis is only well defined if the integral in \eqref{eq:coeffND} converges for a bi-local configuration $\eta_{ND}(|x_{12}|)$, but this integral in fact diverges in the $|x_{12}|\to0$ regime for \eqref{bilocalmass} and $\Delta\in\gamma_0$. We can regularize this divergence by shifting $|x_{12}|\to |x_{12}|+\eps$ when expanding $\eta_{ND}(|x_{12}|)$ in $d_\Delta$, and then setting $\eps\to0$ after we then relate $d_\Delta$ to $\Phi_{ND}(z)$ as in \eqref{eq:coeffNDB} to derive the non-decaying CFT-to-AdS map. The resulting $\Phi_{ND}(z)$ is finite as $\eps\to0$, since \eqref{bilocalmass} is finite, and both the CFT and bulk expansions in terms of $d_\Delta$ are just the same Mellin transform, up to a shift in the contour and the factor of $f_{\tilde\Delta}$.

\begin{figure}[]
\begin{center}
        \includegraphics[width=.7\textwidth]{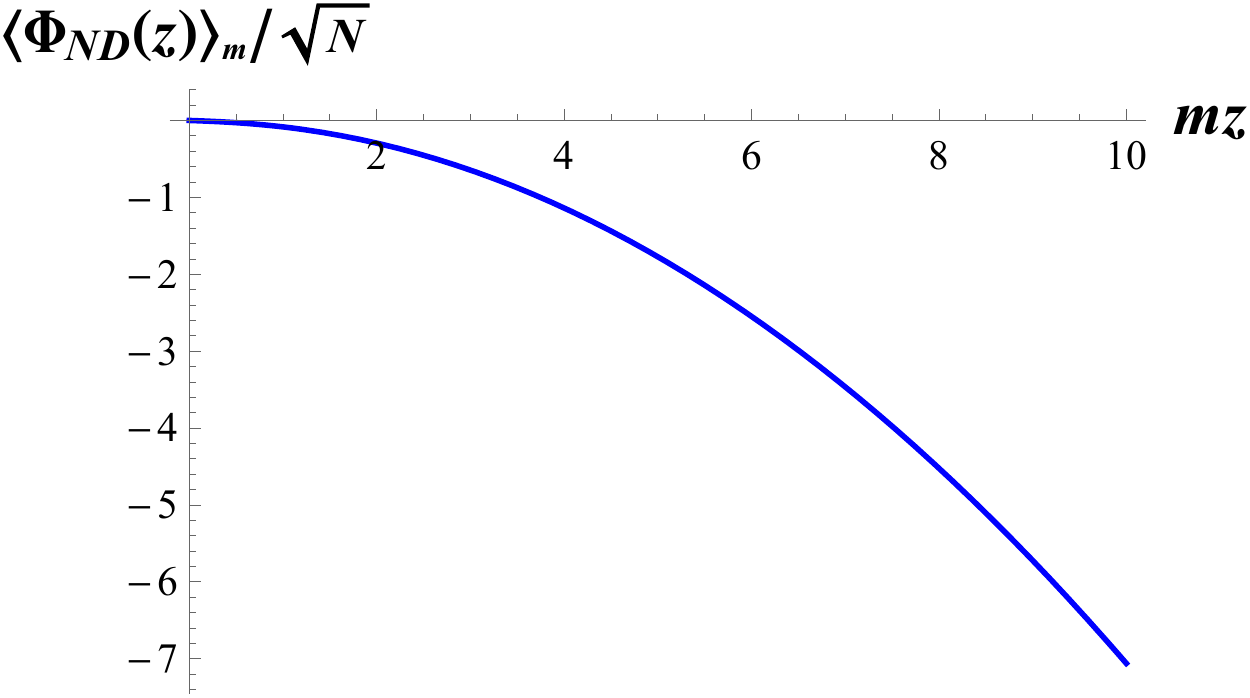}
\caption{The VEV of the non-decaying scalar bulk field in the presence of a finite mass $m$ deformation as a function of $mz$, where $z$ is the radial direction, for $d=3$.
}
\label{mzplot}
\end{center}
\end{figure}

An equivalent and computationally simpler way of regularizing the CFT-to-AdS map of \eqref{bilocalmass} is to simply shift the contour from $\gamma_0$ to 
\es{bargamma}{
\bar\gamma_0:\qquad \Delta=r+is\,,\qquad 2<r<3\,,\quad s\in\mathbb{R}\,,
}
so that in \eqref{cfttoadsND} the $|x_{12}|$ integral converges and takes the form
\es{xint}{
z^{3-\Delta} \int_0^\infty d|x_{12}|\, \frac{\exp(-m|x_{12}|)-1}{4\pi|x_{12}|^{{4-\Delta}}}=(mz)^{3-\Delta}\frac{\Gamma(\Delta-3)}{4\pi}\,.
}
We then explicitly subtract any contributions to the $\Delta$ integral in \eqref{cfttoadsND} from poles and branch cuts that were crossed when shifting the contour. For $f^{\text{local}}_{\Delta,0}$ we get for the resulting mass-deformed bulk VEV
\es{finitemassresults}{
\frac{\langle\Phi_{ND}(z)\rangle_m}{\sqrt{N}}&=\int_{\bar\gamma_0}\frac{d\Delta}{2\pi i} f^\text{local}_{\tilde\Delta,0} (mz)^{3-\Delta}\frac{\Gamma(\Delta-3)}{4\pi}-  \frac{1}{4\pi}\text{Res}\Big[f^\text{local}_{\tilde\Delta,0} (mz)^{3-\Delta}\Gamma(\Delta-3)\Big]_{\Delta=1} \,, \\
}
where the second term corresponds to the pole in $\Gamma(\Delta-3)$ allowed by $\gamma_0$, since $ f^\text{local}_{\tilde\Delta,0} $ has neither poles nor branch cuts in this regime. The integral in the first term is complicated due to the infinite branch cuts in $f^{\text{local}}_{\tilde\Delta,0}$, but it can be easily computed numerically for any $mz$. The result goes from zero at small $mz$ to a constant at large $mz$, so that the large $mz$ scaling is given by the sum over poles. We plot $\langle\Phi_{ND}(z)\rangle_m$ as a function of $mz$ in Figure \ref{mzplot}, which has linear growth at large $mz$. It would be interesting to compare this result to the analogous mass deformed solution to Vasiliev's equations \cite{Iazeolla:2007wt,Sezgin:2005hf}.


\section{The double trace deformation and the critical theory} \label{sec:bulk_deform}

In this section we discuss the deformation of the free scalar CFT by a $(\sum_I |\phi_I|^2)^2$ term, which in the bi-local theory is a double-trace deformation $\lambda NG^2(x,x)$, and the critical theory that it leads to in the low-energy ($\lambda\to\infty$) limit for $2<d<4$. We start by reviewing the double-trace deformation of the free $U(N)$ or $O(N)$ CFT, following \cite{deMelloKoch:1996mj,Mulokwe:2018czu,Gubser:2002vv,Fei:2014yja}. We will then discuss the bulk dual of the critical theory, using the off-shell mapping of $\eta(x,x)$ to the boundary limit of $\Phi_0(x,z)$ \eqref{toshow2}. We will show how the bulk duals of the free and critical theory only differ by the boundary conditions of the bulk scalar two-point function, as anticipated in \cite{Klebanov:1999tb,Hartman:2006dy,Giombi:2011ya}. In this section, we do not use embedding space for clarity.

\subsection{The deformation in the bi-local CFT}
\label{critUN}

Consider the deformation of the bi-local $U(N)$ action $S[\eta]$ in \eqref{eq:bilocalaction} by the double trace scalar deformation, such that the CFT path integral is 
\es{eq:double_trace_defCFT}{
Z_\lambda &= \int D\eta(x,x')\text{exp}\left[-S[\eta] -\frac{\lambda N}{4} \int d^dx G^2(x,x)\right]\\ 
&= \int D\eta(x,x')\text{exp}\left[-S[\eta] -\frac{\lambda }{4} \int d^dx \Big(\eta^2(x,x)+NG_0^2(x,x)+2\sqrt{N}\eta(x,x)G_0(x,x)\Big)\right]\,.\\ 
}
The second term in the integral is just a numerical constant, while the last term is a divergent mass term that can be trivially cancelled at each order by a mass counterterm, so we will ignore both of these terms in the following, and just focus on the nontrivial $\eta^2(x,x)$ term. 

We can equivalently introduce an extra path integral over the Hubbard-Stratonovich field $\sigma(x)$ to get a linear deformation
\begin{equation}\label{eq:double_trace_def_HSCFT}
\begin{split}
Z_\lambda = \int D\sigma(x)D\eta(x,x')\text{exp}\left(-S[\eta] -\int d^dx \left( \frac12\sigma(x) \eta(x,x)- \frac{1}{4\lambda} \sigma^2(x) \right)\right)\,,
\end{split}
\end{equation}
such that performing the gaussian $\sigma(x)$ path integral sets $\sigma(x)=\lambda\eta(x,x)$ and gives us back \eqref{eq:double_trace_defCFT}, up to an overall numerical constant and the trivial mass term. In the latter formulation, we compute correlators of $\sigma(x)$ and $\eta(x_1,x_2)$ in a large $N$ expansion using the same Feynman rules introduced in section \ref{sec:pert_bilocal} for $\eta(x_1,x_2)$, along with the new vertex $\frac12\sigma(x) \eta(x,x)$, and the propagator $\contraction{}{\sigma(x_1)}{}{\sigma(x_2)}\sigma(x_1)\sigma(x_2)=-2\lambda\delta(x_{12})$. For instance, for the $\sigma$ two-point function we get an infinite set of diagrams at leading order in $1/N$:
\es{sig2}{
\langle\sigma(x_1)\sigma(x_2)\rangle_\lambda=&\Big[-2\lambda\delta(x_{12})+(-2\lambda)^2\frac{G_0(x_1,x_2)^2}{4}+(-2\lambda)^3\int d^dy\frac{G_0(x_1,y)^2}{4}\frac{G_0(y,x_2)^2}{4} +\dots\Big]+O(N^{-1})\\
=&-\int \frac{d^d{k}}{(2\pi)^d}\frac{2\lambda e^{ik\cdot x_{12}}}{1-2\lambda \frac{
	|k|^{d-4}2^{1-2d}\pi^{\frac{3-d}{2}}}{\Gamma(\frac{d-1}{2})\sin(\frac{\pi d}{2})}}+O(N^{-1})\,,
}
where in the second equality we performed the geometric series of infinite diagrams in momentum space, using the Fourier transform of the explicit $G_0(x,x')^2/4$ in \eqref{eq:gvev}. For the bi-local two-point function, the free theory answer at $\lambda=0$ is corrected by the same infinite series of diagrams to get
\es{bi2finlam}{
&\langle\eta(x_1,x_2)\eta(x_3,x_4)\rangle_\lambda=G_0(x_1,x_4)G_0(x_2,x_3)\\
&\qquad+\frac14\int d^dy d^dy' G_0(x_1,y) G_0(x_2,y) \langle\sigma(y)\sigma(y')\rangle_\lambda G_0(y',x_3) G_0(y',x_4)+O(N^{-1}) \,.
}
We can then equivalently compute correlators of $\eta(x_1,x_2)$ using the $\lambda\eta(x,x)^2$ formulation in \eqref{eq:double_trace_defCFT}, with no $\sigma(x)$, where the Feynman rules are now given by those in section \ref{sec:pert_bilocal} except that the bi-local propagator is now \eqref{bi2finlam} (with the $\sigma$ two-point function given by \eqref{sig2}). 

For $d<4$, $\lambda\eta(x,x)^2$ is a relevant deformation, and we can go to the critical theory in the IR by taking $\lambda\to\infty$. In the $\sigma(x)$ formulation \eqref{eq:double_trace_def_HSCFT}, we can then drop the $\sigma^2/\lambda$ term in \eqref{eq:double_trace_def_HSCFT}, so that $\sigma(x)$ becomes a Lagrange multiplier setting $\eta(x,x)=0$. We can also now perform the momentum space integral in \eqref{sig2} to get
\es{sig3}{
\langle\sigma(x_1)\sigma(x_2)\rangle_\infty=&\frac{B_d}{x_{12}^4}+O(N^{-1})\,,\qquad B_d=\frac{2^{d+3} \sin \left(\frac{\pi  d}{2}\right) \Gamma (\frac{d-1}{2})}{\pi^{\frac32}  \Gamma
   \left(\frac{d}{2}-2\right)}\,,
}
which is the 2-point function of a conformal primary $\sigma(x)$ with $\Delta_\sigma=2+O(1/N)$. We thus see that to leading order in $1/N$, the critical theory is the same as the free theory, but with the $\frac{\phi_I^*\phi_I(x)}{\sqrt{N}}\equiv \eta(x,x)$ operator with $\Delta_{\phi^2}=d-2$ replaced by the shadow operator $\sigma(x)$ with the shadow dimension $\Delta_\sigma$. For the $O(N)$ theory, we have extra factors of $\frac12$ in \eqref{sig2} and \eqref{sig3}, while in \eqref{bi2finlam} the second term is now multiplied by $\frac12$ instead of $\frac14$ and we should add the symmetrized $x_3\leftrightarrow x_4$ term.


\subsection{The bulk dual}
\label{critUNbulk}

We can map the double trace deformation in \eqref{eq:double_trace_defCFT} and \eqref{eq:double_trace_def_HSCFT} to the bulk using the off-shell mapping \eqref{toshow2} that holds for $2<d<4$. We get 
\es{bulkLam}{
Z_\lambda = \int D\Phi_J(x,z)\text{exp}\left[-S[\Phi] -\frac{\lambda}{4 f_{d-2,0}^2} \int d^dx \left(\lim_{\eps\to0}\eps^{2-d}\Phi_0(x,\eps)\right)^2\right],
}
or equivalently (up to an overall numerical constant)
\es{bulkSig}{
Z_\lambda = \int D\sigma(x)D\Phi_J(x,z)\text{exp}\left[-S[\Phi] -\int d^dx \left( \frac{\sigma(x) }{2f_{d-2,0}}\lim_{\eps\to0}\eps^{2-d}\Phi_0(x,\eps)- \frac{\sigma^2(x)}{4\lambda}  \right)\right]\,.
}
As in the mass deformation, we see that only the boundary of AdS is affected. Note that the only effect of the deformation is to change the scalar bulk propagator, just as in the CFT only the bi-local 2-point function was changed (at leading order in $1/N$). We can compute this as in our computation of the mass deformation VEV \eqref{vev}, to get
\es{bulkCrit2p}{
&\langle  \Phi_0(x_1,z_1)\Phi_0(x_2,z_2\rangle_\lambda=\langle  \Phi_0(x_1,z_1)\Phi_0(x_2,z_2)\rangle_0\\& \quad+ \frac{f_{2,0}^2\Gamma(d-2)^2}{1024\pi^{2d}}
\int d^dyd^dy' G_{d-2,0}(x_1,z_1|y)\langle\sigma(y)\sigma(y')\rangle_\lambda G_{d-2,0}(x_2,z_2|y')+O(N^{-1})\,,
}
where in the second term we resummed the same geometric series of diagrams that appeared in the CFT calculation. 

For the critical theory at $\lambda\to\infty$, we obtain when we choose the local mapping $f^\text{local}_{\Delta,J}$:
\es{bulkCrit2p2}{
&\langle  \Phi_0(x_1,z_1)\Phi_0(x_2,z_2\rangle_\infty=\langle  \Phi_0(x_1,z_1)\Phi_0(x_2,z_2)\rangle_0+\frac{B_d(f^\text{local}_{2,0})^2\Gamma(d-2)^2}{1024\pi^{2d}} \\
&\qquad\qquad\qquad\qquad\qquad\qquad \times\int d^dyd^dy' \frac{G_{d-2,0}(x_1,z_1|y) G_{d-2,0}(x_2,z_2|y')}{|y-y'|^4}+O(N^{-1})\\
&\qquad\qquad\qquad\qquad\quad\quad=\langle  \Phi_0(x_1,z_1)\Phi_0(x_2,z_2)\rangle_0+\frac{B_d(f^\text{local}_{2,0})^2\Gamma(d-2)^2S_B^{d-2,0}N_{2,0}\alpha_0(d-4)}{2048\pi^{2d}}\\
&\qquad\qquad\qquad\qquad\qquad\qquad\times (\Pi^{TT}_{d-2,0}(x_1,z_1;x_2,z_2)-\Pi^{TT}_{2,0}(x_1,z_1;x_2,z_2))+O(N^{-1})\\
&\qquad\qquad\qquad\qquad\quad\quad=\frac{\alpha_0/2}{M_{d,0}^2-M_{d-2,0}^2}(\Pi^{TT}_{2,0}(x_1,z_1;x_2,z_2)-\Pi^{TT}_{d,0}(x_1,z_1;x_2,z_2))+O(N^{-1})\,,\\
}
where in the first equality we used \eqref{sig3}, in the second we used the bulk shadow transform \eqref{eq:shadow_trans_bulk} and the expression \eqref{omega2} for the AdS harmonic function \eqref{eq:omega_def} in terms of the bulk-to-bulk propagator,\footnote{For $J=0$, recall that $\Pi^{TT}_{\Delta,0}(X_1,X_2;W_1,W_2)=\Pi_{\Delta,0}(X_1,X_2;W_1,W_2)$. The identity \eqref{omega2} for the $J=0$ case was also used to relate the bulk dual of the free and critical theories in \cite{Giombi:2011ya,Hartman:2006dy}.} and in the third we used the free theory expression for the bulk propagator in \eqref{eq:2p_local}, and the explicit expressions for the various constants in the previous sections. For the $O(N)$ theory, we get the same result with the usual extra factor of 2. For the more general $f_{\Delta,J}$ defined in section \ref{sec:quadratic_action}, which has the same physical propagator as $f_{\Delta,J}^\text{local}$, we can similarly compute the result in the limit where one point goes to the boundary, to get
\es{bulkCrit2p22}{
&\lim_{z_1\to0}\langle  \Phi_0(x_1,z_1)\Phi_0(x_2,z_2)\rangle_\infty=\frac{\alpha_0f_{2,0}f_{d-2,0}/2}{M_{d,0}^2-M_{d-2,0}^2}\lim_{z_1\to0}\Pi^{TT}_{2,0}(x_1,z_1;x_2,z_2)+O(N^{-1})\,.\\
}
In all these cases, we see that the physical propagator $\Pi^{TT}_{d-2,0}$ for $\Phi_0$ has been replaced by the shadow propagator $\Pi^{TT}_{2,0}$. Recall that in general both $\Pi^{TT}_{\Delta,J}$
and $\Pi^{TT}_{\tilde\Delta,J}$ are defined by the same bulk differential equation \eqref{eq:bulk_bulk_DE_2}, and differ only by the boundary condition \eqref{eq:BB_limit}. Since the only difference between the free and critical bulk theories was this scalar propagator, we see that to all orders in $1/N$ the only difference between the free and critical bulk theories is the boundary condition for the bulk scalar, as discussed in \cite{Klebanov:1999tb,Giombi:2011ya,Hartman:2006dy}.

We can now compute bulk correlators with the Feynman rules discussed in section \ref{sec:bulk_theory}, except with the $\Phi_0$ bulk propagator now given by \eqref{bulkCrit2p2}. As shown in \cite{Giombi:2011ya}, since the difference
\es{difference}{
\langle  \Phi_0(x_1,z_1)\Phi_0(x_2,z_2\rangle_\infty-\langle  \Phi_0(x_1,z_1)\Phi_0(x_2,z_2\rangle_0\propto \Pi^{TT}_{d-2,0}(x_1,z_1;x_2,z_2)-\Pi^{TT}_{2,0}(x_1,z_1;x_2,z_2)
}
is free of short distance singularities, and since we already showed that the free bulk dual gave the expected bulk diagrams dual to the free CFT (with appropriate cancellations of all UV divergences), the modification of the free bulk Feynman rules by replacing $\Pi^{TT}_{d-2,0}\to \Pi^{TT}_{2,0}$ leads to the expected bulk dual of the critical CFT for all correlators at all orders in $1/N$. Note that the argument in \cite{Giombi:2011ya} did not assume any specific bulk action, and only required that the difference between the free and critical bulk Feynman rules be implemented by $\Pi^{TT}_{d-2,0}\to \Pi^{TT}_{2,0}$, so the fact that our bulk theory has a non-standard off-shell kinetic term does not matter.  

The mapping of single-trace local operators in the CFT, as described in section \ref{sec:single_trace_dual}, is now different for the $J=0$ case. Note that for $\lambda\to\infty$, the $\sigma(x)$ field acts as a Lagrange multiplier in \eqref{bulkSig} that sets $\lim_{\eps\to0}\eps^{2-d}\Phi_0(x,\eps)=0$ off-shell, just as in the CFT it set $\eta(x,x)=0$ off-shell. Thus, the off-shell relation \eqref{toshow2} becomes trivial in the critical theory. Since the $z^{d-2}$ mode vanishes, $\Phi_0(x,z)$ now has the same small $z$ boundary condition as all other $J>0$ bulk fields described in section \ref{sec:bulk_bc}, namely it scales as $z^{d/2}$, which is the real part of the principal series contour. On-shell, we find that \eqref{toshow} for $J=0$ is replaced by
\es{toshowcrit}{
\langle \sigma(x)\dots\rangle_\eqref{bulkSig}=\frac{32\pi^{\frac d2}}{\Gamma(\frac d2-2)f_{2,0}}\lim_{\eps\to0}\eps^{-2}\langle\Phi_0(x,\eps)\dots\rangle_\eqref{bulkLam}\,,
}
where on the left-hand side we compute correlators using the $\lambda\to\infty$ Feynman rules from \eqref{bulkSig}, with the free bulk Feynman rules for $\Phi_J$ and the $\sigma$ propagator in \eqref{sig3}, while the right-hand side uses the $\lambda\to\infty$ Feynman rules from \eqref{bulkLam}, with the free bulk Feynman rules for $\Phi_{J>0}$ and  \eqref{bulkCrit2p2} for $\Phi_0$. We can see this on-shell relation by computing the contraction between $\sigma(x)$ and $\Phi_0(x,z)$ as
\es{toshowcrit2}{
\langle \sigma(x_1)\Phi_0(x_2,z) \rangle_\eqref{bulkSig}&=\lim_{\eps\to0}\frac{ \eps^{2-d} }{2f_{d-2,0}} \int d^dy\langle\sigma(x_1)\sigma(y)\rangle_\infty\langle\Phi_0(y,\epsilon)\Phi_0(x_2,z) \rangle_0+O(N^{-1})\\
&=\frac{\mathcal{C}_{d-2,0}\alpha_0 f_{2,0}B_d}{4(M_{d,0}^2-M_{d-2,0}^2)}\int d^dy\frac{G_{d-2,0}(x_2,z|y)}{|x_1-y|^4}+O(N^{-1})\\
&=\frac{S_B^{d-2,0}\mathcal{C}_{d-2,0}\alpha_0 f_{2,0}B_d}{4(M_{d,0}^2-M_{d-2,0}^2)}{G_{2,0}(x_2,z|x_1)}+O(N^{-1})\\
&=\frac{32\pi^{\frac d2}}{\Gamma(\frac d2-2)f_{2,0}}\lim_{\eps\to0}\eps^{-2}\langle\Phi_0(x_1,\eps)\Phi_0(x_2,z)\rangle_\infty+O(N^{-1})\,,
}
where in the first equality we used the $\sigma\Phi_0$ vertex in \eqref{bulkSig}, in the second we used the definitions \eqref{sig3} and \eqref{eq:2p_localf}, in the third we used the bulk shadow transform \eqref{eq:shadow_trans_bulk}, and in the last we used \eqref{bulkCrit2p22}, \eqref{eq:bulk_norm_shadow_cond}, and the explicit expressions for the various constants. In the small $z$ limit, we similarly recover the relation between $\langle\sigma(x_1)\sigma(x_2)\rangle$ and $\langle\Phi_0(x_1,\eps)\Phi_0(x_2,\eps)\rangle$ required for \eqref{toshowcrit}.

\section*{Acknowledgements}

We would like to thank D.~Anninos, M.~Berkooz, S.~Das, S.~Pufu, C.~Sleight, and M.~Taronna for useful discussions, and to especially thank V.~Rosenhaus for many useful discussions, and T.~Solberg for collaboration on parts of this project and on a related project.
This work was supported in part  by an Israel Science Foundation center for excellence grant (grant number 2289/18), by grant no. 2018068 from the United States-Israel Binational Science Foundation (BSF), and by the Minerva foundation with funding from the Federal German Ministry for Education and Research. OA is the Samuel Sebba Professorial Chair of Pure and Applied Physics. SMC is supported in part by a Zuckerman STEM leadership fellowship.

\pagebreak
\appendix

\section{The embedding space formalism} \label{app:emb_form}
It will be convenient for us to use the embedding space formalism, both for CFT computations and for computations in anti-de Sitter (AdS) space.
	This section is mostly copying formulas and notations from~\cite{Costa:2014kfa} and \cite{Costa:2011mg}.
	The idea is to use the embedding of AdS$_{d+1}$ in $\mathbb{R}^{d+1,1}$ as the hyperboloid $X^2=-1$, and of $d$-dimensional flat space (in the CFT, or on the boundary of AdS) in $\mathbb{R}^{d+1,1}$ as the conformal light-cone $P^2=0$. In both cases we denote the coordinates of $\mathbb{R}^{d+1,1}$ by $X=\left(X^{+},X^{-},X^{i}\right)$ ($X^{i}=\left(X^{1},\cdots,X^{d}\right)$), with the metric
	\begin{equation}
		X^{2}=-X^{-}X^{+}+X^{i}X^{i}.
	\end{equation}
	In the bigger space the isometries/conformal symmetries $\text{SO}(d+1,1)$ are linear and therefore writing covariant expressions is simpler. As we discuss traceless symmetric tensors, we will also have vectors in the space as ``spin holders". We usually denote the flat-space spin by $Z$ and the bulk AdS spin by $W$.

	\subsection{Flat-space notations}
	We need two constraints to get a $d$-dimensional flat space, and we take:
	\begin{enumerate}
		\item $P^2=0$.
		\item $P\equiv\lambda P$.
	\end{enumerate}
	We can therefore use
	\begin{equation} \label{eq:slice}
		P=\left(1,y^{2},y^{i}\right)
	\end{equation}
	as the relation between the embedding coordinate $P$ and a standard coordinate $y$ in $d$-dimensional flat space. Whenever we write an integral $dP$ it can be interpreted as an integral over the $y^i$ in this slice \eqref{eq:slice}.

	Flat-space symmetric and traceless tensors can be written as a polynomial in a new vector $Z$:
	\begin{equation}
		H(P,Z) = H_{A_1,...,A_J}(P) Z^{A_1}...Z^{A_J}.
	\end{equation}
	For the tensor to be traceless we require $Z^2=0$. For the tensor to be defined only at directions parallel to the light-cone $P^2=0$ we require further:
	\begin{enumerate}
		\item $H(P,Z+\alpha P) = H(P,Z)$. Now the tensor is only parallel to the light-cone.
		\item $Z\cdot P=0$. Because $P\equiv\lambda P$, the tensor direction $H^{A...}=P^{A}$ should also give a zero projection.
	\end{enumerate}
	Given $H_{A_1,...,A_J}(P)$ we can find the actual flat-space tensor $h_{i_1,...,i_J}(y)$ simply by a projection. In order to extract $H_{A_1,...,A_J}(P)$ from $H(P,Z)$ we define the operator
	\begin{equation}
		D_{Z}^{A}=\left(\frac{d}{2}-1+Z\cdot\frac{\partial}{\partial Z}\right)\frac{\partial}{\partial Z_{A}}-\frac{1}{2}Z^{A}\frac{\partial^{2}}{\partial Z\cdot\partial Z},
	\end{equation}
	so that
	\begin{equation}
		H^{A_{1}...A_{J}}(P)=\frac{1}{J!\left(\frac{d}{2}-1\right)_{J}}D_{Z}^{A_{1}}...D_{Z}^{A_{J}}H\left(P,Z\right). \label{eq:boundary_extract_tensor}
	\end{equation}
	Note that using $D_Z$, the original tensor composing $H(P,Z)$ does not need to be traceless (or symmetric) but the outcome of \eqref{eq:boundary_extract_tensor} will be.

	Finally, a flat-space contraction of indices between two spin-$J$ traceless-symmetric tensors can be written as
	\begin{equation}
		f^{i_{1}...i_{J}}\left(y\right)g_{i_{1}...i_{J}}\left(y\right)=\frac{1}{J!\left(\frac{d}{2}-1\right)_{J}}F\left(P,D_{Z}\right)G\left(P,Z\right). \label{eq:boundary_trace}
	\end{equation}

	\subsection{Bulk notations}
	As we said before, $\text{AdS}_{d+1}$ can be embedded as the hyperboloid $X^2=-1$. Denoting the upper half space coordinates $x^\mu=(z,x^i)$ (with the standard metric $ds^2 = (dz^2+dx^i dx^i) / z^2$ that gives AdS space in Poincar\'e coordinates) we have
	\begin{equation}
		X=\frac{1}{z} \left(1,z^2+x^{2},x^{i}\right),
	\end{equation}
	where $x^2 = \sum_{i=1}^d x_i^2$. As $z\to 0$, the bulk point approaches $\frac{1}{z}$ times a null vector $P$ which can be identified with a point on the boundary of AdS (using the flat-space embedding space formalism of the previous subsection for the boundary).
	
	A bulk symmetric and traceless tensor can be written as a polynomial in a new vector $W$:
	\begin{equation}
		H(X,W) = H_{A_1,...,A_J}(X) W^{A_1}...W^{A_J}.
	\end{equation}
	For the tensor to be traceless we require $W^2=0$. For the tensor to be defined only at directions parallel to the hyperboloid we also require $X\cdot W =0$ (equivalently $X^A H_{A,...}=0$). Again, given $H_{A_1,...,A_J}$ we can find the actual AdS tensor $h_{\mu_1,...,\mu_J}$ simply by a projection. To extract the tensor $H_{A_1,...,A_J}$ out of a field $H(Z)$ we need again a differential operator that would respect the transversality, symmetry and tracelessness:
	\begin{equation}\begin{split}
		K_{A}^{W} & =\frac{d-1}{2} \left( \frac{\partial}{\partial W^A} + X_A \left(X\cdot \frac{\partial}{\partial W} \right) \right) + \left(W\cdot \frac{\partial}{\partial W}\right) \frac{\partial}{\partial W^A} \\
		& + X_A \left( W\cdot \frac{\partial}{\partial W} \right) \left( X \cdot \frac{\partial}{\partial W} \right)
		-\frac{1}{2} W_A \left( \frac{\partial^2}{\partial W\cdot\partial W} + \left( X\cdot \frac{\partial}{\partial W} \right) \left( X\cdot \frac{\partial}{\partial W} \right)\right). 
	\end{split}\end{equation}
	This operator is indeed transverse ($X^A K_A = 0 $), symmetric ($K_A K_B = K_B K_A$) and traceless $K_A K^A =0$, so that
	\begin{equation}
		H^{A_{1}...A_{J}}(X)=\frac{1}{J!\left(\frac{d-1}{2}\right)_{J}}K_{W}^{A_{1}}...K_{W}^{A_{J}}H\left(X,W\right). \label{eq:bulk_extract_tensor}
	\end{equation}
	For points on the embedded AdS (with $X^2=-1$, $W^2=W\cdot X=0$),  it is enough to use
	\begin{equation}
		K_{A}^{W} = \left( \frac{d-1}{2} + W\cdot \frac{\partial}{\partial W} \right) \frac{\partial}{\partial W^A}. \label{eq:bulk_extract_operator}
	\end{equation}
	Again, the original tensor composing $H(X,W)$ does not need to be traceless (or symmetric), but using $K_A$'s the outcome of \eqref{eq:bulk_extract_tensor} will be.

	A bulk contraction of indices between two spin-$J$ traceless-symmetric tensors can be written as
	\begin{equation}
		f^{\mu_{1}...\mu_{J}}\left(z,x^i\right)g_{\mu_{1}...\mu_{J}}\left(z,x^i\right)=\frac{1}{J!\left(\frac{d-1}{2}\right)_{J}}F\left(X,K_{W}\right)G\left(X,W\right). \label{eq:bulk_trace}
	\end{equation}

	Finally, the AdS gradient operator may be written in embedding space as
	\begin{equation}
		\nabla_A = \frac{\partial}{\partial X^A} + X_A \left( X\cdot \frac{\partial}{\partial X}\right) + W_A \left( X\cdot \frac{\partial}{\partial W}\right), \label{eq:bulk_d_operator}
	\end{equation}
	such that after the projection the resulting vector is $D_\mu$.

\section{The $2$-point function of $\eta$ in the $C_{\Delta,J}$ formalism} \label{sec:free_4_point}

In this appendix we use the $C_{\Delta,J}$ variables of section \ref{sec:bi_local_to_conformal} to compute the tree-level two-point function of $\eta$ (or the four-point function of the $\phi_I$), and confirm that we obtain the same result as in section \ref{sec:local_to_bilocal}. We will convert the embedding space results back to coordinate space, to make the calculations more transparent. We will perform the calculation assuming $d>4$, so that we can just use the principal series contour. The results for $d\leq4$ can then be obtained by analytic continuation (or could be computed directly in those dimensions by using the more complicated contours discussed in the main text).

Using the decomposition \eqref{eq:eta_comp} and \eqref{eq:cdelta_const} as written in coordinate space, we get for the $U(N)$ theory (keeping just the quadratic term in the action, and with $\Delta_0=\frac{d-2}{2}$)
\begin{equation}\begin{split}
	\left<\eta(x_1,x_2)\eta(x_3,x_4)\right>_\text{tree} &= \sum_{J,J^\prime=0}^\infty \int_{\gamma_J}\frac{d\Delta}{2\pi i}\int_{\gamma_{J^\prime}}\frac{d\Delta^\prime}{2\pi i} 
	\int d^d y\int d^d y^\prime \\
	 &\quad \times \, 
	 \thpnorm{x_1}{x_2}{\Delta,J}{y} \, \thpnorm{x_3}{x_4}{\tilde\Delta^\prime,J^\prime}{y^\prime}\\
	  & \quad \times (-1)^{J^\prime}\int D C_{\Delta,J}(y) \, C_{\Delta,J}\left(y\right)C^*_{\Delta^\prime,J^\prime}\left(y^\prime\right) e^{-S^{(2)}[C_{\Delta,J}]}\,. \label{eq:4_point_gauss_int}
\end{split}\end{equation}
The path integral is Gaussian. It is useful to define 
 the conformal partial wave (CPW) \cite{Simmons-Duffin:2017nub,Karateev:2018oml}\footnote{The two papers differ by a factor of $2^{-J}$ on their definition of the conformal block. We use the one of \cite{Karateev:2018oml}.}
\begin{equation}\begin{split}
	\Psi_{\Delta,J}^{\Delta_0}(x_i) & \equiv \int d^d y \thpnorm{x_1}{x_2}{\Delta,J}{y}\thpnorm{x_3}{x_4}{\tilde\Delta,J}{y}\\
	   & = \frac{1}{|x_{12}|^{d-2}|x_{34}|^{d-2}}\left[S_{\Delta_0,\Delta_0}^{(\tilde\Delta,J)} G_{\Delta,J}(U,V) 
	   + S_{\Delta_0,\Delta_0}^{(\Delta,J)} G_{\tilde\Delta,J}(U,V)\right] \label{eq:CPW_integral}
\end{split}\end{equation}
where as usual we omit the index contractions, and the conformal cross ratios
\es{UV}{
U=\frac{|x_{12}|^2|x_{34}|^2}{|x_{13}|^2|x_{24}|^2}\,,\qquad V=\frac{|x_{23}|^2|x_{14}|^2}{|x_{13}|^2|x_{24}|^2}\,.
}
Then, using \eqref{eq:c_action} the Gaussian path integral gives\footnote{There are two important distinctions between our expression and that of \cite{Karateev:2018oml}. Firstly, their contour is only for the upper half plane of the principal series, which is why we have a 2 in \eqref{eq:CPW_decompo} relative to their expression. Also, their operators are ordered differently from ours so that our integrand has an extra $(-1)^J$ relative to theirs.}
\es{eq:CPW_decompo}{
	\left<\eta(x_1,x_2)\eta(x_3,x_4)\right>_\text{tree}  =& \sum_{J=0}^\infty \int_{\gamma_J}\frac{(-1)^J d\Delta}{4\pi i\, \lambda_{\Delta,J} N_{\Delta,J}} 
	 \int d^d y 
	 \thpnorm{x_1}{x_2}{\Delta,J}{y} \\
	 &\qquad\qquad\qquad\qquad\qquad\times\thpnorm{x_3}{x_4}{\tilde\Delta,J}{y}\\
	  =&\sum_{J=0}^\infty \int_{\gamma_J}\frac{d\Delta}{4\pi i} 
	\frac{(-1)^J}{\lambda_{\Delta,J} N_{\Delta,J}} \Psi^{\Delta_0}_{\Delta,J}(x_i)\\
	= & \frac{1}{|x_{12}|^{d-2}|x_{34}|^{d-2}}\sum_{J=0}^\infty \int_{\gamma_J}\frac{d\Delta}{2\pi i} 
	\frac{ (-1)^J S_{\Delta_0,\Delta_0}^{(\tilde\Delta,J)}}{\lambda_{\Delta,J} N_{\Delta,J}}  G_{\Delta,J}(U,V)\,.
}
In the last equality of \eqref{eq:CPW_decompo} we used the fact that the rest of the integrand (and the contour itself, including our deformations) is invariant under $\Delta \iff \tilde\Delta$. The coefficients of the conformal blocks have poles at $\Delta=d-2+2n+J$ for $n=0$ with residues:\footnote{As we can see in \eqref{eq:free_lambda}, the function $\frac{1}{\lambda_{\Delta,J}}$ has poles for both $n=0$ ($\Delta=d+J-2$) and $n=1$ ($\Delta = d+J$). So in a way both are ``physical operators". But from the free theory we know that while the first is an actual operator (the mass operator $\phi^2(x)$ and the conserved spin $J$ currents), the second vanishes because of the equation of motion $:\phi \partial^J \square \phi: = 0$. The reason $n=1$ doesn't contribute to the correlation function at the end of the day is because we take residues of $\frac{ S_{\Delta_0,\Delta_0}^{(\tilde\Delta,J)}}{\lambda_{\Delta,J} N_{\Delta,J}}$. In $S_{\Delta_0,\Delta_0}^{(\tilde\Delta,J)}$ we have a term $\Gamma^{-2}\left( \frac{\tilde\Delta +J}{2}\right)$ (see \eqref{eq:shadow_coeff_2}) with a double zero at $\Delta=d+J$ that kills the pole from $\frac{1}{\lambda_{\Delta,J}}$. 
The conclusion is that the $\Delta=d+J$ operators are ``physical" at the level of the action, but don't contribute to any physical OPE. }
\es{R}{
	R^{\left(J\right)} \equiv \text{Res}\left[\frac{(-1)^J  S_{\Delta_0,\Delta_0}^{(\tilde\Delta,J)}}{\lambda_{\Delta,J} N_{\Delta,J}} \right]=-(-1)^J\frac{\pi ^{-d} 2^{J-4} (d+2 J-3) \Gamma \left(\frac{d}{2}+J-1\right)^2 \Gamma
   (d+J-3)}{\Gamma (J+1) \Gamma (d+2 J-2)}
   \,.
}
To compute \eqref{eq:CPW_decompo} we thus need to perform the sum
\begin{equation}
	-\frac{1}{|x_{12}|^{d-2}|x_{34}|^{d-2}}\sum_{J=0}^\infty R^{(J)} G_{d + J - 2,J}(U,V) \,,
\end{equation}
where the minus sign comes from closing the contour to the right, and the conformal blocks $G_{\Delta,J}$ are normalized with the small $U$ expansion
\es{blockU}{
G_{\Delta,J}(U,V)=\frac{1}{(-2)^J}U^{\frac{\Delta-J}{2}}(1-V)^J{}_2F_1\left(\frac{\Delta+J}{2},\frac{\Delta+J}{2},{\Delta+J},1-V\right)+O(U^{\frac{\Delta-J}{2}+1})\,.
}
Since for the free theory we only consider the lowest twist operators, we can actually ignore the sub-leading powers of $U$, which must identically cancel. We thus compute
\es{sumblock}{
&\sum_{J=0}^\infty \frac{  (d+2 J-3) \Gamma \left(\frac{d}{2}+J-1\right)^2 \Gamma
   (d+J-3)}{2^{4-J}\pi ^{d}\Gamma (J+1) \Gamma (d+2 J-2)}\frac{U^{\frac{d-2}{2}}}{2^J}(1-V)^J{}_2F_1\left[\frac{d-2}{2}+J,\frac{d-2}{2}+J,{d-2+2J},1-V\right]\\
&=U^{\frac{d}{2}-1}\int_0^1 dt\sum_{J=0}^\infty \frac{\pi ^{-d} (d+2 J-3) (1-v)^J ((1-t) t)^{\frac{d}{2}+J-2} \Gamma (d+J-3) (t
   (v-1)+1)^{-\frac{d}{2}-J+1}}{16 \Gamma (J+1)}\\
   &=\left[\frac{\Gamma(d/2-1)}{4\pi^{d/2}}\right]^2(U/V)^{\frac{d}{2}-1}\,,
}
so that the resulting (connected part of the) correlation function is
\begin{equation}
	\label{4pointfree}
	\langle\eta(x_1,x_2)\eta(x_3,x_4)\rangle=
	\left[\frac{\Gamma(d/2-1)}{4\pi^{d/2}}\right]^2\frac{1}{|x_{14}|^{d-2}|x_{23}|^{d-2}}\,,
\end{equation}
as we expect from Wick contractions for the canonically normalized scalar with propagator \eqref{eq:gvev}. For the $O(N)$ theory, we sum only over even spins and multiply the residues by two since the quadratic action for the $O(N)$ theory is half that of the $U(N)$. We then get $\left[\frac{\Gamma(d/2-1)}{4\pi^{d/2}}\right]^2({U^{\frac{d}{2}-1}+(U/V)^{\frac{d}{2}-1}})$, and so the correlation function is
\begin{equation}\begin{split}
\label{4pointfreeON}
\langle\eta(x_1,x_2)\eta(x_3,x_4)\rangle =
\left[\frac{\Gamma(d/2-1)}{4\pi^{d/2}}\right]^2\left[\frac{1}{|x_{13}|^{d-2}|x_{24}|^{d-2}}+\frac{1}{|x_{14}|^{d-2}|x_{23}|^{d-2}}\right]\,,
\end{split}\end{equation}
as expected.

\section{Bulk identities} \label{bulkIdentities}
	
	In this appendix we prove some key identities used in the main text to analyze the bulk physics.
	
	\subsection{Proof of \eqref{eq:Theta_Omega4_diag}}

	 The first identity we consider is \eqref{eq:Theta_Omega4_diag}, which can be written explicitly as
\es{eq:Theta_Omega4App}{
	&\sum_J \frac{1}{\left(J! \left(\frac{d-1}{2}\right)_J\right)^2}
	 \int dX   \int dX'   \left(W_1 \cdot \nabla_{X_1}\right)^{J_1} 
	\Omega_0  (X_1,X) \left(K_W \cdot \nabla_{X}\right)^{J} \Omega_0(X_2,X)\\
	&\times \Theta_J(X,X';W,W') \left(K_{W'} \cdot \nabla_{X'}\right)^{J}
	\Omega_0 (X',X_3)
	\left(W_4 \cdot \nabla_{X_4}\right)^{J_4} \Omega_0(X',X_4) 
	\\ 
	&\qquad=  \left(W_1 \cdot \nabla_{X_1}\right)^{J_1} \Omega_0(X_1,X_3)
	\left(W_4 \cdot \nabla_{X_4}\right)^{J_4} \Omega_0(X_2,X_4) \,.
}

As a first step, we will show a simpler identity: we will interpret the conformal basis completeness relation \eqref{eq:3p_comp_rel} as an identity about $\Theta$. We have
\begin{equation}\begin{split}\label{completelyugly}
	\delta(P_1,P_3)\delta(P_2,P_4) & = \frac{1}{2}
	\sum_J \int 
	\frac{dP}{J! \left(\frac{d}{2}-1\right)_J}
	 \int \frac{d\Delta}{2\pi i N_{\Delta,J}} \\
	 &
	\times \thpnorm{P_1}{P_2}{\Delta,J}{P,D_Z}\thptild{P_3}{P_4}{\tilde\Delta,J}{P,Z}\\
	 & = \frac{1}{2}
	\sum_J \frac{1}{\left(J! \left(\frac{d-1}{2}\right)_J\right)^2}
	\int \frac{d\Delta}{2\pi i N_{\Delta,J} b^{(\Delta,J)}_{\Delta_0,\Delta_0} { b}^{(\tilde \Delta,J)}_{\tilde\Delta_0,\tilde\Delta_0} }\\
	 & \times \int dX 
	 (W\cdot \nabla)^J G_{\Delta_0,0}(X,P_2)
	 G_{\Delta_0,0}(X,P_1)
	 \int dX' 
	 (W'\cdot \nabla)^J G_{\tilde\Delta_0,0}(X',P_3)
	 G_{\tilde\Delta_0,0}(X',P_4) \\
	 & \qquad \int \frac{dP}{J! \left(\frac{d}{2}-1\right)_J} G_{\Delta,J}(X,P; K_{W},D_Z) G_{\tilde\Delta,J}(X',P; K_{W'},Z)\\
	 & = \sum_J \frac{1}{\left(J! \left(\frac{d-1}{2}\right)_J\right)^2} \int dX \int dX'   (K_W\cdot \nabla)^J G_{\Delta_0,0}(X,P_2)
	 G_{\Delta_0,0}(X,P_1) \\
	 & \qquad\qquad  \times\Theta(X,X';W,W')(K_{W'}\cdot \nabla)^J G_{\tilde\Delta_0,0}(X',P_3)
	 G_{\tilde\Delta_0,0}(X',P_4)\,,
\end{split}\end{equation}
where in the second equality we used \eqref{eq:penedones_3point} twice, while in the third equality we used \eqref{eq:omega_def} and \eqref{eq:Theta_def}. We can draw this result as
\begin{equation}\label{eq:comp_rel_witten_diag}
	\vcenter{\hbox{\includegraphics{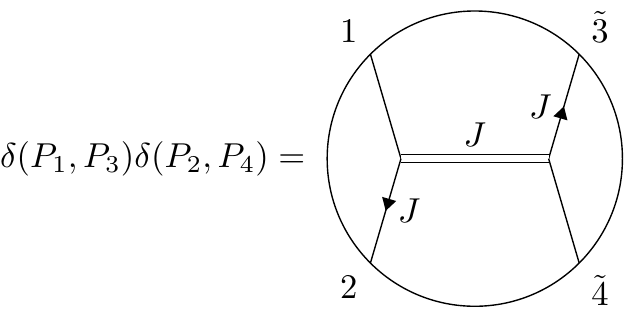}}}
\end{equation}
where the boundary legs are the bulk-to-boundary propagators with $\Delta=\Delta_0$ (and its shadow). Similar to Table \ref{tb:feynman_lines}, the arrows signify the derivatives (and spin contractions).
We can now finish the proof:
\begin{equation}\begin{split}
	\Omega_{0}(X_1,X_3) &
	\Omega_{0}(X_2,X_4) =  
	\int dP_1 dP_2 dP_3 dP_4 \delta(P_1,P_3)
	\delta(P_2,P_4)\\
	& \qquad\qquad\qquad \times G_{\tilde\Delta_0,0}(X_1,P_1) G_{\tilde\Delta_0,0}(X_2,P_2) G_{\Delta_0,0}(X_3,P_3) G_{\Delta_0,0}(X_4,P_4) \\
	= & \sum_J \frac{1}{\left(J! \left(\frac{d-1}{2}\right)_J\right)^2} \int dX \int dX'  \int dP_1 G_{\tilde\Delta_0,0}(X_1,P_1) G_{\Delta_0,0}(X,P_1) \\
	 & \times\int dP_2 (K_{W}\cdot \nabla)^J G_{\Delta_0,0}(X,P_2) G_{\tilde\Delta_0,0}(X_2,P_2)  \Theta(X,X';W,W') \\
	 &\times \int dP_3 G_{\Delta_0,0}(X_3,P_3) (K_{W'}\cdot \nabla)^J G_{\tilde\Delta_0,0}(X',P_3)\int dP_4 G_{\Delta_0,0}(X_4,P_4) G_{\tilde\Delta_0,0}(X',P_4)\,.
\end{split}\end{equation}
Applying $(W_1 \cdot D_{X_1})^{J_1}(W_4 \cdot D_{X_4})^{J_4}$ on both sides gives exactly \eqref{eq:Theta_Omega4_diag}.

\subsection{Proof of \eqref{eq:Theta_loop} and bulk regularization} \label{app:bulk_dive}

We start by considering the completeness relation \eqref{completelyugly}, which we wrote diagrammatically in \eqref{eq:comp_rel_witten_diag}. Suppose that we connect $P_2$ to a bulk point $X_2$ with a bulk-to-boundary propagator $G_{\tilde\Delta_0,0}$, that we connect $P_4$ to a bulk point $X_1$ with a bulk-to-boundary propagator $G_{\Delta_0,0}$, that we identify $P_1=P_3$, and that we integrate over $P_1, P_2$ and $P_4$. Finally we also operate $ (W_1 \cdot D_{X_1})^{J_1}$ on both sides. The equality we then obtain takes the form
\begin{equation}
	\vcenter{\hbox{\includegraphics{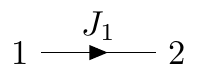}}} \cdot ``\int dP \delta(0)'' = 
	\vcenter{\hbox{\includegraphics{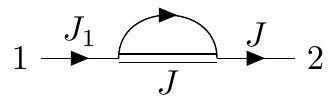}}}
	\,.\label{eq:Theta_loop_proof}
\end{equation}
Evidently, this equality needs a regularization. The $d$-dimensional integral on the left-hand side is the same one that we obtained when we discussed loop integrals in the bi-local formalism in section \ref{sec:bilocal_1_loop}. There it was regularized by putting the theory on a finite lattice, and it was set equal to the number of lattice points $V$ (which then appeared in our counter-terms). More generally, with some high-momentum cutoff $|k|<\frac{1}{\varepsilon}$ and an IR cutoff $|x|<L$, this integral would be regularized to $V\sim L^d/\varepsilon^d$. The bulk integrals on the right-hand side similarly diverge in the UV and in the IR (after the implicit summation over $J$). We can regularize them in the bulk by introducing a cutoff on the radial direction $z>\varepsilon$, and an IR cutoff for the space-time directions $|x|<L$. We then expect the regularized integral to give \eqref{eq:Theta_loop} with $V\sim L^d/\varepsilon^d$, up to corrections that go to zero when we take $\varepsilon \to 0$ and $L\to \infty$. We stress that the sum over the internal spin $J$ needs to be taken before we remove the regulators. The reason is that the dominant spins in the sum with a given regulator scale as $J\sim \frac{L^2}{\eps |x_{12}|}$ \cite{Komargodski:2012ek}. Since all loop divergences in our computations reduce to \eqref{eq:Theta_loop}, this regularization scheme should regularize systematically all the bulk integrals.



\subsection{Modifications for $d<4$} \label{app:d4_mod}

In this appendix we discuss how the arguments of section \ref{sec:bulk_theory} must be modified for $d<4$. The only modification of the inverse mapping is for $J=0$, as given by \eqref{eq:ads_to_cft_free}:
\begin{equation}
\begin{split}
	{\cal M}_{0}^{-1}\left(P_{1},P_{2}\mid X\right) = \int_{\gamma_0}\frac{d\Delta}{2\pi i} 
	\frac{1}{\alpha_0 N_{\Delta,0} \lambda_{\Delta,0} f_{\Delta,0}}
	\int dP 
	\thpnorm{P_1}{P_2}{\Delta,0}{P}
	G_{\tilde{\Delta},0}\left(X,P\right) \\
	\times 
	\left(\nabla_{X}^{2}-M_{d-2,0}^{2}\right)\left(\nabla_{X}^{2}-M_{d,0}^{2}\right)\,.
\end{split}
\end{equation}
Using \eqref{eq:penedones_3point} we can write it as
\begin{equation}
	{\cal M}_{0}^{-1}\left(P_{1},P_{2}\mid X\right) = \int dX'
	G_{\Delta_0,0}(X',P_2) G_{\Delta_0,0}(X',P_1) \theta_0(X,X')\,,
\end{equation}
where we now define (different than \eqref{eq:theta_def})
\begin{equation} \label{eq:d4_theta0}
	\theta_0(X',X)=
	\int_{\gamma_0}\frac{d\Delta}{2\pi i} 
	\frac{1}{\lambda_{\Delta,0} f_{\Delta,0}b^{(\Delta,0)}_{\Delta_0,\Delta_0}}
	\Omega_{\Delta}(X',X)
	\left(\nabla_{X}^{2}-M_{d-2,0}^{2}\right)\left(\nabla_{X}^{2}-M_{d,0}^{2}\right)\,.
\end{equation}
Notice that now $\theta_0$ (and the mappings themselves) is not merely a function (or a distribution) but a differential operator.
For $J>0$ the inverse mapping remains \eqref{eq:ads_to_cft}. 

Next, consider the direct mapping, which for $d<4$ is given by \eqref{eq:free_cft_to_ads_2}
\begin{equation}
\begin{split}
	{\cal M}_{J}\left(X,W \mid P_{1},P_{2}\right)
	& = \frac12 \int_{\gamma_J} \frac{d\Delta}{2\pi i}  \int \frac{dP}{J!\left(\frac{d}{2}-1\right)_{J}} \frac{f_{\Delta,J}}{\lambda_{\Delta,J}N_{\Delta,J}} \\
		& \times  
		G_{\Delta,J}\left(X,P;W,D_Z\right)
		\thpnorm{P_1}{P_2}{\tilde\Delta,J}{P,Z} \, \nabla^2_{P_1}\nabla^2_{P_2}
		.
\end{split}
\end{equation}
Again, using \eqref{eq:penedones_3point} we can write this as
\begin{equation}
\begin{split}
	{\cal M}_{J}\left(X,W \mid P_{1},P_{2}\right)
	& = \int \frac{dX'}{J! \left(\frac{d-1}{2}\right)_J} G_{\Delta_0,0}(X',P_2) (K_{W'}\cdot \nabla_{X'})^J G_{\Delta_0,0}(X',P_1)
	\tilde\theta_J(X,X';W,W') \nabla^2_{P_1}\nabla^2_{P_2}\,,
\end{split}
\end{equation}
where we now define
\begin{equation}
	\tilde\theta_J(X,X';W,W') = \int_{\gamma_J} \frac{d\Delta}{2\pi i}  \frac{\alpha_J f_{\Delta,J}}{2b_{\Delta_0,\Delta_0}^{(\tilde\Delta,J)} \lambda_{\Delta,J}} \\
		\Omega_{\Delta,J}\left(X,X';W,W'\right).
\end{equation}

We would like to show that all the identities of section \ref{sec:bulk_theory} have their version also for these new $\theta,\tilde\theta$. We start with the definition of $\Theta_J$ \eqref{eq:Theta_def_diag}. For $J=0$ we now have
\begin{equation} \label{eq:ordered}
\begin{split}
	\Theta_0(X,X') &= \int dY \theta_0(X,Y) \tilde\theta_0(Y,X') \\
	&=  \int dY \int_{\gamma_0}\frac{d\Delta}{2\pi i} 
	\frac{1}{\lambda_{\Delta,0} f_{\Delta,0}b^{(\Delta,0)}_{\Delta_0,\Delta_0}}
	\Omega_{\Delta,0}(X,Y)
	\left(\nabla_{Y}^{2}-M_{d-2,0}^{2}\right)\left(\nabla_{Y}^{2}-M_{d,0}^{2}\right) \\
	& \quad\times \int_{\gamma_0} \frac{d\Delta'}{2\pi i}  \frac{\alpha_0 f_{\Delta',0}}{2b_{\Delta_0,\Delta_0}^{(\tilde\Delta',0)} \lambda_{\Delta',0}}
		\Omega_{\Delta',0}\left(Y,X'\right)\\
	& = \int_{\gamma_0}\frac{d\Delta}{2\pi i} 
	\frac{\alpha_0}{\lambda_{\Delta,0} 2b_{\Delta_0,\Delta_0}^{(\tilde\Delta',0)}b^{(\Delta,0)}_{\Delta_0,\Delta_0}}
	\Omega_{\Delta,0}(X,X')\,.
\end{split}
\end{equation}
Note that $\theta_0$ and $\tilde\theta_0$ are now operators, so they now do not obviously commute as they did for $d>4$ in \eqref{eq:Theta_def_diag}. We will use the specific ordering of \eqref{eq:ordered}.
For $J>0$ we have
\begin{equation}
\begin{split}
	\Theta_J(X_1,X_2;W_1,W_2) & = \int  \frac{dX}{J! \left(\frac{d-1}{2}\right)_J} \theta_J(X_1,X;W_1,K_{W}) {\tilde \theta}_J(X,X_2;W,W_2)\\
	& = 
	\int_{\gamma_J}\frac{d\Delta}{2\pi i} 
	\frac{\alpha_J}{2 \lambda_{\Delta,J} b_{\Delta_0,\Delta_0}^{(\tilde\Delta,J)} b^{(\Delta,J)}_{\Delta_0,\Delta_0} } 
	\Omega_{\Delta,J}(X_1,X_2;W_1,W_2)\,,
\end{split}
\end{equation}
which is similar to the $J=0$ case.

Next, we would like to show the $d<4$ version of \eqref{eq:invertibility}:
\begin{equation}\begin{split}
	\delta^{TT} & (X_1,X_2) W_{12}^{J_1} \delta_{J_1,J_2} = \int dP_1 \int dP_2 \ {\cal M}_{J_1}\left(X_1,W_1 | P_1,P_2 \right) 
	{\cal M}_{J_2}^{-1}\left(P_1,P_2 | X_2,W_2\right)\\
	& = 
	\int dP_1 \int dP_2
	\int \frac{dX_1'}{{J_1}! \left(\frac{d-1}{2}\right)_{J_1}} 
	G_{\Delta_0,0}(X_1',P_2) (K_{W_1'}\cdot \nabla_{X_1'})^{J_1} G_{\Delta_0,0}(X_1',P_1)
	\tilde\theta_{J_1}(X_1,X_1';W_1,W_1') \\
	& \nabla^2_1\nabla^2_2
	\int  \frac{dX_2' }{J_2! \left(\frac{d-1}{2}\right)_{J_2}} (K_{W_2'} \cdot \nabla_{X_2'})^J G_{\Delta_0,0}(X_2',P_1) G_{\Delta_0,0}(X_2',P_2) 
	\theta_{J_2}(X_2',X_2;W_2',W_2)\\
	& = a_{\Delta_0}^2\int dP_1 \int dP_2
	\int \frac{dX_1'}{{J_1}! \left(\frac{d-1}{2}\right)_{J_1}} 
	G_{\Delta_0,0}(X_1',P_2) (K_{W'_1}\cdot \nabla_{X_1'})^{J_1} G_{\Delta_0,0}(X_1',P_1)
	\tilde\theta_{J_1}(X_1,X_1';W_1,W_1') \\
	& \int  \frac{dX_2' }{J_2! \left(\frac{d-1}{2}\right)_{J_2}} (K_{W_2'} \cdot \nabla_{X_2'})^J G_{\tilde\Delta_0,0}(X_2',P_1) G_{\tilde\Delta_0,0}(X_2',P_2) 
	\theta_{J_2}(X_2',X_2;W_2',W_2)
	\\
	& =a_{\Delta_0}^2
	\int \frac{dX_1'}{{J_1}! \left(\frac{d-1}{2}\right)_{J_1}}
	\int  \frac{dX_2' }{J_2! \left(\frac{d-1}{2}\right)_{J_2}} 
	(K_{W_2'} \cdot \nabla_{X_2'})^J  \Omega_{0}(X_2',X_1')
	(K_{W_1'}\cdot \nabla_{X_1'})^{J_1} \Omega_{0}(X_2',X_1')
	 \\
	&\tilde\theta_{J_1}(X_1,X_1';W_1,W_1')
	\theta_{J_2}(X_2',X_2;W_2',W_2)\,,
\end{split}\end{equation}
where we used \eqref{eq:a_def} in the last line. We see a factor of $a_{\Delta_0}^2$ compared to \eqref{eq:invertibility}. 

We can now derive the $d<4$ version of \eqref{eq:Theta_Omega4_diag} as we did in the previous subsection for $d>4$, by starting with \eqref{eq:3p_comp_rel} and then applying \eqref{eq:penedones_3point}. The difference for $d<4$ is that now the completeness relation for $d<4$ is altered \eqref{eq:conf_comp_rel_free}, so that we now have
\begin{equation}\label{eq:d4_Theta_Omega}
\begin{split}
	\delta(P_1,P_3)\delta(P_2,P_4) =& \sum_{J=0}^\infty \int_{\gamma_J}\frac{d\Delta}{2\pi i} \int 
	\frac{dP}{J! \left(\frac{d}{2}-1\right)_J}  \, 	
	\frac{1}{2N_{\Delta,J}\lambda_{\Delta,J}}\cdot  \\ & \qquad
	\thpnorm{P_1}{P_2}{\Delta,J}{P,D_Z}
	\thpnorm{P_3}{P_4}{\tilde\Delta,J}{P,Z}
	 \nabla_{P_3}^{2}\nabla_{P_4}^{2}\\
	 &= \sum_{J=0}^\infty \int \frac{dX_1'}{J! \left(\frac{d-1}{2}\right)_J} 
	\int \frac{dX_2'}{J! \left(\frac{d-1}{2}\right)_J} 
	 \Theta_J(X_1',X_2'; K_{W_1'},K_{W_2'})\\
	& \quad G_{\Delta_0,0}(X_1',P_1) (W_1'\cdot \nabla_{X_1'})^J G_{\Delta_0,0}(X_1',P_2)\\
	& \quad
	G_{\Delta_0,0}(X_2',P_4)
	(W_2'\cdot \nabla_{X_2'})^J G_{\Delta_0,0}(X_2',P_3)
	\nabla_{P_3}^{2}\nabla_{P_4}^{2}\,.
\end{split}
\end{equation}
Multiplying by $G_{\tilde\Delta_0,0}(X_1,P_1) G_{\tilde\Delta_0,0}(X_2,P_2) G_{\Delta_0,0}(X_3,P_3) G_{\Delta_0,0}(X_4,P_4)$ and integrating over $P_1,P_2,P_3,P_4$ gives (using \eqref{eq:a_def})
\begin{equation}
\begin{split}
	\Omega_0(X_1,X_3) \Omega_0(X_2,X_4) &= a_{\Delta_0}^2  \sum_{J=0}^\infty \int \frac{dX_1'}{J! \left(\frac{d-1}{2}\right)_J} 
	\int \frac{dX_2'}{J! \left(\frac{d-1}{2}\right)_J} 
	 \Theta_J(X_1',X_2'; K_{W_1'},K_{W_2'})\\
	& \quad \Omega_{0}(X_1',X_1) (W_1'\cdot \nabla_{X_1'})^J \Omega_{0}(X_1',X_2)\\
	& \quad	\Omega_0(X_2',X_4) (W_2'\cdot \nabla_{X_2'})^J \Omega_{0}(X_2',X_3)\,.
\end{split}
\end{equation}
We can then apply $(W_1 \cdot D_{X_1})^{J_1}(W_4 \cdot D_{X_4})^{J_4}$ on both sides to get  $a_{\Delta_0}^2\cdot$\eqref{eq:Theta_Omega4_diag} (with our modified $\Theta$).

Next, we can derive the modified version of \eqref{eq:Theta_loop} for $d<4$ by multiplying \eqref{eq:d4_Theta_Omega} by
$G_{\tilde{\Delta}_{0}}\left(X_{2},P_{2}\right)
G_{\Delta_{0}}\left(X_{1},P_4\right)
\delta\left(P_{1},P_{3}\right)$ and integrating over $P_1,P_2,P_3,P_4$ to get
\begin{equation}
\begin{split}
	\Omega_{0}\left(X_{1},X_{2}\right) V &= \sum_{J=0}^\infty \int \frac{dX'}{J! \left(\frac{d-1}{2}\right)_J} 
	\int \frac{dX}{J! \left(\frac{d-1}{2}\right)_J} 
	 \Theta_J(X',X;K_{W'},K_{W})\\
	& \quad \int dP_1 dP_2 dP_3 dP_4 G_{\Delta_0,0}(X',P_1) (W'\cdot \nabla_{X'})^J G_{\Delta_0,0}(X',P_2)\\
	& \quad
	G_{\Delta_0,0}(X,P_4)
	(W\cdot \nabla_{X})^J G_{\Delta_0,0}(X,P_3)\\
	& \quad\nabla_{P_3}^{2}\nabla_{P_4}^{2} 
	\left(G_{\tilde{\Delta}_{0},0}\left(X_{2},P_{2}\right)
G_{\Delta_{0},0}\left(X_{1},P_4\right)
\delta\left(P_{1},P_{3}\right)\right)\\
	& = a_{\Delta_0}\cdot \sum_{J=0}^\infty \int \frac{dX'}{J! \left(\frac{d-1}{2}\right)_J}
	\int \frac{dX}{J! \left(\frac{d-1}{2}\right)_J} 
	 \Theta_J(X',X;K_{W'},K_{W})\\
	& \quad   (W'\cdot \nabla_{X'})^J \Omega_{0}(X_2,X') \Omega_{0}(X_1,X)\\
	& \quad \int dP_1 dP_3
	(W\cdot \nabla_{X})^J G_{\Delta_0,0}(X,P_3)
	G_{\Delta_{0},0}\left(X',P_1\right)
	\nabla_{P_3}^{2}\delta\left(P_{1},P_{3}\right)\,,
\end{split}
\end{equation}
which we can integrate by parts to get the right-hand side
\begin{equation}
\begin{split}
	a_{\Delta_0}^2\cdot \sum_{J=0}^\infty \int \frac{dX'}{J! \left(\frac{d-1}{2}\right)_J} 
	\int \frac{dX}{J! \left(\frac{d-1}{2}\right)_J} 
	 \Theta_J(X',X;K_{W'},K_{W})\\
	\quad   (W'\cdot \nabla_{X'})^J \Omega_{0}(X_2,X') \Omega_{0}(X_1,X)
	(W\cdot \nabla_{X})^J \Omega_{0}(X,X')\,.
\end{split}
\end{equation}
Applying $(W_1\cdot D_{X_1})^{J_1}$ to both sides then gives $a_{\Delta_0}^2$ times \eqref{eq:Theta_loop}.

The two-point function in the bulk stays the same as discussed in section \ref{sec:mapping_to_the_bulk}. The Feynman rules are produced by applying the inverse mapping to the bi-local action. Written in terms of our modified $\theta_J,\tilde\theta_J$ it has the same form (the same diagrams). Because the operators we defined satisfy the same conditions (up to the $a_{\Delta_0}^2$ that cancel between the uses of the identities) we get the same cancellations, as in section  \ref{sec:bulk_theory}.

We would like to write exactly the same Feynman rules as we found for $d>4$. The only modification of the vertex has to do with the modification of $\theta_J$ described above, where only the $J=0$ term \eqref{eq:d4_theta0} was changed by acting with $\left(\nabla_{X}^{2}-M_{d-2,0}^{2}\right)\left(\nabla_{X}^{2}-M_{d,0}^{2}\right)$ and inserting a $1/\lambda_{\Delta,J}$ in the $\Delta$ integral. The question then is whether we are allowed to perform integration-by-parts to get the original Feynman rules. Inside a Feynman diagram computation, the differential operator will act on a bulk 2-point function from the Feynman rules. Applying the operator we would get exactly a  bulk delta-function, so the bulk integral is trivial, and we will have the same properties as for $d>4$. On the other hand, if we try to perform an integration by parts directly on the vertex (to bring it to the same form as the $d>4$ vertex), we obtain when applying the Feynman rules an integral involving $\int dX\Omega_{\Delta}(X',X)\Pi_{d-2}(X,X'')$, which in the small $z$ limit behaves as $z^{d/2+d-2-d}\sim z^{(d-4)/2}$, and so diverges for $d<4$.
So the difference between the Feynman rules of $d<4$ and those of $d>4$ are the diverging boundary terms from the integration by parts. We can think about them as UV counter terms, made to regularize the naive action because of the unconventional boundary condition for $\Phi_0$ in this case. The bottom line is that we need to use for $d<4$ the modified Feynman rules of this section, but the expressions we get after using them are the same as we obtain by continuing the $d>4$ results to $d<4$.  

As explained in section \ref{sec:bulk_deform}, in the dual to the critical theory the boundary condition for $\Phi_0$ (with $d<4$) does not allow for a $z^{d-2}$ term. Hence, in this case no divergences will appear when integrating by parts the action, so we can simply use precisely the same bulk action and Feynman rules as in section \ref{sec:bulk_theory}.

\bibliographystyle{ssg}
\bibliography{ref}

\begingroup\raggedright\begin{thebibliography}{10}

\bibitem{Maldacena:1997re}
J.~M. Maldacena, ``{The Large N limit of superconformal field theories and
  supergravity},'' {\em Int. J. Theor. Phys.} {\bf 38} (1999) 1113--1133,
  \href{https://arxiv.org/abs/hep-th/9711200}{{\tt hep-th/9711200}}. [Adv.
  Theor. Math. Phys.2,231(1998)].

\bibitem{Witten:1998qj}
E.~Witten, ``{Anti-de Sitter space and holography},'' {\em Adv. Theor. Math.
  Phys.} {\bf 2} (1998) 253--291,
  \href{https://arxiv.org/abs/hep-th/9802150}{{\tt hep-th/9802150}}.

\bibitem{Gubser:1998bc}
S.~Gubser, I.~R. Klebanov, and A.~M. Polyakov, ``{Gauge theory correlators from
  noncritical string theory},'' {\em Phys. Lett. B} {\bf 428} (1998) 105--114,
  \href{https://arxiv.org/abs/hep-th/9802109}{{\tt hep-th/9802109}}.

\bibitem{Chester:2019jas}
S.~M. Chester, M.~B. Green, S.~S. Pufu, Y.~Wang, and C.~Wen, ``{Modular
  Invariance in Superstring Theory From ${\cal N} = 4$ Super-Yang-Mills},''
  \href{https://arxiv.org/abs/1912.13365}{{\tt 1912.13365}}.

\bibitem{Binder:2019mpb}
D.~J. Binder, S.~M. Chester, and S.~S. Pufu, ``{AdS$_{4}$/CFT$_{3}$ from weak
  to strong string coupling},'' {\em JHEP} {\bf 01} (2020) 034,
  \href{https://arxiv.org/abs/1906.07195}{{\tt 1906.07195}}.

\bibitem{Gromov:2009zb}
N.~Gromov, V.~Kazakov, and P.~Vieira, ``{Exact Spectrum of Planar ${\cal N}=4$
  Supersymmetric Yang-Mills Theory: Konishi Dimension at Any Coupling},'' {\em
  Phys. Rev. Lett.} {\bf 104} (2010) 211601,
  \href{https://arxiv.org/abs/0906.4240}{{\tt 0906.4240}}.

\bibitem{Eberhardt:2019ywk}
L.~Eberhardt, M.~R. Gaberdiel, and R.~Gopakumar, ``{Deriving the
  AdS$_{3}$/CFT$_{2}$ correspondence},'' {\em JHEP} {\bf 02} (2020) 136,
  \href{https://arxiv.org/abs/1911.00378}{{\tt 1911.00378}}.

\bibitem{Dei:2020zui}
A.~Dei, M.~R. Gaberdiel, R.~Gopakumar, and B.~Knighton, ``{Free field
  world-sheet correlators for ${\rm AdS}_3$},''
  \href{https://arxiv.org/abs/2009.11306}{{\tt 2009.11306}}.

\bibitem{Eberhardt:2020akk}
L.~Eberhardt, ``{AdS$_{3}$/CFT$_{2}$ at higher genus},'' {\em JHEP} {\bf 05}
  (2020) 150, \href{https://arxiv.org/abs/2002.11729}{{\tt 2002.11729}}.

\bibitem{Saad:2019lba}
P.~Saad, S.~H. Shenker, and D.~Stanford, ``{JT gravity as a matrix integral},''
  \href{https://arxiv.org/abs/1903.11115}{{\tt 1903.11115}}.

\bibitem{Almheiri:2019qdq}
A.~Almheiri, T.~Hartman, J.~Maldacena, E.~Shaghoulian, and A.~Tajdini,
  ``{Replica Wormholes and the Entropy of Hawking Radiation},'' {\em JHEP} {\bf
  05} (2020) 013, \href{https://arxiv.org/abs/1911.12333}{{\tt 1911.12333}}.

\bibitem{Klebanov:2002ja}
I.~R. Klebanov and A.~M. Polyakov, ``{AdS dual of the critical O(N) vector
  model},'' {\em Phys. Lett.} {\bf B550} (2002) 213--219,
  \href{https://arxiv.org/abs/hep-th/0210114}{{\tt hep-th/0210114}}.

\bibitem{Giombi:2009wh}
S.~Giombi and X.~Yin, ``{Higher Spin Gauge Theory and Holography: The
  Three-Point Functions},'' {\em JHEP} {\bf 09} (2010) 115,
  \href{https://arxiv.org/abs/0912.3462}{{\tt 0912.3462}}.

\bibitem{Giombi:2011kc}
S.~Giombi, S.~Minwalla, S.~Prakash, S.~P. Trivedi, S.~R. Wadia, and X.~Yin,
  ``{Chern-Simons Theory with Vector Fermion Matter},'' {\em Eur. Phys. J. C}
  {\bf 72} (2012) 2112, \href{https://arxiv.org/abs/1110.4386}{{\tt
  1110.4386}}.

\bibitem{Aharony:2011jz}
O.~Aharony, G.~Gur-Ari, and R.~Yacoby, ``{d=3 Bosonic Vector Models Coupled to
  Chern-Simons Gauge Theories},'' {\em JHEP} {\bf 03} (2012) 037,
  \href{https://arxiv.org/abs/1110.4382}{{\tt 1110.4382}}.

\bibitem{Vasiliev:1990en}
M.~A. Vasiliev, ``{Consistent equation for interacting gauge fields of all
  spins in (3+1)-dimensions},'' {\em Phys. Lett.} {\bf B243} (1990) 378--382.

\bibitem{Vasiliev:1992av}
M.~A. Vasiliev, ``{More on equations of motion for interacting massless fields
  of all spins in (3+1)-dimensions},'' {\em Phys. Lett. B} {\bf 285} (1992)
  225--234.

\bibitem{Vasiliev:1995dn}
M.~A. Vasiliev, ``{Higher spin gauge theories in four-dimensions,
  three-dimensions, and two-dimensions},'' {\em Int. J. Mod. Phys. D} {\bf 5}
  (1996) 763--797, \href{https://arxiv.org/abs/hep-th/9611024}{{\tt
  hep-th/9611024}}.

\bibitem{Bekaert:2014cea}
X.~Bekaert, J.~Erdmenger, D.~Ponomarev, and C.~Sleight, ``{Towards holographic
  higher-spin interactions: Four-point functions and higher-spin exchange},''
  {\em JHEP} {\bf 03} (2015) 170, \href{https://arxiv.org/abs/1412.0016}{{\tt
  1412.0016}}.

\bibitem{Bekaert:2016ezc}
X.~Bekaert, J.~Erdmenger, D.~Ponomarev, and C.~Sleight, ``{Bulk quartic
  vertices from boundary four-point correlators},'' in {\em {International
  Workshop on Higher Spin Gauge Theories}}, pp.~291--303, 2017.
\newblock \href{https://arxiv.org/abs/1602.08570}{{\tt 1602.08570}}.

\bibitem{Skvortsov:2018uru}
E.~Skvortsov, ``{Light-Front Bootstrap for Chern-Simons Matter Theories},''
  {\em JHEP} {\bf 06} (2019) 058, \href{https://arxiv.org/abs/1811.12333}{{\tt
  1811.12333}}.

\bibitem{Sleight:2016dba}
C.~Sleight and M.~Taronna, ``{Higher Spin Interactions from Conformal Field
  Theory: The Complete Cubic Couplings},'' {\em Phys. Rev. Lett.} {\bf 116}
  (2016), no.~18 181602, \href{https://arxiv.org/abs/1603.00022}{{\tt
  1603.00022}}.

\bibitem{Boulanger:2015ova}
N.~Boulanger, P.~Kessel, E.~Skvortsov, and M.~Taronna, ``{Higher spin
  interactions in four-dimensions: Vasiliev versus Fronsdal},'' {\em J. Phys.
  A} {\bf 49} (2016), no.~9 095402,
  \href{https://arxiv.org/abs/1508.04139}{{\tt 1508.04139}}.

\bibitem{Sleight:2017pcz}
C.~Sleight and M.~Taronna, ``{Higher-Spin Gauge Theories and Bulk Locality},''
  {\em Phys. Rev. Lett.} {\bf 121} (2018), no.~17 171604,
  \href{https://arxiv.org/abs/1704.07859}{{\tt 1704.07859}}.

\bibitem{Neiman:2015wma}
Y.~Neiman, ``{Higher-spin gravity as a theory on a fixed (anti) de Sitter
  background},'' {\em JHEP} {\bf 04} (2015) 144,
  \href{https://arxiv.org/abs/1502.06685}{{\tt 1502.06685}}.

\bibitem{Das:2003vw}
S.~R. Das and A.~Jevicki, ``{Large N collective fields and holography},'' {\em
  Phys. Rev.} {\bf D68} (2003) 044011,
  \href{https://arxiv.org/abs/hep-th/0304093}{{\tt hep-th/0304093}}.

\bibitem{Koch:2010cy}
R.~de~Mello~Koch, A.~Jevicki, K.~Jin, and J.~P. Rodrigues, ``{$AdS_4/CFT_3$
  Construction from Collective Fields},'' {\em Phys. Rev. D} {\bf 83} (2011)
  025006, \href{https://arxiv.org/abs/1008.0633}{{\tt 1008.0633}}.

\bibitem{Koch:2014aqa}
R.~de~Mello~Koch, A.~Jevicki, J.~a.~P. Rodrigues, and J.~Yoon, ``{Canonical
  Formulation of $O(N)$ Vector/Higher Spin Correspondence},'' {\em J. Phys. A}
  {\bf 48} (2015), no.~10 105403, \href{https://arxiv.org/abs/1408.4800}{{\tt
  1408.4800}}.

\bibitem{Koch:2014mxa}
R.~de~Mello~Koch, A.~Jevicki, J.~P. Rodrigues, and J.~Yoon, ``{Holography as a
  Gauge Phenomenon in Higher Spin Duality},'' {\em JHEP} {\bf 01} (2015) 055,
  \href{https://arxiv.org/abs/1408.1255}{{\tt 1408.1255}}.

\bibitem{deMelloKoch:2012vc}
R.~de~Mello~Koch, A.~Jevicki, K.~Jin, J.~P. Rodrigues, and Q.~Ye, ``{S=1 in
  O(N)/HS duality},'' {\em Class. Quant. Grav.} {\bf 30} (2013) 104005,
  \href{https://arxiv.org/abs/1205.4117}{{\tt 1205.4117}}.

\bibitem{deMelloKoch:2018ivk}
R.~de~Mello~Koch, A.~Jevicki, K.~Suzuki, and J.~Yoon, ``{AdS Maps and Diagrams
  of Bi-local Holography},'' {\em JHEP} {\bf 03} (2019) 133,
  \href{https://arxiv.org/abs/1810.02332}{{\tt 1810.02332}}.

\bibitem{Dobrev:1977qv}
V.~K. Dobrev, G.~Mack, V.~B. Petkova, S.~G. Petrova, and I.~T. Todorov,
  ``{Harmonic Analysis on the n-Dimensional Lorentz Group and Its Application
  to Conformal Quantum Field Theory},'' {\em Lect. Notes Phys.} {\bf 63} (1977)
  1--280.

\bibitem{Dobrev:1976vr}
V.~K. Dobrev, G.~Mack, I.~T. Todorov, V.~B. Petkova, and S.~G. Petrova, ``{On
  the Clebsch-Gordan Expansion for the Lorentz Group in n Dimensions},'' {\em
  Rept. Math. Phys.} {\bf 9} (1976) 219--246.

\bibitem{Costa:2014kfa}
M.~S. Costa, V.~Gon{\c c}alves, and J.~Penedones, ``{Spinning AdS
  Propagators},'' {\em JHEP} {\bf 09} (2014) 064,
  \href{https://arxiv.org/abs/1404.5625}{{\tt 1404.5625}}.

\bibitem{Hamilton:2006az}
A.~Hamilton, D.~N. Kabat, G.~Lifschytz, and D.~A. Lowe, ``{Holographic
  representation of local bulk operators},'' {\em Phys. Rev. D} {\bf 74} (2006)
  066009, \href{https://arxiv.org/abs/hep-th/0606141}{{\tt hep-th/0606141}}.

\bibitem{Klebanov:1999tb}
I.~R. Klebanov and E.~Witten, ``{AdS / CFT correspondence and symmetry
  breaking},'' {\em Nucl. Phys.} {\bf B556} (1999) 89--114,
  \href{https://arxiv.org/abs/hep-th/9905104}{{\tt hep-th/9905104}}.

\bibitem{Giombi:2011ya}
S.~Giombi and X.~Yin, ``{On Higher Spin Gauge Theory and the Critical O(N)
  Model},'' {\em Phys. Rev. D} {\bf 85} (2012) 086005,
  \href{https://arxiv.org/abs/1105.4011}{{\tt 1105.4011}}.

\bibitem{Shenker:2011zf}
S.~H. Shenker and X.~Yin, ``{Vector Models in the Singlet Sector at Finite
  Temperature},'' \href{https://arxiv.org/abs/1109.3519}{{\tt 1109.3519}}.

\bibitem{Douglas:2010rc}
M.~R. Douglas, L.~Mazzucato, and S.~S. Razamat, ``{Holographic dual of free
  field theory},'' {\em Phys. Rev.} {\bf D83} (2011) 071701,
  \href{https://arxiv.org/abs/1011.4926}{{\tt 1011.4926}}.

\bibitem{Leigh:2014tza}
R.~G. Leigh, O.~Parrikar, and A.~B. Weiss, ``{Holographic geometry of the
  renormalization group and higher spin symmetries},'' {\em Phys. Rev. D} {\bf
  89} (2014), no.~10 106012, \href{https://arxiv.org/abs/1402.1430}{{\tt
  1402.1430}}.

\bibitem{Leigh:2014qca}
R.~G. Leigh, O.~Parrikar, and A.~B. Weiss, ``{Exact renormalization group and
  higher-spin holography},'' {\em Phys. Rev.} {\bf D91} (2015), no.~2 026002,
  \href{https://arxiv.org/abs/1407.4574}{{\tt 1407.4574}}.

\bibitem{Mintun:2014gua}
E.~Mintun and J.~Polchinski, ``{Higher Spin Holography, RG, and the Light
  Cone},'' \href{https://arxiv.org/abs/1411.3151}{{\tt 1411.3151}}.

\bibitem{tomer}
O.~Aharony, S.~M. Chester, T.~Solberg, and E.~Y. Urbach, ``{To Appear},''.

\bibitem{Rosenhaus:2018dtp}
V.~Rosenhaus, ``{An introduction to the SYK model},'' {\em J. Phys. A} {\bf 52}
  (2019) 323001, \href{https://arxiv.org/abs/1807.03334}{{\tt 1807.03334}}.

\bibitem{Jevicki:2016bwu}
A.~Jevicki, K.~Suzuki, and J.~Yoon, ``{Bi-Local Holography in the SYK Model},''
  {\em JHEP} {\bf 07} (2016) 007, \href{https://arxiv.org/abs/1603.06246}{{\tt
  1603.06246}}.

\bibitem{Giombi:2013fka}
S.~Giombi and I.~R. Klebanov, ``{One Loop Tests of Higher Spin AdS/CFT},'' {\em
  JHEP} {\bf 12} (2013) 068, \href{https://arxiv.org/abs/1308.2337}{{\tt
  1308.2337}}.

\bibitem{Giombi:2014iua}
S.~Giombi, I.~R. Klebanov, and B.~R. Safdi, ``{Higher Spin AdS$_{d+1}$/CFT$_d$
  at One Loop},'' {\em Phys. Rev.} {\bf D89} (2014), no.~8 084004,
  \href{https://arxiv.org/abs/1401.0825}{{\tt 1401.0825}}.

\bibitem{Skvortsov:2017ldz}
E.~D. Skvortsov and T.~Tran, ``{AdS/CFT in Fractional Dimension and Higher Spin
  Gravity at One Loop},'' {\em Universe} {\bf 3} (2017), no.~3 61,
  \href{https://arxiv.org/abs/1707.00758}{{\tt 1707.00758}}.

\bibitem{Anninos:2011ui}
D.~Anninos, T.~Hartman, and A.~Strominger, ``{Higher Spin Realization of the
  dS/CFT Correspondence},'' {\em Class. Quant. Grav.} {\bf 34} (2017), no.~1
  015009, \href{https://arxiv.org/abs/1108.5735}{{\tt 1108.5735}}.

\bibitem{Jevicki:1979mb}
A.~Jevicki and B.~Sakita, ``{The Quantum Collective Field Method and Its
  Application to the Planar Limit},'' {\em Nucl. Phys.} {\bf B165} (1980) 511.

\bibitem{Maldacena:2011jn}
J.~Maldacena and A.~Zhiboedov, ``{Constraining Conformal Field Theories with A
  Higher Spin Symmetry},'' {\em J. Phys.} {\bf A46} (2013) 214011,
  \href{https://arxiv.org/abs/1112.1016}{{\tt 1112.1016}}.

\bibitem{Binder:2019zqc}
D.~J. Binder and S.~Rychkov, ``{Deligne Categories in Lattice Models and
  Quantum Field Theory, or Making Sense of $O(N)$ Symmetry with Non-integer
  $N$},'' \href{https://arxiv.org/abs/1911.07895}{{\tt 1911.07895}}.

\bibitem{Karateev:2018oml}
D.~Karateev, P.~Kravchuk, and D.~Simmons-Duffin, ``{Harmonic Analysis and Mean
  Field Theory},'' {\em JHEP} {\bf 10} (2019) 217,
  \href{https://arxiv.org/abs/1809.05111}{{\tt 1809.05111}}.

\bibitem{Simmons-Duffin:2017nub}
D.~Simmons-Duffin, D.~Stanford, and E.~Witten, ``{A spacetime derivation of the
  Lorentzian OPE inversion formula},'' {\em JHEP} {\bf 07} (2018) 085,
  \href{https://arxiv.org/abs/1711.03816}{{\tt 1711.03816}}.

\bibitem{Caron-Huot:2017vep}
S.~Caron-Huot, ``{Analyticity in Spin in Conformal Theories},'' {\em JHEP} {\bf
  09} (2017) 078, \href{https://arxiv.org/abs/1703.00278}{{\tt 1703.00278}}.

\bibitem{Craigie:1983fb}
N.~Craigie, V.~Dobrev, and I.~Todorov, ``{Conformally Covariant Composite
  Operators in Quantum Chromodynamics},'' {\em Annals Phys.} {\bf 159} (1985)
  411--444.

\bibitem{Gaberdiel:2010ar}
M.~R. Gaberdiel, R.~Gopakumar, and A.~Saha, ``{Quantum $W$-symmetry in
  $AdS_3$},'' {\em JHEP} {\bf 02} (2011) 004,
  \href{https://arxiv.org/abs/1009.6087}{{\tt 1009.6087}}.

\bibitem{Gaberdiel:2010xv}
M.~R. Gaberdiel, D.~Grumiller, and D.~Vassilevich, ``{Graviton 1-loop partition
  function for 3-dimensional massive gravity},'' {\em JHEP} {\bf 11} (2010)
  094, \href{https://arxiv.org/abs/1007.5189}{{\tt 1007.5189}}.

\bibitem{Gupta:2012he}
R.~K. Gupta and S.~Lal, ``{Partition Functions for Higher-Spin theories in
  AdS},'' {\em JHEP} {\bf 07} (2012) 071,
  \href{https://arxiv.org/abs/1205.1130}{{\tt 1205.1130}}.

\bibitem{Giombi:2016ejx}
S.~Giombi, ``{Higher Spin --- CFT Duality},'' in {\em {Proceedings, Theoretical
  Advanced Study Institute in Elementary Particle Physics: New Frontiers in
  Fields and Strings (TASI 2015): Boulder, CO, USA, June 1-26, 2015}},
  pp.~137--214, 2017.
\newblock \href{https://arxiv.org/abs/1607.02967}{{\tt 1607.02967}}.

\bibitem{Freedman:1998tz}
D.~Z. Freedman, S.~D. Mathur, A.~Matusis, and L.~Rastelli, ``{Correlation
  functions in the CFT(d) / AdS(d+1) correspondence},'' {\em Nucl. Phys. B}
  {\bf 546} (1999) 96--118, \href{https://arxiv.org/abs/hep-th/9804058}{{\tt
  hep-th/9804058}}.

\bibitem{Iazeolla:2007wt}
C.~Iazeolla, E.~Sezgin, and P.~Sundell, ``{Real forms of complex higher spin
  field equations and new exact solutions},'' {\em Nucl. Phys. B} {\bf 791}
  (2008) 231--264, \href{https://arxiv.org/abs/0706.2983}{{\tt 0706.2983}}.

\bibitem{Sezgin:2005hf}
E.~Sezgin and P.~Sundell, ``{On an exact cosmological solution of higher spin
  gauge theory},'' {\em Bulg. J. Phys.} {\bf 33} (2006), no.~s1 506--519,
  \href{https://arxiv.org/abs/hep-th/0511296}{{\tt hep-th/0511296}}.

\bibitem{deMelloKoch:1996mj}
R.~de~Mello~Koch and J.~P. Rodrigues, ``{Systematic 1/N corrections for bosonic
  and fermionic vector models without auxiliary fields},'' {\em Phys. Rev. D}
  {\bf 54} (1996) 7794--7814, \href{https://arxiv.org/abs/hep-th/9605079}{{\tt
  hep-th/9605079}}.

\bibitem{Mulokwe:2018czu}
M.~Mulokwe and J.~a.~P. Rodrigues, ``{Large N bilocals at the infrared fixed
  point of the three dimensional O(N) invariant vector theory with a quartic
  interaction},'' {\em JHEP} {\bf 11} (2018) 047,
  \href{https://arxiv.org/abs/1808.00042}{{\tt 1808.00042}}.

\bibitem{Gubser:2002vv}
S.~S. Gubser and I.~R. Klebanov, ``{A Universal result on central charges in
  the presence of double trace deformations},'' {\em Nucl. Phys. B} {\bf 656}
  (2003) 23--36, \href{https://arxiv.org/abs/hep-th/0212138}{{\tt
  hep-th/0212138}}.

\bibitem{Fei:2014yja}
L.~Fei, S.~Giombi, and I.~R. Klebanov, ``{Critical $O(N)$ models in
  $6-\epsilon$ dimensions},'' {\em Phys. Rev. D} {\bf 90} (2014), no.~2 025018,
  \href{https://arxiv.org/abs/1404.1094}{{\tt 1404.1094}}.

\bibitem{Hartman:2006dy}
T.~Hartman and L.~Rastelli, ``{Double-trace deformations, mixed boundary
  conditions and functional determinants in AdS/CFT},'' {\em JHEP} {\bf 01}
  (2008) 019, \href{https://arxiv.org/abs/hep-th/0602106}{{\tt
  hep-th/0602106}}.

\bibitem{Costa:2011mg}
M.~S. Costa, J.~Penedones, D.~Poland, and S.~Rychkov, ``{Spinning Conformal
  Correlators},'' {\em JHEP} {\bf 11} (2011) 071,
  \href{https://arxiv.org/abs/1107.3554}{{\tt 1107.3554}}.

\bibitem{Komargodski:2012ek}
Z.~Komargodski and A.~Zhiboedov, ``{Convexity and Liberation at Large Spin},''
  {\em JHEP} {\bf 11} (2013) 140, \href{https://arxiv.org/abs/1212.4103}{{\tt
  1212.4103}}.

\end{thebibliography}\endgroup
\end{document}